\documentclass[aps,prb,twocolumn,superscriptaddress]{revtex4-2}
\usepackage[dvipdfmx]{graphicx}
\usepackage{color}
\usepackage{ulem}
\begin{document}

\title{Large anomalous Hall effect induced by weak ferromagnetism \\in the noncentrosymmetric antiferromagnet $\mathbf{Co}\mathbf{Nb}_3\mathbf{S}_6$}

\author{Hiroaki Tanaka}
\affiliation{Institute for Solid State Physics, The University of Tokyo, Kashiwa, Chiba 277-8581, Japan}

\author{Shota Okazaki}
\affiliation{Materials and Structures Laboratory, Tokyo Institute of Technology, Yokohama, Kanagawa 226-8503, Japan}

\author{Kenta Kuroda}
\email{kuroken224@hiroshima-u.ac.jp}
\affiliation{Institute for Solid State Physics, The University of Tokyo, Kashiwa, Chiba 277-8581, Japan}

\author{Ryo Noguchi}
\affiliation{Institute for Solid State Physics, The University of Tokyo, Kashiwa, Chiba 277-8581, Japan}
\affiliation{Center for Correlated Electron Systems, Institute for Basic Science, Seoul 08826, Republic of Korea}
\affiliation{Department of Physics and Astronomy, Seoul National University, Seoul 08826, Republic of Korea}

\author{Yosuke Arai}
\affiliation{Institute for Solid State Physics, The University of Tokyo, Kashiwa, Chiba 277-8581, Japan}

\author{Susumu Minami}
\affiliation{Department of Physics, The University of Tokyo, Tokyo 113-0033, Japan}

\author{Shinichiro Ideta}
\affiliation{UVSOR Facility, Institute for Molecular Science, Okazaki, Aichi 444-8585, Japan}

\author{Kiyohisa Tanaka}
\affiliation{UVSOR Facility, Institute for Molecular Science, Okazaki, Aichi 444-8585, Japan}

\author{Donghui Lu}
\affiliation{Stanford Synchrotron Radiation Lightsource, SLAC National Accelerator Laboratory, Menlo Park, CA 94025, United States}

\author{Makoto Hashimoto}
\affiliation{Stanford Synchrotron Radiation Lightsource, SLAC National Accelerator Laboratory, Menlo Park, CA 94025, United States}

\author{Viktor Kandyba}
\affiliation{Elettra Sincrotrone Trieste, 34149 Basovizza, Trieste, Italy}

\author{Mattia Cattelan}
\affiliation{Elettra Sincrotrone Trieste, 34149 Basovizza, Trieste, Italy}

\author{Alexei Barinov}
\affiliation{Elettra Sincrotrone Trieste, 34149 Basovizza, Trieste, Italy}

\author{Takayuki Muro}
\affiliation{Japan Synchrotron Radiation Research Institute (JASRI), Sayo, Hyogo 679-5198, Japan}

\author{Takao Sasagawa}
\email{sasagawa@msl.titech.ac.jp}
\affiliation{Materials and Structures Laboratory, Tokyo Institute of Technology, Yokohama, Kanagawa 226-8503, Japan}

\author{Takeshi Kondo}
\email{kondo1215@issp.u-tokyo.ac.jp}
\affiliation{Institute for Solid State Physics, The University of Tokyo, Kashiwa, Chiba 277-8581, Japan}
\affiliation{Trans-scale Quantum Science Institute, The University of Tokyo, Tokyo 113-0033, Japan}

\date{\today}

\begin{abstract}
We study the mechanism of the exceptionally large anomalous Hall effect (AHE) in the noncentrosymmetric antiferromagnet $\mathrm{Co}\mathrm{Nb}_3\mathrm{S}_6$ by angle-resolved photoemission spectroscopy (ARPES) and magnetotransport measurements.
From ARPES measurements of $\mathrm{Co}\mathrm{Nb}_3\mathrm{S}_6$ and its family compounds ($\mathrm{Fe}\mathrm{Nb}_3\mathrm{S}_6$ and $\mathrm{Ni}\mathrm{Nb}_3\mathrm{S}_6$), we find a band dispersion unique to the Co intercalation existing near the Fermi level.
We further demonstrate that a slight deficiency of sulfur in $\mathrm{Co}\mathrm{Nb}_3\mathrm{S}_6$ eliminates the ferromagnetism and the AHE simultaneously while hardly changing the band structure, indicating that the weak ferromagnetism is responsible for the emergence of the large AHE.
Based on our results, we propose Weyl points near the Fermi level to cause the large AHE.
\end{abstract}

\maketitle

\begin{figure*}
\includegraphics[width=178mm]{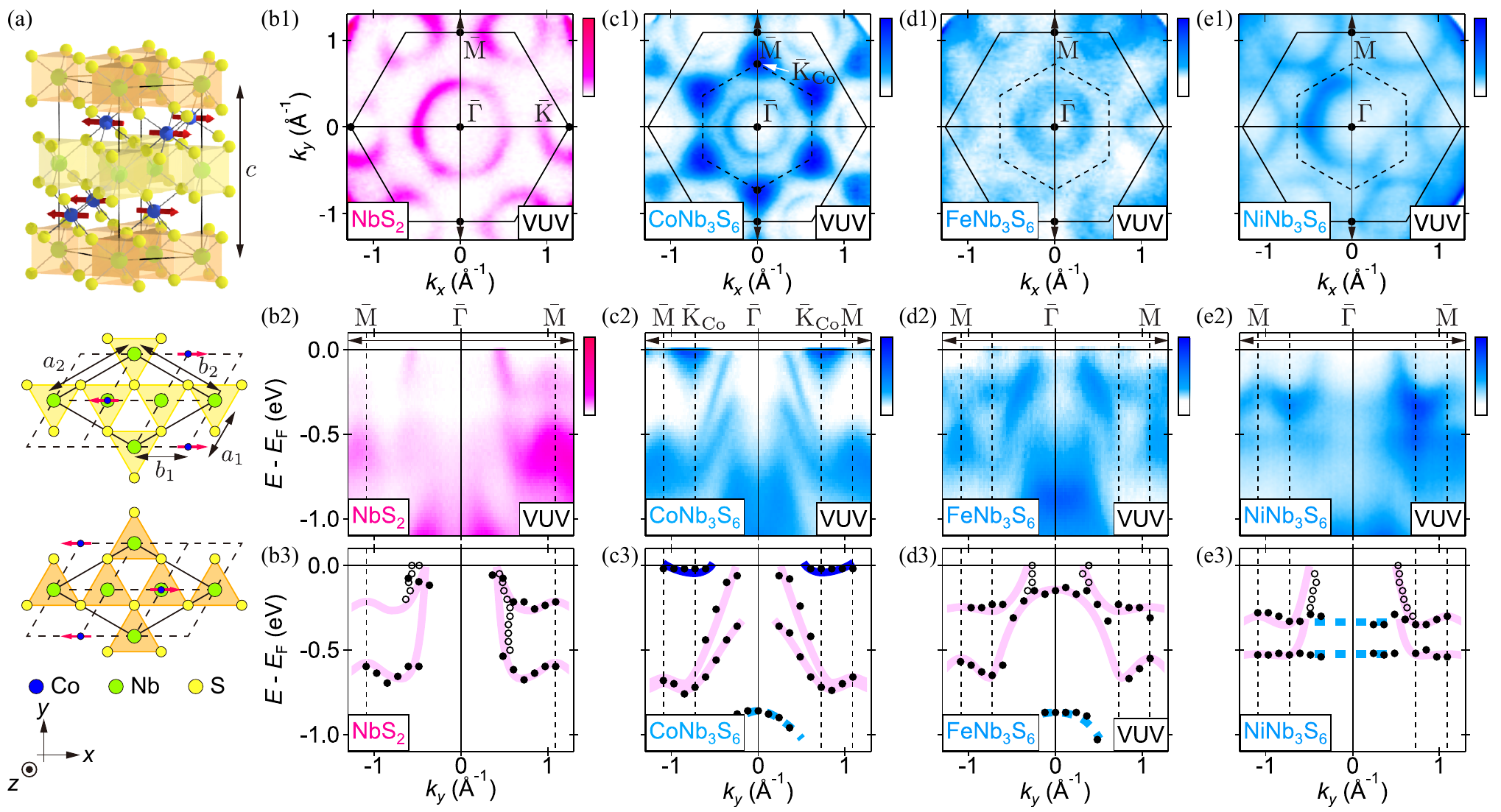}
\caption{\label{Fig: Structure_and_VUV} Crystal structure and band dispersion of $\mathrm{Co}\mathrm{Nb}_3\mathrm{S}_6$ obtained by VUV-ARPES ($h\nu=120\ \mathrm{eV}$). (a) Crystal structure of $\mathrm{Co}\mathrm{Nb}_3\mathrm{S}_6$. Two $\mathrm{Nb}\mathrm{S}_2$ layers of different orientations are represented by orange and yellow. In the bottom panels, the crystallographic unit cells of $\mathrm{Nb}\mathrm{S}_2$ and $\mathrm{Co}\mathrm{Nb}_3\mathrm{S}_6$ are represented by the dashed and solid rhomboids respectively. The red arrows represent antiferromagnetic moments of Co atoms. (b1), (c1) Fermi surfaces of $\mathrm{Nb}\mathrm{S}_2$ and $\mathrm{Co}\mathrm{Nb}_3\mathrm{S}_6$. (b2), (c2) Band dispersions of $\mathrm{Nb}\mathrm{S}_2$ and $\mathrm{Co}\mathrm{Nb}_3\mathrm{S}_6$ along the $k_y$ direction [vertical arrows in (b1) and (c1)]. (b3), (c3) Schematics of the band dispersions of $\mathrm{Nb}\mathrm{S}_2$ and $\mathrm{Co}\mathrm{Nb}_3\mathrm{S}_6$. The black filled and open dots represent peak positions extracted from energy and momentum distribution curves [Figs.\ S7 and S8 \cite{Supplemental}], respectively. Band dispersions originating from $\mathrm{Nb}\mathrm{S}_2$ are drawn by pink and additional dispersions in (c3) are marked by blue or light blue. (d1)-(e3) Fermi surfaces and band dispersions of $\mathrm{Fe}\mathrm{Nb}_3\mathrm{S}_6$ and $\mathrm{Ni}\mathrm{Nb}_3\mathrm{S}_6$. In the Fermi surface maps, the solid and dashed hexagons correspond to the Brillouin zones of $\mathrm{Nb}\mathrm{S}_2$ and $M\mathrm{Nb}_3\mathrm{S}_6$ respectively.}
\end{figure*}

Transition metal dichalcogenides $MX_2$ ($M$: transition metal, $X$: chalcogen) exhibit various types of layered structures \cite{Kolobov2016} and have gained attention as a platform to investigate enriched properties, such as superconductivity \cite{Yokoya2518}, charge density wave states \cite{Wilson1975}, and band topology \cite{Soluyanov2015}.
In these, both antiferromagnetism and ferromagnetism can be induced by intercalation of $3d$ magnetic elements \cite{Friend1977, Parkin1980}.
Among these materials, great interest has been recently given to the noncentrosymmetric antiferromagnet $\mathrm{Co}\mathrm{Nb}_3\mathrm{S}_6$ \cite{Fn2}, in which Co ions intercalated into $2H$-$\mathrm{Nb}\mathrm{S}_2$ break the inversion symmetry [Fig.\ \ref{Fig: Structure_and_VUV}(a)].
The neutron scattering measurements revealed the collinear antiferromagnetic structure along the in-plane direction [Refs.\ \cite{Parkin1983,MAGNDATA} and Supplemental Material Fig.\ S1(a) \cite{Supplemental}], which keeps the time-reversal symmetry.

Recently, weak ferromagnetism along the out-of-plane direction and large anomalous Hall effect (AHE) have been observed in $\mathrm{Co}\mathrm{Nb}_3\mathrm{S}_6$ \cite{Ghimire2018}; interestingly, the anomalous Hall response is exceptionally large compared to the small ferromagnetic moment.
When the intrinsic AHE \cite{PhysRevLett.49.405,Onoda2002} derived from the Berry curvature \cite{Berry1984} is considered, the Hall resistivity can be decomposed as $\rho_{xy}=R_0\mu_0H+R_s\mu_0 m$ \cite{RevModPhys.82.1539}: here, $H$ and $m$ are the magnetic field and magnetic moment density, whereas $R_0$ and $R_s$ are coefficients for the ordinal Hall effect and AHE, respectively.
$R_s$ can be at most hundreds of times larger than $R_0$ [Refs.\ \cite{PhysRev.36.1503,PhysRevB.94.075135,Ye2018,Liu2018,Dijkstra_1989,PhysRevB.77.014433} and Supplemental Material Note 1 \cite{Supplemental}].
For $\mathrm{Co}\mathrm{Nb}_3\mathrm{S}_6$, the ratio of $R_s$ and $R_0$ is reported to be $5.9\times10^4$, much larger than usual cases.
The relation between the weak ferromagnetism and the large AHE has been unclear.
This issue has been all the more controversial because noncollinear antiferromagnetic structures \cite{Ghimire2018, PhysRevResearch.2.023051} and emergent time-reversal symmetry breaking [Ref.\ \cite{Smejkal2020} and Note 2 \cite{Supplemental}] have been proposed as alternative origins of the large AHE rather than the weak ferromagnetism.

In this Letter, we used various techniques of angle-resolved photoemission spectroscopy (ARPES) with a wide range of photon energies from vacuum ultraviolet (VUV) light and soft x ray (SX) to study the electronic structure of $\mathrm{Co}\mathrm{Nb}_3\mathrm{S}_6$, and compared it with those of series materials $\mathrm{Nb}\mathrm{S}_2$, $\mathrm{Fe}\mathrm{Nb}_3\mathrm{S}_6$, $\mathrm{Ni}\mathrm{Nb}_3\mathrm{S}_6$, and $\mathrm{Co}\mathrm{Nb}_3\mathrm{S}_{6-x}$.
Our measurements unveil two unique properties of $\mathrm{Co}\mathrm{Nb}_3\mathrm{S}_6$.
Firstly, there is a specific band dispersion due to the Co intercalation, located very close to the Fermi level.
Secondly, comparing magnetotransport properties, we revealed that the weak ferromagnetism and AHE observed in $\mathrm{Co}\mathrm{Nb}_3\mathrm{S}_6$ simultaneously disappeared in $\mathrm{Co}\mathrm{Nb}_3\mathrm{S}_{6-x}$ with a sulfur deficiency, providing experimental evidence that the weak ferromagnetism is responsible for the emergence of the AHE.
Focusing on the noncentrosymmetric nature of $\mathrm{Co}\mathrm{Nb}_3\mathrm{S}_6$, we propose that it enhances the AHE even under weak ferromagnetism because the broken time-reversal and inversion symmetries lead to Weyl points near the Fermi level originating from Kramers-Weyl nodes \cite{Chang2018}.

Single crystals of $2H$-$\mathrm{Nb}\mathrm{S}_2$, $M\mathrm{Nb}_3\mathrm{S}_6\ (M=\mathrm{Fe},\ \mathrm{Co},\ \mathrm{Ni})$, and $\mathrm{Co}\mathrm{Nb}_3\mathrm{S}_{6-x}$ were grown by the chemical vapor transport method from polycrystals with iodine as the transport agent (see Note 3 \cite{Supplemental} for details).
VUV-ARPES measurements were performed at BL5U of UVSOR and BL5-2 of SSRL, nano-VUV-ARPES was at Spectromicroscopy-3.2L beamline of Elettra Sincrotrone Trieste \cite{Dudin2010}, and SX-ARPES was at BL25SU of SPring-8 \cite{Senba2016, Muro:ok5049} (see Note 4 \cite{Supplemental} for details).
The measurement temperature was kept lower than 20 K in most measurements, other than temperature dependence (up to 200 K) and nano-VUV-ARPES measurements (30 K due to the machine limitation).
In \textit{ab initio} calculations based on the density functional theory (DFT) without spin, we used {\sc quantum espresso} code \cite{Giannozzi2009,Giannozzi2017} and ultrasoft pseudopotentials \cite{PhysRevB.41.1227, Pseudopotentials}.
Wannier90 \cite{Pizzi2020} was also used to obtain the maximally localized Wannier functions.

We compare Fermi surface maps [Figs.\ \ref{Fig: Structure_and_VUV}(b1)-(e1)] and band dispersions [Figs.\ \ref{Fig: Structure_and_VUV}(b2)-(e2)] of $\mathrm{Co}\mathrm{Nb}_3\mathrm{S}_6$ and its series materials obtained by VUV-ARPES.
$\mathrm{Nb}\mathrm{S}_2$ has two Fermi pockets around the $\bar{\Gamma}$ and $\bar{K}$ points [Fig.\ \ref{Fig: Structure_and_VUV}(b1)].
The pocket around the $\bar{\Gamma}$ point is hole-type containing two bands, drawn by the pink curves in the schematic [Fig.\ \ref{Fig: Structure_and_VUV}(b3)] (see Fig.\ S4(b) \cite{Supplemental} center panel for the pocket around the $\bar{K}$ point).
We found that the hole pocket around the $\bar{\Gamma}$ point became smaller in $\mathrm{Co}\mathrm{Nb}_3\mathrm{S}_6$ [Fig.\ \ref{Fig: Structure_and_VUV}(c1)].
This change can be observed as a band shift in the dispersion map [pink curves in Figs.\ \ref{Fig: Structure_and_VUV} (b3) and (c3)], which is consistent with our XPS measurements (Note 5 \cite{Supplemental}) and other intercalated materials \cite{Battaglia2007, PhysRevB.94.075141}.

Furthermore, we observed some additional bands in VUV-ARPES measurements of $\mathrm{Co}\mathrm{Nb}_3\mathrm{S}_6$ as illustrated by the blue and light blue curves in Fig.\ \ref{Fig: Structure_and_VUV}(c3).
Among them, the tiny electron pockets at the corners of the Brillouin zone [blue curves in Fig.\ \ref{Fig: Structure_and_VUV}(c3)], are the most intriguing, because they can contribute to the AHE.
While $\mathrm{Fe}\mathrm{Nb}_3\mathrm{S}_6$ and $\mathrm{Ni}\mathrm{Nb}_3\mathrm{S}_6$ also have additional occupied states [light blue dashed curves in Figs.\ \ref{Fig: Structure_and_VUV}(d3) and (e3); see Note 9 \cite{Supplemental} for the details of these states], these two materials have similar electronic structure as $\mathrm{Nb}\mathrm{S}_2$ without the intercalation at the Fermi level.
Although there seems to be intensity modulation near the Fermi level, especially in $\mathrm{Fe}\mathrm{Nb}_3\mathrm{S}_6$ [Figs.\ \ref{Fig: Structure_and_VUV}(d1) and (d2)], energy distribution curves of them [Figs.\ S7(c) and (d) \cite{Supplemental}] have only small peaks.
While they may indicate the existence of additional electronic states above the Fermi level, the additional band dispersion crossing the Fermi level is unique to the Co intercalation.

The additional electron pockets in $\mathrm{Co}\mathrm{Nb}_3\mathrm{S}_6$ originate from the bulk electronic structure and are dominated by Co atoms.
SX-ARPES measurements of $\mathrm{Co}\mathrm{Nb}_3\mathrm{S}_6$ also captured these additional bands [Figs.\ S5(b) and (c) \cite{Supplemental}] and $k_z$ dependence of them [Fig.\ S5(d) \cite{Supplemental}].
Since SX-ARPES is bulk-sensitive \cite{STROCOV2019} and does not capture surface states such as topological Fermi arcs \cite{Xu613}, these results show the bulk origin of these bands.
For the second point, such dispersions do not exist in $\mathrm{Nb}\mathrm{S}_2$ even if we considered unoccupied states calculated for $\mathrm{Nb}\mathrm{S}_2$ \cite{Fn1}, so this additional dispersion seems to come from Co atoms.
This interpretation about the unique band was validated by surface-selective measurements using nano-focused light (Note 10 \cite{Supplemental}) and the intensity enhancement near the Fermi level at the energy on Co resonance [Figs.\ S5(f) and (g) \cite{Supplemental}]; the suppressed spectral intensities of this band at elevated temperatures also may be due to the disordered magnetic structure of Co atoms above $T_\mathrm{N}$ (Note 8 \cite{Supplemental}).
We note that we carefully chose measurement positions in VUV-ARPES measurements [Fig.\ \ref{Fig: Structure_and_VUV}] to reduce the effect of termination-surface dependence (Note 11 \cite{Supplemental}).


\begin{figure}
\includegraphics{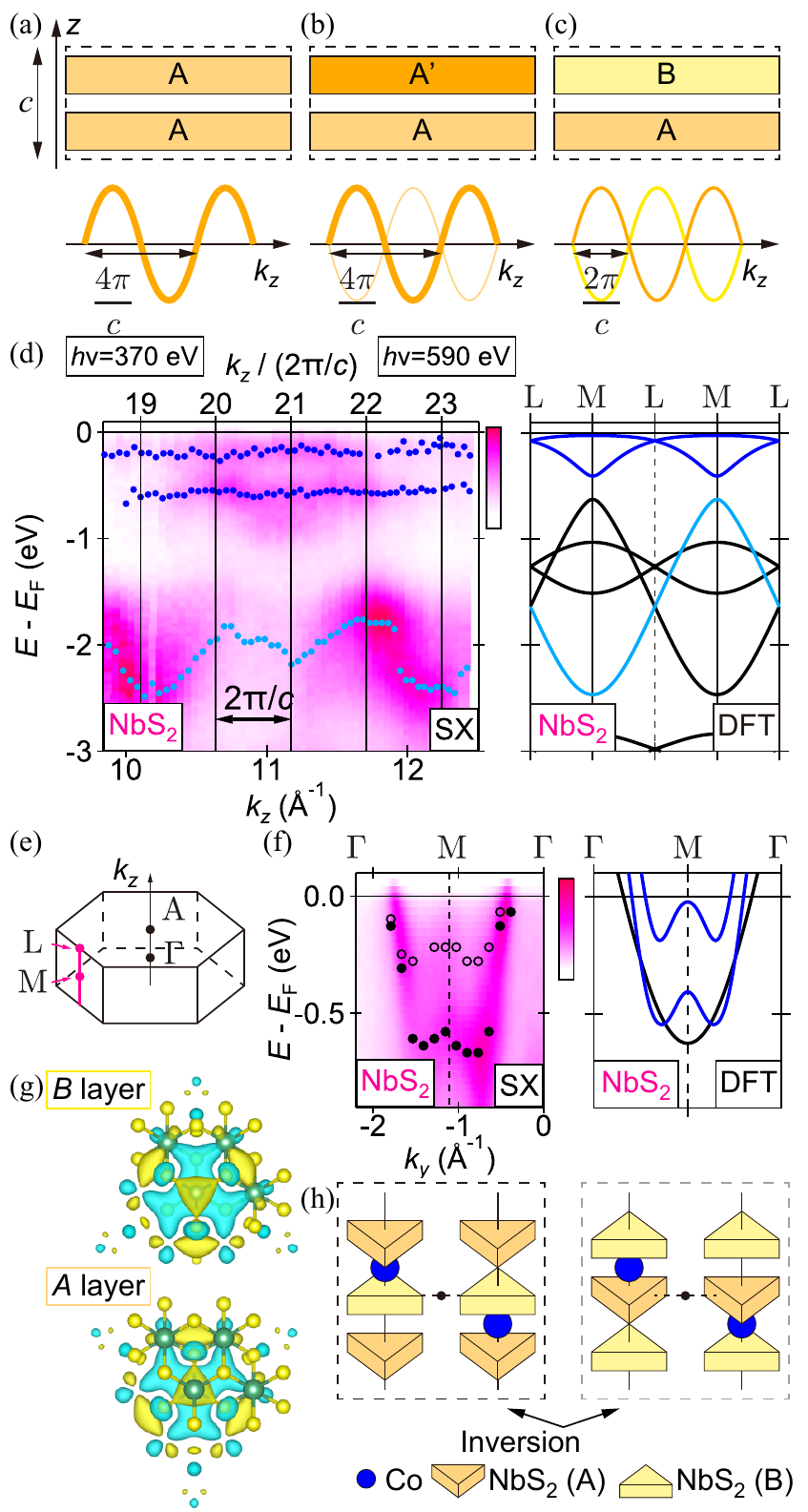}
\caption{\label{Fig: Inversion_broken_kz} Evaluation of the broken inversion symmetry in $\mathrm{Co}\mathrm{Nb}_3\mathrm{S}_6$ by the $k_z$ dispersion of $2H$-$\mathrm{Nb}\mathrm{S}_2$. (a)-(c) Schematics of layered materials and estimated ARPES spectra along the $k_z$ direction. (d) $k_z$ dispersions of $\mathrm{Nb}\mathrm{S}_2$ along the $M$-$L$ path by SX-ARPES and DFT calculations. In the left panel, peak positions extracted from energy distribution curves [Fig.\ S12(a) \cite{Supplemental}] are shown by the blue and light blue points. In the right panel, band dispersions corresponding to those observed by SX-ARPES are highlighted by the same colors. (e) Brillouin zone of $\mathrm{Nb}\mathrm{S}_2$. (f) Band dispersions of SX-ARPES ($h\nu=525\ \mathrm{eV}$) and DFT calculations along the $k_y$ direction. Blue curves in the right panel correspond to those in (d). Filled and open circles in the left panel represent peak positions extracted from energy distribution curves (EDCs) [Figs.\ S14(b) and (c)], while the open circles are those determined from EDC spectra in $k_y>0$ region and symmetrized with respect to the $\Gamma$ point. (g) Maximally localized Wannier functions of two bands highlighted by blue in (d) and (f). (h) Schematic of the electronic structure near the Fermi level. Left and right schematics are exchanged by the inversion for the black point, which keeps Co atoms unchanged.}
\end{figure}

\begin{figure*}
\includegraphics{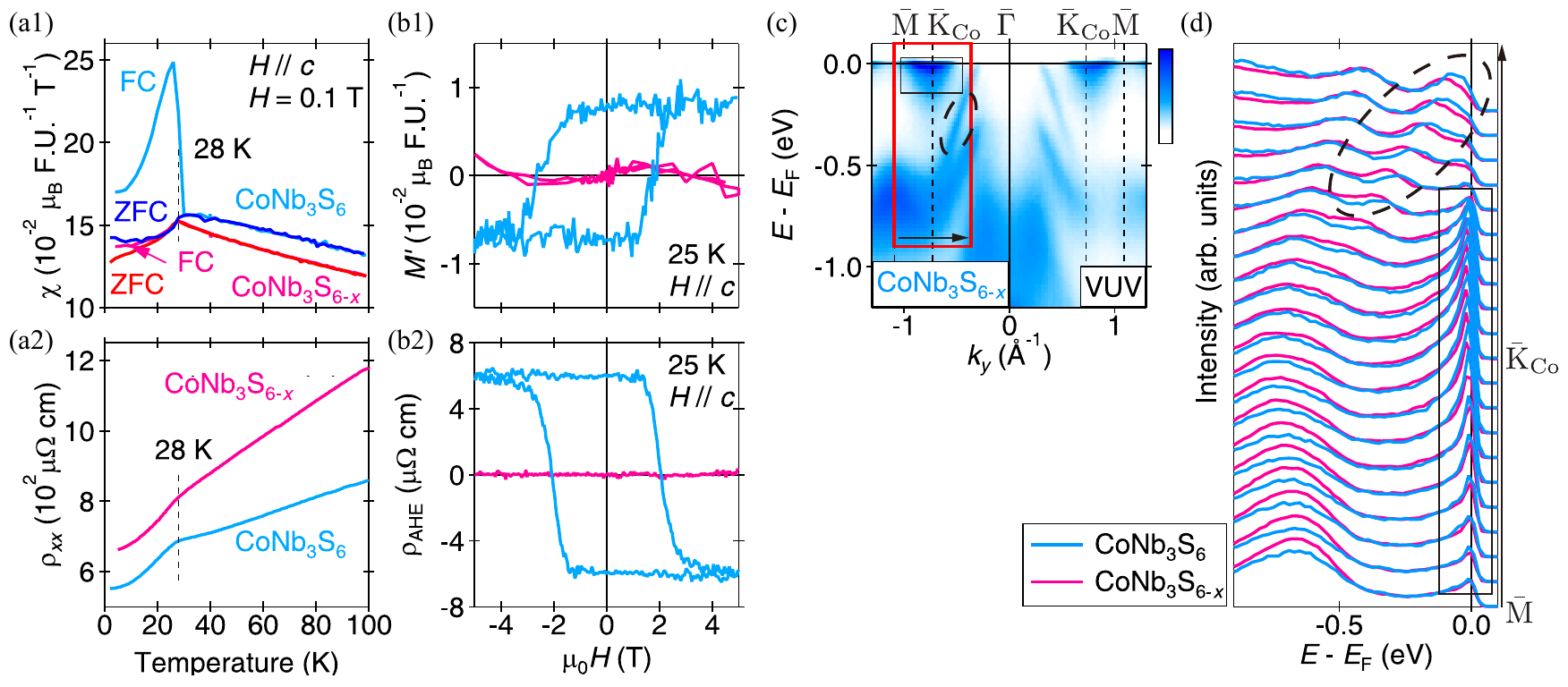}
\caption{\label{Fig: Magnetotransport} Magnetic and magnetotransport properties of $\mathrm{Co}\mathrm{Nb}_3\mathrm{S}_6$. (a1) Magnetic susceptibility $\chi$ of $\mathrm{Co}\mathrm{Nb}_3\mathrm{S}_6$ (light blue and blue) and $\mathrm{Co}\mathrm{Nb}_3\mathrm{S}_{6-x}$ (pink and red), obtained by field-cooled (FC) and zero-field-cooled (ZFC) processes. (a2) Temperature dependence of the longitudinal resistivity $\rho_{xx}$ in $\mathrm{Co}\mathrm{Nb}_3\mathrm{S}_6$ (light blue) and $\mathrm{Co}\mathrm{Nb}_3\mathrm{S}_{6-x}$ (pink). (b1) Out-of-plane ferromagnetic moment extracted from $M$-$H$ curves like Fig.\ S15(b) \cite{Supplemental}. (b2) Anomalous Hall resistivity $\rho_{\mathrm{AHE}}$ extracted from $\rho_{xy}$-$H$ curves like Fig.\ S15(c) \cite{Supplemental}. (c) Band dispersion of $\mathrm{Co}\mathrm{Nb}_3\mathrm{S}_{6-x}$ along the $k_y$ direction taken by the 120 eV VUV light. (d) Energy distribution curves of $\mathrm{Co}\mathrm{Nb}_3\mathrm{S}_6$ and $\mathrm{Co}\mathrm{Nb}_3\mathrm{S}_{6-x}$ in the red rectangular area in (c). The dashed ovals and solid rectangles in (c) and (d) correspond each other.}
\end{figure*}

Next we argue the effect of the broken inversion symmetry on wavefunctions in $\mathrm{Co}\mathrm{Nb}_3\mathrm{S}_6$, originating from the $AB$-stacking of the original material $2H$-$\mathrm{Nb}\mathrm{S}_2$ (see Note 12 \cite{Supplemental} for details).
Since the inversion operation exchanges $A$ and $B$ layers of $2H$-$\mathrm{Nb}\mathrm{S}_2$ [orange and yellow layers in Fig.\ \ref{Fig: Structure_and_VUV}(a)], we can consider the difference of wavefunctions on $A$ and $B$ layers as the strength of the noncentrosymmetric property for those in $\mathrm{Co}\mathrm{Nb}_3\mathrm{S}_6$.
We claim that we can evaluate it from the ARPES spectrum intensities along the $k_z$ direction; we focus on $4\pi/c$-periodic $k_z$ dispersion, reported in ARPES measurements of $AB$-stacked materials \cite{PhysRevLett.119.026403, PhysRevB.97.045430}.
First, we consider the $k_z$ dispersions of one-layered ($AA$-) and two-layered ($AB$-) materials; the $k_z$ dispersions contain one and two bands, respectively [Figs.\ \ref{Fig: Inversion_broken_kz}(a) and (c)]. 
On the other hand, in a quasi-two-layered ($AA^\prime$-) structure [Fig.\ \ref{Fig: Inversion_broken_kz}(b)], where wavefunctions on two layers are only slightly different, the ARPES spectrum is expected to change from Fig.\ \ref{Fig: Inversion_broken_kz}(a) only slightly, and therefore the spectrum intensities of two bands can be completely different [Fig.\ \ref{Fig: Inversion_broken_kz}(b) bottom panel].
In general, the weak spectrum is experimentally undetectable, and the resulting $4\pi/c$ periodicity of the $k_z$ dispersion is observed in most $AB$-stacked materials \cite{PhysRevLett.119.026403, PhysRevB.97.045430}.
The one-dimensional tight-binding model with two orbitals [Ref.\ \cite{MOSER2017} and Note 13 \cite{Supplemental}] can discuss this argument more quantitatively.

Following the above argument, we performed bulk-sensitive SX-ARPES of $\mathrm{Nb}\mathrm{S}_2$ and investigated the $k_z$ dispersion [Fig.\ \ref{Fig: Inversion_broken_kz}(d)].
The dispersion around $E-E_\mathrm{F}=-2\ \mathrm{eV}$, highlighted by light blue, is one of the paired dispersions represented in the DFT calculations.
This situation is the case where these two wavefunctions form a quasi-two-layered structure [Fig.\ \ref{Fig: Inversion_broken_kz}(b)], similarly to previous research of $AB$-stacked materials \cite{PhysRevLett.119.026403, PhysRevB.97.045430}.
On the other hand, two dispersions highlighted by blue form a pair, as confirmed by the inverted M-shaped bands along the $k_y$ direction [Fig.\ \ref{Fig: Inversion_broken_kz}(f)], which is the case of Fig.\ \ref{Fig: Inversion_broken_kz}(c).
This situation is consistent with the maximally localized Wannier functions of these bands, having triangular shapes and resulting in small overlap with each other [Fig.\ \ref{Fig: Inversion_broken_kz}(g)].
Therefore, in $\mathrm{Co}\mathrm{Nb}_3\mathrm{S}_6$ these dispersions contribute to the substantially broken inversion symmetry of the electronic structure near the Fermi level [Fig.\ \ref{Fig: Inversion_broken_kz}(h)].

As a minor difference of SX-ARPES measurements and DFT calculations, we did not observe the band degenerations at $L$ points.
Since they are associated with $2_1$ screw rotational symmetry \cite{PhysRevB.93.085427, Funada2019}, this result can be because, while SX-ARPES is bulk sensitive, it can detect the electronic structure in few layers near the surface, where the screw rotational symmetry is incomplete.
Although this observation does not affect much our discussion of the broken inversion symmetry, we need further experiments and calculations to explain it in more detail.

\begin{figure*}
\includegraphics{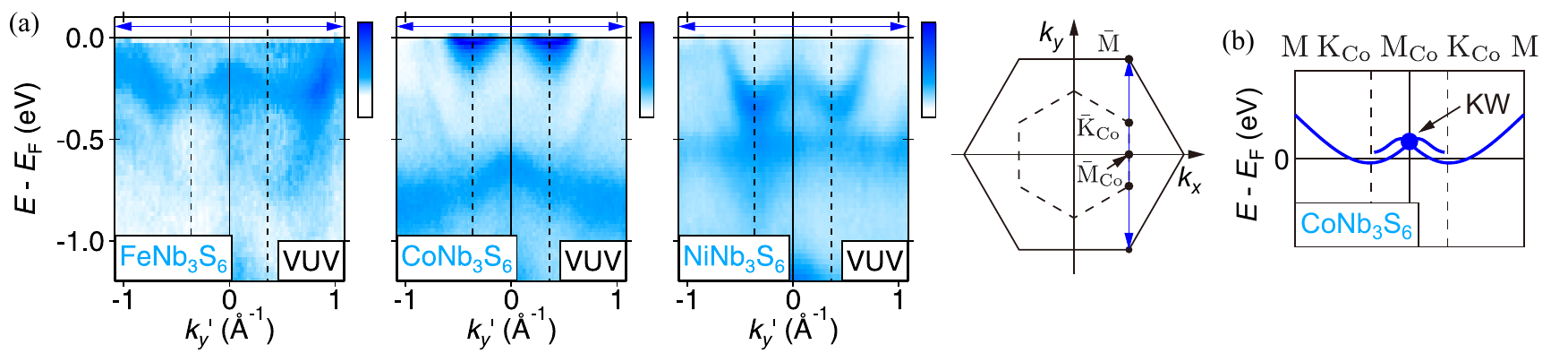}
\caption{\label{Fig: KW} Possibility of Kramers-Weyl nodes near the Fermi level in $\mathrm{Co}\mathrm{Nb}_3\mathrm{S}_6$. (a) Band dispersions of $M\mathrm{Nb}_3\mathrm{S}_6\ (M=\mathrm{Fe},\ \mathrm{Co},\ \mathrm{Ni})$ along the blue arrow in the right schematic of the Brillouin zones. (b) Schematic of band dispersions near the Fermi level for $\mathrm{Co}\mathrm{Nb}_3\mathrm{S}_6$. The Kramers-Weyl node is marked by the blue circle and labeled as ``KW".}
\end{figure*}

To better understand the mechanism of the large AHE in $\mathrm{Co}\mathrm{Nb}_3\mathrm{S}_6$, we prepared $\mathrm{Co}\mathrm{Nb}_3\mathrm{S}_{6-x}$ with a sulfur deficiency, and compared their electronic structures and magnetotransport properties.
We observed kinks at $T_\mathrm{N}\simeq28\ \mathrm{K}$ in the magnetic susceptibility and longitudinal resistivity measurements of not only $\mathrm{Co}\mathrm{Nb}_3\mathrm{S}_6$ \cite{Ghimire2018,PhysRevResearch.2.023051} but also $\mathrm{Co}\mathrm{Nb}_3\mathrm{S}_{6-x}$ [Figs.\ \ref{Fig: Magnetotransport}(a1) and (a2)].
However, only $\mathrm{Co}\mathrm{Nb}_3\mathrm{S}_6$ exhibited the abrupt jump of the magnetic susceptibility at $T_\mathrm{N}$, associated with out-of-plane ferromagnetic moment.
This difference is also observed by magnetic-field dependence measurements of the out-of-plane ferromagnetic moment [Fig.\ \ref{Fig: Magnetotransport}(b1)].
For $\mathrm{Co}\mathrm{Nb}_3\mathrm{S}_6$, the ferromagnetic moment and AHE showed hysteresis curves with the same coercive magnetic field [light blue curves in Figs.\ \ref{Fig: Magnetotransport}(b1) and (b2)], as previously reported \cite{Ghimire2018}.
Although the ferromagnetic moment is larger than previous research, the ratio of $R_s$ and $R_0$ is still exceptionally large [Note 1 \cite{Supplemental}].
On the other hand, such hysteresis curves completely vanished in $\mathrm{Co}\mathrm{Nb}_3\mathrm{S}_{6-x}$ [pink curves in Figs.\ \ref{Fig: Magnetotransport}(b1) and (b2)].

In contrast to the clear difference in the magnetotransport properties, VUV-ARPES measurements of $\mathrm{Co}\mathrm{Nb}_3\mathrm{S}_{6-x}$ revealed that the band dispersion is almost the same as that of $\mathrm{Co}\mathrm{Nb}_3\mathrm{S}_6$ [Figs.\ \ref{Fig: Structure_and_VUV}(c2) and \ref{Fig: Magnetotransport}(c)], as well as the unique band near the Fermi level.
As a clear difference, we only found band shift as small as 50 meV in the area represented by the dashed oval in Fig.\ \ref{Fig: Magnetotransport}(d) [see Fig.\ S16(b) \cite{Supplemental} for $k_y>0$], which corresponds to the band dispersion originating from $\mathrm{Nb}\mathrm{S}_2$ layers.
While this variation can be explained by electron doping due to the vacancy of electronegative sulfur atoms, negligible change of the unique band originating from Co atoms, highlighted by the solid rectangle in Fig.\ \ref{Fig: Magnetotransport}(d), show that Co atoms are more weakly affected by the electron doping than $\mathrm{Nb}\mathrm{S}_2$ layers.
Such a slight difference cannot account for the complete disappearance of the AHE, so we conclude that the weak ferromagnetism is responsible for the emergence of the AHE in $\mathrm{Co}\mathrm{Nb}_3\mathrm{S}_6$ (see Note 15 \cite{Supplemental} for other magnetic and magnetotransport properties).

We propose that Weyl points near the Fermi level may cause the large AHE even under the weak ferromagnetism.
Without the ferromagnetic moment, this system should have Kramers-Weyl nodes, which are Weyl points at time-reversal invariant momenta (TRIMs) as a result of the noncentrosymmetric crystal structure \cite{Chang2018}.
The Kramers-Weyl node at the $M_\mathrm{Co}$ point for the unique band is expected to be very close to the Fermi level [Fig.\ \ref{Fig: KW}(a) center panel and (b)]; on the contrary, $\mathrm{Fe}\mathrm{Nb}_3\mathrm{S}_6$ and $\mathrm{Ni}\mathrm{Nb}_3\mathrm{S}_6$ have band dispersions of $\mathrm{Nb}\mathrm{S}_2$ and occupied Fe/Ni bands in Fig.\ \ref{Fig: KW}(a).
Although these occupied states also might have Kramers-Weyl nodes, they are unrelated to magnetotransport properties such as the AHE.
With broken time-reversal symmetry by the ferromagnetic moment, such a system emerges nonzero anomalous Hall conductance (Note 17 \cite{Supplemental}).
In this case, Weyl points should exist near the Fermi level, so the resultant anomalous Hall conductance is expected to be large even if the ferromagnetic moment is small, as discussed in other magnetic Weyl semimetals with the inversion symmetry \cite{Yang2017, Kuroda2017,Liu2018,Liu2019}.
Since the inversion symmetry is evaluated to be substantially broken by analyzing the $k_z$ dispersion [Fig.\ \ref{Fig: Inversion_broken_kz}], the spin splitting of the noncentrosymmetric system should not be ignored, which is another preferable condition for the emergence of the large AHE.
These Weyl points associated with Kramers-Weyl nodes can be possible only in materials with both the broken time-reversal and inversion symmetries.
Since most magnetic Weyl semimetals are centrosymmetric such as $\mathrm{Mn}_3\mathrm{Sn}$ \cite{Yang2017, Kuroda2017} and $\mathrm{Co}_3\mathrm{Sn}_2\mathrm{S}_2$ \cite{Liu2018,Liu2019}, the noncentrosymmetric $\mathrm{Co}\mathrm{Nb}_3\mathrm{S}_6$ could be a new type of magnetic Weyl semimetals showing the large AHE.

In conclusion, we experimentally studied the electronic structure and the large AHE of $\mathrm{Co}\mathrm{Nb}_3\mathrm{S}_6$ via the comparison with its series materials.
VUV- and SX-ARPES measurements revealed the unique band dispersion near the Fermi level derived from the Co intercalation.
SX-ARPES spectra of $\mathrm{Nb}\mathrm{S}_2$ along the $k_z$ direction indicated that the electronic structure of $\mathrm{Co}\mathrm{Nb}_3\mathrm{S}_6$ near the Fermi level substantially breaks the inversion symmetry.
While the possibility of a noncollinear magnetic structure has been proposed to explain the large AHE \cite{Ghimire2018, PhysRevResearch.2.023051}, our magnetic and magnetotransport measurements of $\mathrm{Co}\mathrm{Nb}_3\mathrm{S}_6$ revealed that the weak ferromagnetism is responsible for the emergence of the AHE.
Based on our results, we propose the existence of Weyl points near the Fermi level, associated with the Kramers-Weyl node system in noncentrosymmetric crystal structures \cite{Chang2018}, which could cause the large AHE even if the ferromagnetism is weak.

\nocite{SAITOH2005}
\nocite{PhysRevB.89.041407}
\nocite{PhysRevB.91.041116}
\nocite{Lee2013}
\nocite{PhysRevLett.98.157002,Kang2020}
\nocite{PhysRevB.98.081112,PhysRevB.99.140510,Nakayama2019,XU2020}
\nocite{Sirica2020}

\begin{acknowledgments}
We thank Y.\ Ishida for supporting the analysis of ARPES data \cite{Ishida2018}, and M.\ Hirayama, T.\ Nomoto, and R.\ Arita for fruitful discussions.
R.N. was supported by the Institute for Basic Science in Korea (Grant No.\ IBS-R009-G2).
This work is also supported by Grants-in-Aid for JSPS Fellows (Grant No.\ JP21J20657), Grants-in-Aid for Scientific Research(B) (Grant No.\ 19H02683), Grants-in-Aid for Scientific Research(A) (Grants Nos. 20H00337, 21H04439, and 21H04652), the Murata Science Foundation, Photon and Quantum Basic Research Coordinated Development Program from MEXT, and CREST project (JPMJCR16F2) from JST.
Use of the Stanford Synchrotron Radiation Lightsource, SLAC National Accelerator Laboratory, is supported by the US Department of Energy, Office of Science, Office of Basic Energy Sciences under contract no.\ DE-AC02-76SF00515.
The synchrotron radiation experiments were performed with the approval of UVSOR (Proposals Nos.\ 20-757 and 20-842), Elettra Sincrotrone Trieste (Proposal No.\ 20200395), and JASRI (Proposal No.\ 2020A1181).


H.T. and S.O. contributed equally to this work.
\end{acknowledgments}

\bibliography{CoNb3S6_ARPES_magnetotransport}

\begin{thebibliography}{58}%
\makeatletter
\providecommand \@ifxundefined [1]{%
 \@ifx{#1\undefined}
}%
\providecommand \@ifnum [1]{%
 \ifnum #1\expandafter \@firstoftwo
 \else \expandafter \@secondoftwo
 \fi
}%
\providecommand \@ifx [1]{%
 \ifx #1\expandafter \@firstoftwo
 \else \expandafter \@secondoftwo
 \fi
}%
\providecommand \natexlab [1]{#1}%
\providecommand \enquote  [1]{``#1''}%
\providecommand \bibnamefont  [1]{#1}%
\providecommand \bibfnamefont [1]{#1}%
\providecommand \citenamefont [1]{#1}%
\providecommand \href@noop [0]{\@secondoftwo}%
\providecommand \href [0]{\begingroup \@sanitize@url \@href}%
\providecommand \@href[1]{\@@startlink{#1}\@@href}%
\providecommand \@@href[1]{\endgroup#1\@@endlink}%
\providecommand \@sanitize@url [0]{\catcode `\\12\catcode `\$12\catcode
  `\&12\catcode `\#12\catcode `\^12\catcode `\_12\catcode `\%12\relax}%
\providecommand \@@startlink[1]{}%
\providecommand \@@endlink[0]{}%
\providecommand \url  [0]{\begingroup\@sanitize@url \@url }%
\providecommand \@url [1]{\endgroup\@href {#1}{\urlprefix }}%
\providecommand \urlprefix  [0]{URL }%
\providecommand \Eprint [0]{\href }%
\providecommand \doibase [0]{https://doi.org/}%
\providecommand \selectlanguage [0]{\@gobble}%
\providecommand \bibinfo  [0]{\@secondoftwo}%
\providecommand \bibfield  [0]{\@secondoftwo}%
\providecommand \translation [1]{[#1]}%
\providecommand \BibitemOpen [0]{}%
\providecommand \bibitemStop [0]{}%
\providecommand \bibitemNoStop [0]{.\EOS\space}%
\providecommand \EOS [0]{\spacefactor3000\relax}%
\providecommand \BibitemShut  [1]{\csname bibitem#1\endcsname}%
\let\auto@bib@innerbib\@empty
\bibitem [{\citenamefont {Kolobov}\ and\ \citenamefont
  {Tominaga}(2016)}]{Kolobov2016}%
  \BibitemOpen
  \bibfield  {author} {\bibinfo {author} {\bibfnamefont {A.~V.}\ \bibnamefont
  {Kolobov}}\ and\ \bibinfo {author} {\bibfnamefont {J.}~\bibnamefont
  {Tominaga}},\ }\href
  {https://doi.org/https://doi.org/10.1007/978-3-319-31450-1} {\emph {\bibinfo
  {title} {Two-Dimensional Transition-Metal Dichalcogenides}}}\ (\bibinfo
  {publisher} {Springer},\ \bibinfo {year} {2016})\BibitemShut {NoStop}%
\bibitem [{\citenamefont {Yokoya}\ \emph {et~al.}(2001)\citenamefont {Yokoya},
  \citenamefont {Kiss}, \citenamefont {Chainani}, \citenamefont {Shin},
  \citenamefont {Nohara},\ and\ \citenamefont {Takagi}}]{Yokoya2518}%
  \BibitemOpen
  \bibfield  {author} {\bibinfo {author} {\bibfnamefont {T.}~\bibnamefont
  {Yokoya}}, \bibinfo {author} {\bibfnamefont {T.}~\bibnamefont {Kiss}},
  \bibinfo {author} {\bibfnamefont {A.}~\bibnamefont {Chainani}}, \bibinfo
  {author} {\bibfnamefont {S.}~\bibnamefont {Shin}}, \bibinfo {author}
  {\bibfnamefont {M.}~\bibnamefont {Nohara}},\ and\ \bibinfo {author}
  {\bibfnamefont {H.}~\bibnamefont {Takagi}},\ }\href
  {https://doi.org/10.1126/science.1065068} {\bibfield  {journal} {\bibinfo
  {journal} {Science}\ }\textbf {\bibinfo {volume} {294}},\ \bibinfo {pages}
  {2518} (\bibinfo {year} {2001})}\BibitemShut {NoStop}%
\bibitem [{\citenamefont {Wilson}\ \emph {et~al.}(1975)\citenamefont {Wilson},
  \citenamefont {Salvo},\ and\ \citenamefont {Mahajan}}]{Wilson1975}%
  \BibitemOpen
  \bibfield  {author} {\bibinfo {author} {\bibfnamefont {J.}~\bibnamefont
  {Wilson}}, \bibinfo {author} {\bibfnamefont {F.~D.}\ \bibnamefont {Salvo}},\
  and\ \bibinfo {author} {\bibfnamefont {S.}~\bibnamefont {Mahajan}},\ }\href
  {https://doi.org/10.1080/00018737500101391} {\bibfield  {journal} {\bibinfo
  {journal} {Advances in Physics}\ }\textbf {\bibinfo {volume} {24}},\ \bibinfo
  {pages} {117} (\bibinfo {year} {1975})}\BibitemShut {NoStop}%
\bibitem [{\citenamefont {Soluyanov}\ \emph {et~al.}(2015)\citenamefont
  {Soluyanov}, \citenamefont {Gresch}, \citenamefont {Wang}, \citenamefont
  {Wu}, \citenamefont {Troyer}, \citenamefont {Dai},\ and\ \citenamefont
  {Bernevig}}]{Soluyanov2015}%
  \BibitemOpen
  \bibfield  {author} {\bibinfo {author} {\bibfnamefont {A.~A.}\ \bibnamefont
  {Soluyanov}}, \bibinfo {author} {\bibfnamefont {D.}~\bibnamefont {Gresch}},
  \bibinfo {author} {\bibfnamefont {Z.}~\bibnamefont {Wang}}, \bibinfo {author}
  {\bibfnamefont {Q.}~\bibnamefont {Wu}}, \bibinfo {author} {\bibfnamefont
  {M.}~\bibnamefont {Troyer}}, \bibinfo {author} {\bibfnamefont
  {X.}~\bibnamefont {Dai}},\ and\ \bibinfo {author} {\bibfnamefont {B.~A.}\
  \bibnamefont {Bernevig}},\ }\href {https://doi.org/10.1038/nature15768}
  {\bibfield  {journal} {\bibinfo  {journal} {Nature (London)}\ }\textbf
  {\bibinfo {volume} {527}},\ \bibinfo {pages} {495} (\bibinfo {year}
  {2015})}\BibitemShut {NoStop}%
\bibitem [{\citenamefont {Friend}\ \emph {et~al.}(1977)\citenamefont {Friend},
  \citenamefont {Beal},\ and\ \citenamefont {Yoffe}}]{Friend1977}%
  \BibitemOpen
  \bibfield  {author} {\bibinfo {author} {\bibfnamefont {R.~H.}\ \bibnamefont
  {Friend}}, \bibinfo {author} {\bibfnamefont {A.~R.}\ \bibnamefont {Beal}},\
  and\ \bibinfo {author} {\bibfnamefont {A.~D.}\ \bibnamefont {Yoffe}},\ }\href
  {https://doi.org/10.1080/14786437708232952} {\bibfield  {journal} {\bibinfo
  {journal} {The Philosophical Magazine: A Journal of Theoretical Experimental
  and Applied Physics}\ }\textbf {\bibinfo {volume} {35}},\ \bibinfo {pages}
  {1269} (\bibinfo {year} {1977})}\BibitemShut {NoStop}%
\bibitem [{\citenamefont {Parkin}\ and\ \citenamefont
  {Friend}(1980)}]{Parkin1980}%
  \BibitemOpen
  \bibfield  {author} {\bibinfo {author} {\bibfnamefont {S.~S.~P.}\
  \bibnamefont {Parkin}}\ and\ \bibinfo {author} {\bibfnamefont {R.~H.}\
  \bibnamefont {Friend}},\ }\href {https://doi.org/10.1080/13642818008245370}
  {\bibfield  {journal} {\bibinfo  {journal} {Philosophical Magazine B}\
  }\textbf {\bibinfo {volume} {41}},\ \bibinfo {pages} {65} (\bibinfo {year}
  {1980})}\BibitemShut {NoStop}%
\bibitem [{Fn2()}]{Fn2}%
  \BibitemOpen
  \href@noop {} {}\bibinfo {note} {{{In this report, we use
  ``noncentrosymmetric" instead of ``chiral" in the meaning of the broken
  inversion symmetry, because they are different in some crystallographic
  contexts \cite{BURNS2013}.}}}\BibitemShut {Stop}%
\bibitem [{\citenamefont {Parkin}\ \emph {et~al.}(1983)\citenamefont {Parkin},
  \citenamefont {Marseglia},\ and\ \citenamefont {Brown}}]{Parkin1983}%
  \BibitemOpen
  \bibfield  {author} {\bibinfo {author} {\bibfnamefont {S.~S.~P.}\
  \bibnamefont {Parkin}}, \bibinfo {author} {\bibfnamefont {E.~A.}\
  \bibnamefont {Marseglia}},\ and\ \bibinfo {author} {\bibfnamefont {P.~J.}\
  \bibnamefont {Brown}},\ }\href {https://doi.org/10.1088/0022-3719/16/14/016}
  {\bibfield  {journal} {\bibinfo  {journal} {Journal of Physics C: Solid State
  Physics}\ }\textbf {\bibinfo {volume} {16}},\ \bibinfo {pages} {2765}
  (\bibinfo {year} {1983})}\BibitemShut {NoStop}%
\bibitem [{MAG()}]{MAGNDATA}%
  \BibitemOpen
  \href@noop {} {}\bibinfo {note} {{MAGNDATA data base:
  http://webbdcrista1.ehu.es/magndata/index.php .}}\BibitemShut {Stop}%
\bibitem [{Sup()}]{Supplemental}%
  \BibitemOpen
  \href@noop {} {}\bibinfo {note} {See Supplemental Material at [URL] for
  crystal characterization, additional ARPES data and DFT
  calculations.}\BibitemShut {Stop}%
\bibitem [{\citenamefont {Ghimire}\ \emph {et~al.}(2018)\citenamefont
  {Ghimire}, \citenamefont {Botana}, \citenamefont {Jiang}, \citenamefont
  {Zhang}, \citenamefont {Chen},\ and\ \citenamefont {Mitchell}}]{Ghimire2018}%
  \BibitemOpen
  \bibfield  {author} {\bibinfo {author} {\bibfnamefont {N.~J.}\ \bibnamefont
  {Ghimire}}, \bibinfo {author} {\bibfnamefont {A.~S.}\ \bibnamefont {Botana}},
  \bibinfo {author} {\bibfnamefont {J.~S.}\ \bibnamefont {Jiang}}, \bibinfo
  {author} {\bibfnamefont {J.}~\bibnamefont {Zhang}}, \bibinfo {author}
  {\bibfnamefont {Y.-S.}\ \bibnamefont {Chen}},\ and\ \bibinfo {author}
  {\bibfnamefont {J.~F.}\ \bibnamefont {Mitchell}},\ }\href
  {https://doi.org/10.1038/s41467-018-05756-7} {\bibfield  {journal} {\bibinfo
  {journal} {Nature Communications}\ }\textbf {\bibinfo {volume} {9}},\
  \bibinfo {pages} {3280} (\bibinfo {year} {2018})}\BibitemShut {NoStop}%
\bibitem [{\citenamefont {Thouless}\ \emph {et~al.}(1982)\citenamefont
  {Thouless}, \citenamefont {Kohmoto}, \citenamefont {Nightingale},\ and\
  \citenamefont {den Nijs}}]{PhysRevLett.49.405}%
  \BibitemOpen
  \bibfield  {author} {\bibinfo {author} {\bibfnamefont {D.~J.}\ \bibnamefont
  {Thouless}}, \bibinfo {author} {\bibfnamefont {M.}~\bibnamefont {Kohmoto}},
  \bibinfo {author} {\bibfnamefont {M.~P.}\ \bibnamefont {Nightingale}},\ and\
  \bibinfo {author} {\bibfnamefont {M.}~\bibnamefont {den Nijs}},\ }\href
  {https://doi.org/10.1103/PhysRevLett.49.405} {\bibfield  {journal} {\bibinfo
  {journal} {Phys. Rev. Lett.}\ }\textbf {\bibinfo {volume} {49}},\ \bibinfo
  {pages} {405} (\bibinfo {year} {1982})}\BibitemShut {NoStop}%
\bibitem [{\citenamefont {Onoda}\ and\ \citenamefont
  {Nagaosa}(2002)}]{Onoda2002}%
  \BibitemOpen
  \bibfield  {author} {\bibinfo {author} {\bibfnamefont {M.}~\bibnamefont
  {Onoda}}\ and\ \bibinfo {author} {\bibfnamefont {N.}~\bibnamefont
  {Nagaosa}},\ }\href {https://doi.org/10.1143/JPSJ.71.19} {\bibfield
  {journal} {\bibinfo  {journal} {Journal of the Physical Society of Japan}\
  }\textbf {\bibinfo {volume} {71}},\ \bibinfo {pages} {19} (\bibinfo {year}
  {2002})}\BibitemShut {NoStop}%
\bibitem [{\citenamefont {Berry}(1984)}]{Berry1984}%
  \BibitemOpen
  \bibfield  {author} {\bibinfo {author} {\bibfnamefont {M.~V.}\ \bibnamefont
  {Berry}},\ }\href {https://doi.org/10.1098/rspa.1984.0023} {\bibfield
  {journal} {\bibinfo  {journal} {Proceedings of the Royal Society of London.
  A. Mathematical and Physical Sciences}\ }\textbf {\bibinfo {volume} {392}},\
  \bibinfo {pages} {45} (\bibinfo {year} {1984})}\BibitemShut {NoStop}%
\bibitem [{\citenamefont {Nagaosa}\ \emph {et~al.}(2010)\citenamefont
  {Nagaosa}, \citenamefont {Sinova}, \citenamefont {Onoda}, \citenamefont
  {MacDonald},\ and\ \citenamefont {Ong}}]{RevModPhys.82.1539}%
  \BibitemOpen
  \bibfield  {author} {\bibinfo {author} {\bibfnamefont {N.}~\bibnamefont
  {Nagaosa}}, \bibinfo {author} {\bibfnamefont {J.}~\bibnamefont {Sinova}},
  \bibinfo {author} {\bibfnamefont {S.}~\bibnamefont {Onoda}}, \bibinfo
  {author} {\bibfnamefont {A.~H.}\ \bibnamefont {MacDonald}},\ and\ \bibinfo
  {author} {\bibfnamefont {N.~P.}\ \bibnamefont {Ong}},\ }\href
  {https://doi.org/10.1103/RevModPhys.82.1539} {\bibfield  {journal} {\bibinfo
  {journal} {Rev. Mod. Phys.}\ }\textbf {\bibinfo {volume} {82}},\ \bibinfo
  {pages} {1539} (\bibinfo {year} {2010})}\BibitemShut {NoStop}%
\bibitem [{\citenamefont {Pugh}(1930)}]{PhysRev.36.1503}%
  \BibitemOpen
  \bibfield  {author} {\bibinfo {author} {\bibfnamefont {E.~M.}\ \bibnamefont
  {Pugh}},\ }\href {https://doi.org/10.1103/PhysRev.36.1503} {\bibfield
  {journal} {\bibinfo  {journal} {Phys. Rev.}\ }\textbf {\bibinfo {volume}
  {36}},\ \bibinfo {pages} {1503} (\bibinfo {year} {1930})}\BibitemShut
  {NoStop}%
\bibitem [{\citenamefont {Wang}\ \emph {et~al.}(2016)\citenamefont {Wang},
  \citenamefont {Sun}, \citenamefont {Zhang}, \citenamefont {Pang},\ and\
  \citenamefont {Lei}}]{PhysRevB.94.075135}%
  \BibitemOpen
  \bibfield  {author} {\bibinfo {author} {\bibfnamefont {Q.}~\bibnamefont
  {Wang}}, \bibinfo {author} {\bibfnamefont {S.}~\bibnamefont {Sun}}, \bibinfo
  {author} {\bibfnamefont {X.}~\bibnamefont {Zhang}}, \bibinfo {author}
  {\bibfnamefont {F.}~\bibnamefont {Pang}},\ and\ \bibinfo {author}
  {\bibfnamefont {H.}~\bibnamefont {Lei}},\ }\href
  {https://doi.org/10.1103/PhysRevB.94.075135} {\bibfield  {journal} {\bibinfo
  {journal} {Phys. Rev. B}\ }\textbf {\bibinfo {volume} {94}},\ \bibinfo
  {pages} {075135} (\bibinfo {year} {2016})}\BibitemShut {NoStop}%
\bibitem [{\citenamefont {Ye}\ \emph {et~al.}(2018)\citenamefont {Ye},
  \citenamefont {Kang}, \citenamefont {Liu}, \citenamefont {von Cube},
  \citenamefont {Wicker}, \citenamefont {Suzuki}, \citenamefont {Jozwiak},
  \citenamefont {Bostwick}, \citenamefont {Rotenberg}, \citenamefont {Bell},
  \citenamefont {Fu}, \citenamefont {Comin},\ and\ \citenamefont
  {Checkelsky}}]{Ye2018}%
  \BibitemOpen
  \bibfield  {author} {\bibinfo {author} {\bibfnamefont {L.}~\bibnamefont
  {Ye}}, \bibinfo {author} {\bibfnamefont {M.}~\bibnamefont {Kang}}, \bibinfo
  {author} {\bibfnamefont {J.}~\bibnamefont {Liu}}, \bibinfo {author}
  {\bibfnamefont {F.}~\bibnamefont {von Cube}}, \bibinfo {author}
  {\bibfnamefont {C.~R.}\ \bibnamefont {Wicker}}, \bibinfo {author}
  {\bibfnamefont {T.}~\bibnamefont {Suzuki}}, \bibinfo {author} {\bibfnamefont
  {C.}~\bibnamefont {Jozwiak}}, \bibinfo {author} {\bibfnamefont
  {A.}~\bibnamefont {Bostwick}}, \bibinfo {author} {\bibfnamefont
  {E.}~\bibnamefont {Rotenberg}}, \bibinfo {author} {\bibfnamefont {D.~C.}\
  \bibnamefont {Bell}}, \bibinfo {author} {\bibfnamefont {L.}~\bibnamefont
  {Fu}}, \bibinfo {author} {\bibfnamefont {R.}~\bibnamefont {Comin}},\ and\
  \bibinfo {author} {\bibfnamefont {J.~G.}\ \bibnamefont {Checkelsky}},\ }\href
  {https://doi.org/10.1038/nature25987} {\bibfield  {journal} {\bibinfo
  {journal} {Nature}\ }\textbf {\bibinfo {volume} {555}},\ \bibinfo {pages}
  {638} (\bibinfo {year} {2018})}\BibitemShut {NoStop}%
\bibitem [{\citenamefont {Liu}\ \emph {et~al.}(2018)\citenamefont {Liu},
  \citenamefont {Sun}, \citenamefont {Kumar}, \citenamefont {Muechler},
  \citenamefont {Sun}, \citenamefont {Jiao}, \citenamefont {Yang},
  \citenamefont {Liu}, \citenamefont {Liang}, \citenamefont {Xu}, \citenamefont
  {Kroder}, \citenamefont {S{\"u}{\ss}}, \citenamefont {Borrmann},
  \citenamefont {Shekhar}, \citenamefont {Wang}, \citenamefont {Xi},
  \citenamefont {Wang}, \citenamefont {Schnelle}, \citenamefont {Wirth},
  \citenamefont {Chen}, \citenamefont {Goennenwein},\ and\ \citenamefont
  {Felser}}]{Liu2018}%
  \BibitemOpen
  \bibfield  {author} {\bibinfo {author} {\bibfnamefont {E.}~\bibnamefont
  {Liu}}, \bibinfo {author} {\bibfnamefont {Y.}~\bibnamefont {Sun}}, \bibinfo
  {author} {\bibfnamefont {N.}~\bibnamefont {Kumar}}, \bibinfo {author}
  {\bibfnamefont {L.}~\bibnamefont {Muechler}}, \bibinfo {author}
  {\bibfnamefont {A.}~\bibnamefont {Sun}}, \bibinfo {author} {\bibfnamefont
  {L.}~\bibnamefont {Jiao}}, \bibinfo {author} {\bibfnamefont {S.-Y.}\
  \bibnamefont {Yang}}, \bibinfo {author} {\bibfnamefont {D.}~\bibnamefont
  {Liu}}, \bibinfo {author} {\bibfnamefont {A.}~\bibnamefont {Liang}}, \bibinfo
  {author} {\bibfnamefont {Q.}~\bibnamefont {Xu}}, \bibinfo {author}
  {\bibfnamefont {J.}~\bibnamefont {Kroder}}, \bibinfo {author} {\bibfnamefont
  {V.}~\bibnamefont {S{\"u}{\ss}}}, \bibinfo {author} {\bibfnamefont
  {H.}~\bibnamefont {Borrmann}}, \bibinfo {author} {\bibfnamefont
  {C.}~\bibnamefont {Shekhar}}, \bibinfo {author} {\bibfnamefont
  {Z.}~\bibnamefont {Wang}}, \bibinfo {author} {\bibfnamefont {C.}~\bibnamefont
  {Xi}}, \bibinfo {author} {\bibfnamefont {W.}~\bibnamefont {Wang}}, \bibinfo
  {author} {\bibfnamefont {W.}~\bibnamefont {Schnelle}}, \bibinfo {author}
  {\bibfnamefont {S.}~\bibnamefont {Wirth}}, \bibinfo {author} {\bibfnamefont
  {Y.}~\bibnamefont {Chen}}, \bibinfo {author} {\bibfnamefont {S.~T.~B.}\
  \bibnamefont {Goennenwein}},\ and\ \bibinfo {author} {\bibfnamefont
  {C.}~\bibnamefont {Felser}},\ }\href
  {https://doi.org/10.1038/s41567-018-0234-5} {\bibfield  {journal} {\bibinfo
  {journal} {Nature Physics}\ }\textbf {\bibinfo {volume} {14}},\ \bibinfo
  {pages} {1125} (\bibinfo {year} {2018})}\BibitemShut {NoStop}%
\bibitem [{\citenamefont {Dijkstra}\ \emph {et~al.}(1989)\citenamefont
  {Dijkstra}, \citenamefont {Zijlema}, \citenamefont {van Bruggen},
  \citenamefont {Haas},\ and\ \citenamefont {de~Groot}}]{Dijkstra_1989}%
  \BibitemOpen
  \bibfield  {author} {\bibinfo {author} {\bibfnamefont {J.}~\bibnamefont
  {Dijkstra}}, \bibinfo {author} {\bibfnamefont {P.~J.}\ \bibnamefont
  {Zijlema}}, \bibinfo {author} {\bibfnamefont {C.~F.}\ \bibnamefont {van
  Bruggen}}, \bibinfo {author} {\bibfnamefont {C.}~\bibnamefont {Haas}},\ and\
  \bibinfo {author} {\bibfnamefont {R.~A.}\ \bibnamefont {de~Groot}},\ }\href
  {https://doi.org/10.1088/0953-8984/1/36/005} {\bibfield  {journal} {\bibinfo
  {journal} {Journal of Physics: Condensed Matter}\ }\textbf {\bibinfo {volume}
  {1}},\ \bibinfo {pages} {6363} (\bibinfo {year} {1989})}\BibitemShut
  {NoStop}%
\bibitem [{\citenamefont {Checkelsky}\ \emph {et~al.}(2008)\citenamefont
  {Checkelsky}, \citenamefont {Lee}, \citenamefont {Morosan}, \citenamefont
  {Cava},\ and\ \citenamefont {Ong}}]{PhysRevB.77.014433}%
  \BibitemOpen
  \bibfield  {author} {\bibinfo {author} {\bibfnamefont {J.~G.}\ \bibnamefont
  {Checkelsky}}, \bibinfo {author} {\bibfnamefont {M.}~\bibnamefont {Lee}},
  \bibinfo {author} {\bibfnamefont {E.}~\bibnamefont {Morosan}}, \bibinfo
  {author} {\bibfnamefont {R.~J.}\ \bibnamefont {Cava}},\ and\ \bibinfo
  {author} {\bibfnamefont {N.~P.}\ \bibnamefont {Ong}},\ }\href
  {https://doi.org/10.1103/PhysRevB.77.014433} {\bibfield  {journal} {\bibinfo
  {journal} {Phys. Rev. B}\ }\textbf {\bibinfo {volume} {77}},\ \bibinfo
  {pages} {014433} (\bibinfo {year} {2008})}\BibitemShut {NoStop}%
\bibitem [{\citenamefont {Tenasini}\ \emph {et~al.}(2020)\citenamefont
  {Tenasini}, \citenamefont {Martino}, \citenamefont {Ubrig}, \citenamefont
  {Ghimire}, \citenamefont {Berger}, \citenamefont {Zaharko}, \citenamefont
  {Wu}, \citenamefont {Mitchell}, \citenamefont {Martin}, \citenamefont
  {Forr\'o},\ and\ \citenamefont {Morpurgo}}]{PhysRevResearch.2.023051}%
  \BibitemOpen
  \bibfield  {author} {\bibinfo {author} {\bibfnamefont {G.}~\bibnamefont
  {Tenasini}}, \bibinfo {author} {\bibfnamefont {E.}~\bibnamefont {Martino}},
  \bibinfo {author} {\bibfnamefont {N.}~\bibnamefont {Ubrig}}, \bibinfo
  {author} {\bibfnamefont {N.~J.}\ \bibnamefont {Ghimire}}, \bibinfo {author}
  {\bibfnamefont {H.}~\bibnamefont {Berger}}, \bibinfo {author} {\bibfnamefont
  {O.}~\bibnamefont {Zaharko}}, \bibinfo {author} {\bibfnamefont
  {F.}~\bibnamefont {Wu}}, \bibinfo {author} {\bibfnamefont {J.~F.}\
  \bibnamefont {Mitchell}}, \bibinfo {author} {\bibfnamefont {I.}~\bibnamefont
  {Martin}}, \bibinfo {author} {\bibfnamefont {L.}~\bibnamefont {Forr\'o}},\
  and\ \bibinfo {author} {\bibfnamefont {A.~F.}\ \bibnamefont {Morpurgo}},\
  }\href {https://doi.org/10.1103/PhysRevResearch.2.023051} {\bibfield
  {journal} {\bibinfo  {journal} {Phys. Rev. Research}\ }\textbf {\bibinfo
  {volume} {2}},\ \bibinfo {pages} {023051} (\bibinfo {year}
  {2020})}\BibitemShut {NoStop}%
\bibitem [{\citenamefont {{\v S}mejkal}\ \emph {et~al.}(2020)\citenamefont {{\v
  S}mejkal}, \citenamefont {Gonz{\'a}lez-Hern{\'a}ndez}, \citenamefont
  {Jungwirth},\ and\ \citenamefont {Sinova}}]{Smejkal2020}%
  \BibitemOpen
  \bibfield  {author} {\bibinfo {author} {\bibfnamefont {L.}~\bibnamefont {{\v
  S}mejkal}}, \bibinfo {author} {\bibfnamefont {R.}~\bibnamefont
  {Gonz{\'a}lez-Hern{\'a}ndez}}, \bibinfo {author} {\bibfnamefont
  {T.}~\bibnamefont {Jungwirth}},\ and\ \bibinfo {author} {\bibfnamefont
  {J.}~\bibnamefont {Sinova}},\ }\bibfield  {journal} {\bibinfo  {journal}
  {Science Advances}\ }\textbf {\bibinfo {volume} {6}},\ \href
  {https://doi.org/10.1126/sciadv.aaz8809} {10.1126/sciadv.aaz8809} (\bibinfo
  {year} {2020})\BibitemShut {NoStop}%
\bibitem [{\citenamefont {Chang}\ \emph {et~al.}(2018)\citenamefont {Chang},
  \citenamefont {Wieder}, \citenamefont {Schindler}, \citenamefont {Sanchez},
  \citenamefont {Belopolski}, \citenamefont {Huang}, \citenamefont {Singh},
  \citenamefont {Wu}, \citenamefont {Chang}, \citenamefont {Neupert},
  \citenamefont {Xu}, \citenamefont {Lin},\ and\ \citenamefont
  {Hasan}}]{Chang2018}%
  \BibitemOpen
  \bibfield  {author} {\bibinfo {author} {\bibfnamefont {G.}~\bibnamefont
  {Chang}}, \bibinfo {author} {\bibfnamefont {B.~J.}\ \bibnamefont {Wieder}},
  \bibinfo {author} {\bibfnamefont {F.}~\bibnamefont {Schindler}}, \bibinfo
  {author} {\bibfnamefont {D.~S.}\ \bibnamefont {Sanchez}}, \bibinfo {author}
  {\bibfnamefont {I.}~\bibnamefont {Belopolski}}, \bibinfo {author}
  {\bibfnamefont {S.-M.}\ \bibnamefont {Huang}}, \bibinfo {author}
  {\bibfnamefont {B.}~\bibnamefont {Singh}}, \bibinfo {author} {\bibfnamefont
  {D.}~\bibnamefont {Wu}}, \bibinfo {author} {\bibfnamefont {T.-R.}\
  \bibnamefont {Chang}}, \bibinfo {author} {\bibfnamefont {T.}~\bibnamefont
  {Neupert}}, \bibinfo {author} {\bibfnamefont {S.-Y.}\ \bibnamefont {Xu}},
  \bibinfo {author} {\bibfnamefont {H.}~\bibnamefont {Lin}},\ and\ \bibinfo
  {author} {\bibfnamefont {M.~Z.}\ \bibnamefont {Hasan}},\ }\href
  {https://doi.org/10.1038/s41563-018-0169-3} {\bibfield  {journal} {\bibinfo
  {journal} {Nature Materials}\ }\textbf {\bibinfo {volume} {17}},\ \bibinfo
  {pages} {978} (\bibinfo {year} {2018})}\BibitemShut {NoStop}%
\bibitem [{\citenamefont {Dudin}\ \emph {et~al.}(2010)\citenamefont {Dudin},
  \citenamefont {Lacovig}, \citenamefont {Fava}, \citenamefont {Nicolini},
  \citenamefont {Bianco}, \citenamefont {Cautero},\ and\ \citenamefont
  {Barinov}}]{Dudin2010}%
  \BibitemOpen
  \bibfield  {author} {\bibinfo {author} {\bibfnamefont {P.}~\bibnamefont
  {Dudin}}, \bibinfo {author} {\bibfnamefont {P.}~\bibnamefont {Lacovig}},
  \bibinfo {author} {\bibfnamefont {C.}~\bibnamefont {Fava}}, \bibinfo {author}
  {\bibfnamefont {E.}~\bibnamefont {Nicolini}}, \bibinfo {author}
  {\bibfnamefont {A.}~\bibnamefont {Bianco}}, \bibinfo {author} {\bibfnamefont
  {G.}~\bibnamefont {Cautero}},\ and\ \bibinfo {author} {\bibfnamefont
  {A.}~\bibnamefont {Barinov}},\ }\href
  {https://doi.org/10.1107/S0909049510013993} {\bibfield  {journal} {\bibinfo
  {journal} {Journal of Synchrotron Radiation}\ }\textbf {\bibinfo {volume}
  {17}},\ \bibinfo {pages} {445} (\bibinfo {year} {2010})}\BibitemShut
  {NoStop}%
\bibitem [{\citenamefont {Senba}\ \emph {et~al.}(2016)\citenamefont {Senba},
  \citenamefont {Ohashi}, \citenamefont {Kotani}, \citenamefont {Nakamura},
  \citenamefont {Muro}, \citenamefont {Ohkochi}, \citenamefont {Tsuji},
  \citenamefont {Kishimoto}, \citenamefont {Miura}, \citenamefont {Tanaka},
  \citenamefont {Higashiyama}, \citenamefont {Takahashi}, \citenamefont
  {Ishizawa}, \citenamefont {Matsushita}, \citenamefont {Furukawa},
  \citenamefont {Ohata}, \citenamefont {Nariyama}, \citenamefont {Takeshita},
  \citenamefont {Kinoshita}, \citenamefont {Fujiwara}, \citenamefont {Takata},\
  and\ \citenamefont {Goto}}]{Senba2016}%
  \BibitemOpen
  \bibfield  {author} {\bibinfo {author} {\bibfnamefont {Y.}~\bibnamefont
  {Senba}}, \bibinfo {author} {\bibfnamefont {H.}~\bibnamefont {Ohashi}},
  \bibinfo {author} {\bibfnamefont {Y.}~\bibnamefont {Kotani}}, \bibinfo
  {author} {\bibfnamefont {T.}~\bibnamefont {Nakamura}}, \bibinfo {author}
  {\bibfnamefont {T.}~\bibnamefont {Muro}}, \bibinfo {author} {\bibfnamefont
  {T.}~\bibnamefont {Ohkochi}}, \bibinfo {author} {\bibfnamefont
  {N.}~\bibnamefont {Tsuji}}, \bibinfo {author} {\bibfnamefont
  {H.}~\bibnamefont {Kishimoto}}, \bibinfo {author} {\bibfnamefont
  {T.}~\bibnamefont {Miura}}, \bibinfo {author} {\bibfnamefont
  {M.}~\bibnamefont {Tanaka}}, \bibinfo {author} {\bibfnamefont
  {M.}~\bibnamefont {Higashiyama}}, \bibinfo {author} {\bibfnamefont
  {S.}~\bibnamefont {Takahashi}}, \bibinfo {author} {\bibfnamefont
  {Y.}~\bibnamefont {Ishizawa}}, \bibinfo {author} {\bibfnamefont
  {T.}~\bibnamefont {Matsushita}}, \bibinfo {author} {\bibfnamefont
  {Y.}~\bibnamefont {Furukawa}}, \bibinfo {author} {\bibfnamefont
  {T.}~\bibnamefont {Ohata}}, \bibinfo {author} {\bibfnamefont
  {N.}~\bibnamefont {Nariyama}}, \bibinfo {author} {\bibfnamefont
  {K.}~\bibnamefont {Takeshita}}, \bibinfo {author} {\bibfnamefont
  {T.}~\bibnamefont {Kinoshita}}, \bibinfo {author} {\bibfnamefont
  {A.}~\bibnamefont {Fujiwara}}, \bibinfo {author} {\bibfnamefont
  {M.}~\bibnamefont {Takata}},\ and\ \bibinfo {author} {\bibfnamefont
  {S.}~\bibnamefont {Goto}},\ }\href {https://doi.org/10.1063/1.4952867}
  {\bibfield  {journal} {\bibinfo  {journal} {AIP Conference Proceedings}\
  }\textbf {\bibinfo {volume} {1741}},\ \bibinfo {pages} {030044} (\bibinfo
  {year} {2016})}\BibitemShut {NoStop}%
\bibitem [{\citenamefont {Muro}\ \emph {et~al.}(2021)\citenamefont {Muro},
  \citenamefont {Senba}, \citenamefont {Ohashi}, \citenamefont {Ohkochi},
  \citenamefont {Matsushita}, \citenamefont {Kinoshita},\ and\ \citenamefont
  {Shin}}]{Muro:ok5049}%
  \BibitemOpen
  \bibfield  {author} {\bibinfo {author} {\bibfnamefont {T.}~\bibnamefont
  {Muro}}, \bibinfo {author} {\bibfnamefont {Y.}~\bibnamefont {Senba}},
  \bibinfo {author} {\bibfnamefont {H.}~\bibnamefont {Ohashi}}, \bibinfo
  {author} {\bibfnamefont {T.}~\bibnamefont {Ohkochi}}, \bibinfo {author}
  {\bibfnamefont {T.}~\bibnamefont {Matsushita}}, \bibinfo {author}
  {\bibfnamefont {T.}~\bibnamefont {Kinoshita}},\ and\ \bibinfo {author}
  {\bibfnamefont {S.}~\bibnamefont {Shin}},\ }\href
  {https://doi.org/10.1107/S1600577521007487} {\bibfield  {journal} {\bibinfo
  {journal} {Journal of Synchrotron Radiation}\ }\textbf {\bibinfo {volume}
  {28}} (\bibinfo {year} {2021})}\BibitemShut {NoStop}%
\bibitem [{\citenamefont {Giannozzi}\ \emph {et~al.}(2009)\citenamefont
  {Giannozzi}, \citenamefont {Baroni}, \citenamefont {Bonini}, \citenamefont
  {Calandra}, \citenamefont {Car}, \citenamefont {Cavazzoni}, \citenamefont
  {Ceresoli}, \citenamefont {Chiarotti}, \citenamefont {Cococcioni},
  \citenamefont {Dabo}, \citenamefont {Corso}, \citenamefont {de~Gironcoli},
  \citenamefont {Fabris}, \citenamefont {Fratesi}, \citenamefont {Gebauer},
  \citenamefont {Gerstmann}, \citenamefont {Gougoussis}, \citenamefont
  {Kokalj}, \citenamefont {Lazzeri}, \citenamefont {Martin-Samos},
  \citenamefont {Marzari}, \citenamefont {Mauri}, \citenamefont {Mazzarello},
  \citenamefont {Paolini}, \citenamefont {Pasquarello}, \citenamefont
  {Paulatto}, \citenamefont {Sbraccia}, \citenamefont {Scandolo}, \citenamefont
  {Sclauzero}, \citenamefont {Seitsonen}, \citenamefont {Smogunov},
  \citenamefont {Umari},\ and\ \citenamefont {Wentzcovitch}}]{Giannozzi2009}%
  \BibitemOpen
  \bibfield  {author} {\bibinfo {author} {\bibfnamefont {P.}~\bibnamefont
  {Giannozzi}}, \bibinfo {author} {\bibfnamefont {S.}~\bibnamefont {Baroni}},
  \bibinfo {author} {\bibfnamefont {N.}~\bibnamefont {Bonini}}, \bibinfo
  {author} {\bibfnamefont {M.}~\bibnamefont {Calandra}}, \bibinfo {author}
  {\bibfnamefont {R.}~\bibnamefont {Car}}, \bibinfo {author} {\bibfnamefont
  {C.}~\bibnamefont {Cavazzoni}}, \bibinfo {author} {\bibfnamefont
  {D.}~\bibnamefont {Ceresoli}}, \bibinfo {author} {\bibfnamefont {G.~L.}\
  \bibnamefont {Chiarotti}}, \bibinfo {author} {\bibfnamefont {M.}~\bibnamefont
  {Cococcioni}}, \bibinfo {author} {\bibfnamefont {I.}~\bibnamefont {Dabo}},
  \bibinfo {author} {\bibfnamefont {A.~D.}\ \bibnamefont {Corso}}, \bibinfo
  {author} {\bibfnamefont {S.}~\bibnamefont {de~Gironcoli}}, \bibinfo {author}
  {\bibfnamefont {S.}~\bibnamefont {Fabris}}, \bibinfo {author} {\bibfnamefont
  {G.}~\bibnamefont {Fratesi}}, \bibinfo {author} {\bibfnamefont
  {R.}~\bibnamefont {Gebauer}}, \bibinfo {author} {\bibfnamefont
  {U.}~\bibnamefont {Gerstmann}}, \bibinfo {author} {\bibfnamefont
  {C.}~\bibnamefont {Gougoussis}}, \bibinfo {author} {\bibfnamefont
  {A.}~\bibnamefont {Kokalj}}, \bibinfo {author} {\bibfnamefont
  {M.}~\bibnamefont {Lazzeri}}, \bibinfo {author} {\bibfnamefont
  {L.}~\bibnamefont {Martin-Samos}}, \bibinfo {author} {\bibfnamefont
  {N.}~\bibnamefont {Marzari}}, \bibinfo {author} {\bibfnamefont
  {F.}~\bibnamefont {Mauri}}, \bibinfo {author} {\bibfnamefont
  {R.}~\bibnamefont {Mazzarello}}, \bibinfo {author} {\bibfnamefont
  {S.}~\bibnamefont {Paolini}}, \bibinfo {author} {\bibfnamefont
  {A.}~\bibnamefont {Pasquarello}}, \bibinfo {author} {\bibfnamefont
  {L.}~\bibnamefont {Paulatto}}, \bibinfo {author} {\bibfnamefont
  {C.}~\bibnamefont {Sbraccia}}, \bibinfo {author} {\bibfnamefont
  {S.}~\bibnamefont {Scandolo}}, \bibinfo {author} {\bibfnamefont
  {G.}~\bibnamefont {Sclauzero}}, \bibinfo {author} {\bibfnamefont {A.~P.}\
  \bibnamefont {Seitsonen}}, \bibinfo {author} {\bibfnamefont {A.}~\bibnamefont
  {Smogunov}}, \bibinfo {author} {\bibfnamefont {P.}~\bibnamefont {Umari}},\
  and\ \bibinfo {author} {\bibfnamefont {R.~M.}\ \bibnamefont {Wentzcovitch}},\
  }\href {https://doi.org/10.1088/0953-8984/21/39/395502} {\bibfield  {journal}
  {\bibinfo  {journal} {Journal of Physics: Condensed Matter}\ }\textbf
  {\bibinfo {volume} {21}},\ \bibinfo {pages} {395502} (\bibinfo {year}
  {2009})}\BibitemShut {NoStop}%
\bibitem [{\citenamefont {Giannozzi}\ \emph {et~al.}(2017)\citenamefont
  {Giannozzi}, \citenamefont {Andreussi}, \citenamefont {Brumme}, \citenamefont
  {Bunau}, \citenamefont {Nardelli}, \citenamefont {Calandra}, \citenamefont
  {Car}, \citenamefont {Cavazzoni}, \citenamefont {Ceresoli}, \citenamefont
  {Cococcioni}, \citenamefont {Colonna}, \citenamefont {Carnimeo},
  \citenamefont {Corso}, \citenamefont {de~Gironcoli}, \citenamefont {Delugas},
  \citenamefont {DiStasio}, \citenamefont {Ferretti}, \citenamefont {Floris},
  \citenamefont {Fratesi}, \citenamefont {Fugallo}, \citenamefont {Gebauer},
  \citenamefont {Gerstmann}, \citenamefont {Giustino}, \citenamefont {Gorni},
  \citenamefont {Jia}, \citenamefont {Kawamura}, \citenamefont {Ko},
  \citenamefont {Kokalj}, \citenamefont {K{\"u}{\c{c}}{\"u}kbenli},
  \citenamefont {Lazzeri}, \citenamefont {Marsili}, \citenamefont {Marzari},
  \citenamefont {Mauri}, \citenamefont {Nguyen}, \citenamefont {Nguyen},
  \citenamefont {de-la Roza}, \citenamefont {Paulatto}, \citenamefont
  {Ponc{\'{e}}}, \citenamefont {Rocca}, \citenamefont {Sabatini}, \citenamefont
  {Santra}, \citenamefont {Schlipf}, \citenamefont {Seitsonen}, \citenamefont
  {Smogunov}, \citenamefont {Timrov}, \citenamefont {Thonhauser}, \citenamefont
  {Umari}, \citenamefont {Vast}, \citenamefont {Wu},\ and\ \citenamefont
  {Baroni}}]{Giannozzi2017}%
  \BibitemOpen
  \bibfield  {author} {\bibinfo {author} {\bibfnamefont {P.}~\bibnamefont
  {Giannozzi}}, \bibinfo {author} {\bibfnamefont {O.}~\bibnamefont
  {Andreussi}}, \bibinfo {author} {\bibfnamefont {T.}~\bibnamefont {Brumme}},
  \bibinfo {author} {\bibfnamefont {O.}~\bibnamefont {Bunau}}, \bibinfo
  {author} {\bibfnamefont {M.~B.}\ \bibnamefont {Nardelli}}, \bibinfo {author}
  {\bibfnamefont {M.}~\bibnamefont {Calandra}}, \bibinfo {author}
  {\bibfnamefont {R.}~\bibnamefont {Car}}, \bibinfo {author} {\bibfnamefont
  {C.}~\bibnamefont {Cavazzoni}}, \bibinfo {author} {\bibfnamefont
  {D.}~\bibnamefont {Ceresoli}}, \bibinfo {author} {\bibfnamefont
  {M.}~\bibnamefont {Cococcioni}}, \bibinfo {author} {\bibfnamefont
  {N.}~\bibnamefont {Colonna}}, \bibinfo {author} {\bibfnamefont
  {I.}~\bibnamefont {Carnimeo}}, \bibinfo {author} {\bibfnamefont {A.~D.}\
  \bibnamefont {Corso}}, \bibinfo {author} {\bibfnamefont {S.}~\bibnamefont
  {de~Gironcoli}}, \bibinfo {author} {\bibfnamefont {P.}~\bibnamefont
  {Delugas}}, \bibinfo {author} {\bibfnamefont {R.~A.}\ \bibnamefont
  {DiStasio}}, \bibinfo {author} {\bibfnamefont {A.}~\bibnamefont {Ferretti}},
  \bibinfo {author} {\bibfnamefont {A.}~\bibnamefont {Floris}}, \bibinfo
  {author} {\bibfnamefont {G.}~\bibnamefont {Fratesi}}, \bibinfo {author}
  {\bibfnamefont {G.}~\bibnamefont {Fugallo}}, \bibinfo {author} {\bibfnamefont
  {R.}~\bibnamefont {Gebauer}}, \bibinfo {author} {\bibfnamefont
  {U.}~\bibnamefont {Gerstmann}}, \bibinfo {author} {\bibfnamefont
  {F.}~\bibnamefont {Giustino}}, \bibinfo {author} {\bibfnamefont
  {T.}~\bibnamefont {Gorni}}, \bibinfo {author} {\bibfnamefont
  {J.}~\bibnamefont {Jia}}, \bibinfo {author} {\bibfnamefont {M.}~\bibnamefont
  {Kawamura}}, \bibinfo {author} {\bibfnamefont {H.-Y.}\ \bibnamefont {Ko}},
  \bibinfo {author} {\bibfnamefont {A.}~\bibnamefont {Kokalj}}, \bibinfo
  {author} {\bibfnamefont {E.}~\bibnamefont {K{\"u}{\c{c}}{\"u}kbenli}},
  \bibinfo {author} {\bibfnamefont {M.}~\bibnamefont {Lazzeri}}, \bibinfo
  {author} {\bibfnamefont {M.}~\bibnamefont {Marsili}}, \bibinfo {author}
  {\bibfnamefont {N.}~\bibnamefont {Marzari}}, \bibinfo {author} {\bibfnamefont
  {F.}~\bibnamefont {Mauri}}, \bibinfo {author} {\bibfnamefont {N.~L.}\
  \bibnamefont {Nguyen}}, \bibinfo {author} {\bibfnamefont {H.-V.}\
  \bibnamefont {Nguyen}}, \bibinfo {author} {\bibfnamefont {A.~O.}\
  \bibnamefont {de-la Roza}}, \bibinfo {author} {\bibfnamefont
  {L.}~\bibnamefont {Paulatto}}, \bibinfo {author} {\bibfnamefont
  {S.}~\bibnamefont {Ponc{\'{e}}}}, \bibinfo {author} {\bibfnamefont
  {D.}~\bibnamefont {Rocca}}, \bibinfo {author} {\bibfnamefont
  {R.}~\bibnamefont {Sabatini}}, \bibinfo {author} {\bibfnamefont
  {B.}~\bibnamefont {Santra}}, \bibinfo {author} {\bibfnamefont
  {M.}~\bibnamefont {Schlipf}}, \bibinfo {author} {\bibfnamefont {A.~P.}\
  \bibnamefont {Seitsonen}}, \bibinfo {author} {\bibfnamefont {A.}~\bibnamefont
  {Smogunov}}, \bibinfo {author} {\bibfnamefont {I.}~\bibnamefont {Timrov}},
  \bibinfo {author} {\bibfnamefont {T.}~\bibnamefont {Thonhauser}}, \bibinfo
  {author} {\bibfnamefont {P.}~\bibnamefont {Umari}}, \bibinfo {author}
  {\bibfnamefont {N.}~\bibnamefont {Vast}}, \bibinfo {author} {\bibfnamefont
  {X.}~\bibnamefont {Wu}},\ and\ \bibinfo {author} {\bibfnamefont
  {S.}~\bibnamefont {Baroni}},\ }\href
  {https://doi.org/10.1088/1361-648x/aa8f79} {\bibfield  {journal} {\bibinfo
  {journal} {Journal of Physics: Condensed Matter}\ }\textbf {\bibinfo {volume}
  {29}},\ \bibinfo {pages} {465901} (\bibinfo {year} {2017})}\BibitemShut
  {NoStop}%
\bibitem [{\citenamefont {Rappe}\ \emph {et~al.}(1990)\citenamefont {Rappe},
  \citenamefont {Rabe}, \citenamefont {Kaxiras},\ and\ \citenamefont
  {Joannopoulos}}]{PhysRevB.41.1227}%
  \BibitemOpen
  \bibfield  {author} {\bibinfo {author} {\bibfnamefont {A.~M.}\ \bibnamefont
  {Rappe}}, \bibinfo {author} {\bibfnamefont {K.~M.}\ \bibnamefont {Rabe}},
  \bibinfo {author} {\bibfnamefont {E.}~\bibnamefont {Kaxiras}},\ and\ \bibinfo
  {author} {\bibfnamefont {J.~D.}\ \bibnamefont {Joannopoulos}},\ }\href
  {https://doi.org/10.1103/PhysRevB.41.1227} {\bibfield  {journal} {\bibinfo
  {journal} {Phys. Rev. B}\ }\textbf {\bibinfo {volume} {41}},\ \bibinfo
  {pages} {1227} (\bibinfo {year} {1990})}\BibitemShut {NoStop}%
\bibitem [{Pse()}]{Pseudopotentials}%
  \BibitemOpen
  \href@noop {} {}\bibinfo {note} {We used the pseudopotentials
  Nb.pbe-spn-rrkjus\_psl.1.0.0.UPF and S.pbe-n-rrkjus\_psl.1.0.0.UPF from the
  {\sc quantum espresso} pseudopotential data base:
  http://www.quantum-espresso.org/pseudopotentials .}\BibitemShut {Stop}%
\bibitem [{\citenamefont {Pizzi}\ \emph {et~al.}(2020)\citenamefont {Pizzi},
  \citenamefont {Vitale}, \citenamefont {Arita}, \citenamefont {Bl{\"u}gel},
  \citenamefont {Freimuth}, \citenamefont {G{\'{e}}ranton}, \citenamefont
  {Gibertini}, \citenamefont {Gresch}, \citenamefont {Johnson}, \citenamefont
  {Koretsune}, \citenamefont {Iba{\~{n}}ez-Azpiroz}, \citenamefont {Lee},
  \citenamefont {Lihm}, \citenamefont {Marchand}, \citenamefont {Marrazzo},
  \citenamefont {Mokrousov}, \citenamefont {Mustafa}, \citenamefont {Nohara},
  \citenamefont {Nomura}, \citenamefont {Paulatto}, \citenamefont
  {Ponc{\'{e}}}, \citenamefont {Ponweiser}, \citenamefont {Qiao}, \citenamefont
  {Th{\"o}le}, \citenamefont {Tsirkin}, \citenamefont {Wierzbowska},
  \citenamefont {Marzari}, \citenamefont {Vanderbilt}, \citenamefont {Souza},
  \citenamefont {Mostofi},\ and\ \citenamefont {Yates}}]{Pizzi2020}%
  \BibitemOpen
  \bibfield  {author} {\bibinfo {author} {\bibfnamefont {G.}~\bibnamefont
  {Pizzi}}, \bibinfo {author} {\bibfnamefont {V.}~\bibnamefont {Vitale}},
  \bibinfo {author} {\bibfnamefont {R.}~\bibnamefont {Arita}}, \bibinfo
  {author} {\bibfnamefont {S.}~\bibnamefont {Bl{\"u}gel}}, \bibinfo {author}
  {\bibfnamefont {F.}~\bibnamefont {Freimuth}}, \bibinfo {author}
  {\bibfnamefont {G.}~\bibnamefont {G{\'{e}}ranton}}, \bibinfo {author}
  {\bibfnamefont {M.}~\bibnamefont {Gibertini}}, \bibinfo {author}
  {\bibfnamefont {D.}~\bibnamefont {Gresch}}, \bibinfo {author} {\bibfnamefont
  {C.}~\bibnamefont {Johnson}}, \bibinfo {author} {\bibfnamefont
  {T.}~\bibnamefont {Koretsune}}, \bibinfo {author} {\bibfnamefont
  {J.}~\bibnamefont {Iba{\~{n}}ez-Azpiroz}}, \bibinfo {author} {\bibfnamefont
  {H.}~\bibnamefont {Lee}}, \bibinfo {author} {\bibfnamefont {J.-M.}\
  \bibnamefont {Lihm}}, \bibinfo {author} {\bibfnamefont {D.}~\bibnamefont
  {Marchand}}, \bibinfo {author} {\bibfnamefont {A.}~\bibnamefont {Marrazzo}},
  \bibinfo {author} {\bibfnamefont {Y.}~\bibnamefont {Mokrousov}}, \bibinfo
  {author} {\bibfnamefont {J.~I.}\ \bibnamefont {Mustafa}}, \bibinfo {author}
  {\bibfnamefont {Y.}~\bibnamefont {Nohara}}, \bibinfo {author} {\bibfnamefont
  {Y.}~\bibnamefont {Nomura}}, \bibinfo {author} {\bibfnamefont
  {L.}~\bibnamefont {Paulatto}}, \bibinfo {author} {\bibfnamefont
  {S.}~\bibnamefont {Ponc{\'{e}}}}, \bibinfo {author} {\bibfnamefont
  {T.}~\bibnamefont {Ponweiser}}, \bibinfo {author} {\bibfnamefont
  {J.}~\bibnamefont {Qiao}}, \bibinfo {author} {\bibfnamefont {F.}~\bibnamefont
  {Th{\"o}le}}, \bibinfo {author} {\bibfnamefont {S.~S.}\ \bibnamefont
  {Tsirkin}}, \bibinfo {author} {\bibfnamefont {M.}~\bibnamefont
  {Wierzbowska}}, \bibinfo {author} {\bibfnamefont {N.}~\bibnamefont
  {Marzari}}, \bibinfo {author} {\bibfnamefont {D.}~\bibnamefont {Vanderbilt}},
  \bibinfo {author} {\bibfnamefont {I.}~\bibnamefont {Souza}}, \bibinfo
  {author} {\bibfnamefont {A.~A.}\ \bibnamefont {Mostofi}},\ and\ \bibinfo
  {author} {\bibfnamefont {J.~R.}\ \bibnamefont {Yates}},\ }\href
  {https://doi.org/10.1088/1361-648x/ab51ff} {\bibfield  {journal} {\bibinfo
  {journal} {Journal of Physics: Condensed Matter}\ }\textbf {\bibinfo {volume}
  {32}},\ \bibinfo {pages} {165902} (\bibinfo {year} {2020})}\BibitemShut
  {NoStop}%
\bibitem [{\citenamefont {Battaglia}\ \emph {et~al.}(2007)\citenamefont
  {Battaglia}, \citenamefont {Cercellier}, \citenamefont {Despont},
  \citenamefont {Monney}, \citenamefont {Prester}, \citenamefont {Berger},
  \citenamefont {Forr{\'o}}, \citenamefont {Garnier},\ and\ \citenamefont
  {Aebi}}]{Battaglia2007}%
  \BibitemOpen
  \bibfield  {author} {\bibinfo {author} {\bibfnamefont {C.}~\bibnamefont
  {Battaglia}}, \bibinfo {author} {\bibfnamefont {H.}~\bibnamefont
  {Cercellier}}, \bibinfo {author} {\bibfnamefont {L.}~\bibnamefont {Despont}},
  \bibinfo {author} {\bibfnamefont {C.}~\bibnamefont {Monney}}, \bibinfo
  {author} {\bibfnamefont {M.}~\bibnamefont {Prester}}, \bibinfo {author}
  {\bibfnamefont {H.}~\bibnamefont {Berger}}, \bibinfo {author} {\bibfnamefont
  {L.}~\bibnamefont {Forr{\'o}}}, \bibinfo {author} {\bibfnamefont {M.~G.}\
  \bibnamefont {Garnier}},\ and\ \bibinfo {author} {\bibfnamefont
  {P.}~\bibnamefont {Aebi}},\ }\href
  {https://doi.org/10.1140/epjb/e2007-00188-1} {\bibfield  {journal} {\bibinfo
  {journal} {The European Physical Journal B}\ }\textbf {\bibinfo {volume}
  {57}},\ \bibinfo {pages} {385} (\bibinfo {year} {2007})}\BibitemShut
  {NoStop}%
\bibitem [{\citenamefont {Sirica}\ \emph {et~al.}(2016)\citenamefont {Sirica},
  \citenamefont {Mo}, \citenamefont {Bondino}, \citenamefont {Pis},
  \citenamefont {Nappini}, \citenamefont {Vilmercati}, \citenamefont {Yi},
  \citenamefont {Gai}, \citenamefont {Snijders}, \citenamefont {Das},
  \citenamefont {Vobornik}, \citenamefont {Ghimire}, \citenamefont {Koehler},
  \citenamefont {Li}, \citenamefont {Sapkota}, \citenamefont {Parker},
  \citenamefont {Mandrus},\ and\ \citenamefont
  {Mannella}}]{PhysRevB.94.075141}%
  \BibitemOpen
  \bibfield  {author} {\bibinfo {author} {\bibfnamefont {N.}~\bibnamefont
  {Sirica}}, \bibinfo {author} {\bibfnamefont {S.-K.}\ \bibnamefont {Mo}},
  \bibinfo {author} {\bibfnamefont {F.}~\bibnamefont {Bondino}}, \bibinfo
  {author} {\bibfnamefont {I.}~\bibnamefont {Pis}}, \bibinfo {author}
  {\bibfnamefont {S.}~\bibnamefont {Nappini}}, \bibinfo {author} {\bibfnamefont
  {P.}~\bibnamefont {Vilmercati}}, \bibinfo {author} {\bibfnamefont
  {J.}~\bibnamefont {Yi}}, \bibinfo {author} {\bibfnamefont {Z.}~\bibnamefont
  {Gai}}, \bibinfo {author} {\bibfnamefont {P.~C.}\ \bibnamefont {Snijders}},
  \bibinfo {author} {\bibfnamefont {P.~K.}\ \bibnamefont {Das}}, \bibinfo
  {author} {\bibfnamefont {I.}~\bibnamefont {Vobornik}}, \bibinfo {author}
  {\bibfnamefont {N.}~\bibnamefont {Ghimire}}, \bibinfo {author} {\bibfnamefont
  {M.~R.}\ \bibnamefont {Koehler}}, \bibinfo {author} {\bibfnamefont
  {L.}~\bibnamefont {Li}}, \bibinfo {author} {\bibfnamefont {D.}~\bibnamefont
  {Sapkota}}, \bibinfo {author} {\bibfnamefont {D.~S.}\ \bibnamefont {Parker}},
  \bibinfo {author} {\bibfnamefont {D.~G.}\ \bibnamefont {Mandrus}},\ and\
  \bibinfo {author} {\bibfnamefont {N.}~\bibnamefont {Mannella}},\ }\href
  {https://doi.org/10.1103/PhysRevB.94.075141} {\bibfield  {journal} {\bibinfo
  {journal} {Phys. Rev. B}\ }\textbf {\bibinfo {volume} {94}},\ \bibinfo
  {pages} {075141} (\bibinfo {year} {2016})}\BibitemShut {NoStop}%
\bibitem [{\citenamefont {Strocov}\ \emph {et~al.}(2019)\citenamefont
  {Strocov}, \citenamefont {Lev}, \citenamefont {Kobayashi}, \citenamefont
  {Cancellieri}, \citenamefont {Husanu}, \citenamefont {Chikina}, \citenamefont
  {Schr{\"o}ter}, \citenamefont {Wang}, \citenamefont {Krieger},\ and\
  \citenamefont {Salman}}]{STROCOV2019}%
  \BibitemOpen
  \bibfield  {author} {\bibinfo {author} {\bibfnamefont {V.}~\bibnamefont
  {Strocov}}, \bibinfo {author} {\bibfnamefont {L.}~\bibnamefont {Lev}},
  \bibinfo {author} {\bibfnamefont {M.}~\bibnamefont {Kobayashi}}, \bibinfo
  {author} {\bibfnamefont {C.}~\bibnamefont {Cancellieri}}, \bibinfo {author}
  {\bibfnamefont {M.-A.}\ \bibnamefont {Husanu}}, \bibinfo {author}
  {\bibfnamefont {A.}~\bibnamefont {Chikina}}, \bibinfo {author} {\bibfnamefont
  {N.}~\bibnamefont {Schr{\"o}ter}}, \bibinfo {author} {\bibfnamefont
  {X.}~\bibnamefont {Wang}}, \bibinfo {author} {\bibfnamefont {J.}~\bibnamefont
  {Krieger}},\ and\ \bibinfo {author} {\bibfnamefont {Z.}~\bibnamefont
  {Salman}},\ }\href
  {https://doi.org/https://doi.org/10.1016/j.elspec.2019.06.009} {\bibfield
  {journal} {\bibinfo  {journal} {Journal of Electron Spectroscopy and Related
  Phenomena}\ }\textbf {\bibinfo {volume} {236}},\ \bibinfo {pages} {1}
  (\bibinfo {year} {2019})}\BibitemShut {NoStop}%
\bibitem [{\citenamefont {Xu}\ \emph {et~al.}(2015)\citenamefont {Xu},
  \citenamefont {Belopolski}, \citenamefont {Alidoust}, \citenamefont
  {Neupane}, \citenamefont {Bian}, \citenamefont {Zhang}, \citenamefont
  {Sankar}, \citenamefont {Chang}, \citenamefont {Yuan}, \citenamefont {Lee},
  \citenamefont {Huang}, \citenamefont {Zheng}, \citenamefont {Ma},
  \citenamefont {Sanchez}, \citenamefont {Wang}, \citenamefont {Bansil},
  \citenamefont {Chou}, \citenamefont {Shibayev}, \citenamefont {Lin},
  \citenamefont {Jia},\ and\ \citenamefont {Hasan}}]{Xu613}%
  \BibitemOpen
  \bibfield  {author} {\bibinfo {author} {\bibfnamefont {S.-Y.}\ \bibnamefont
  {Xu}}, \bibinfo {author} {\bibfnamefont {I.}~\bibnamefont {Belopolski}},
  \bibinfo {author} {\bibfnamefont {N.}~\bibnamefont {Alidoust}}, \bibinfo
  {author} {\bibfnamefont {M.}~\bibnamefont {Neupane}}, \bibinfo {author}
  {\bibfnamefont {G.}~\bibnamefont {Bian}}, \bibinfo {author} {\bibfnamefont
  {C.}~\bibnamefont {Zhang}}, \bibinfo {author} {\bibfnamefont
  {R.}~\bibnamefont {Sankar}}, \bibinfo {author} {\bibfnamefont
  {G.}~\bibnamefont {Chang}}, \bibinfo {author} {\bibfnamefont
  {Z.}~\bibnamefont {Yuan}}, \bibinfo {author} {\bibfnamefont {C.-C.}\
  \bibnamefont {Lee}}, \bibinfo {author} {\bibfnamefont {S.-M.}\ \bibnamefont
  {Huang}}, \bibinfo {author} {\bibfnamefont {H.}~\bibnamefont {Zheng}},
  \bibinfo {author} {\bibfnamefont {J.}~\bibnamefont {Ma}}, \bibinfo {author}
  {\bibfnamefont {D.~S.}\ \bibnamefont {Sanchez}}, \bibinfo {author}
  {\bibfnamefont {B.}~\bibnamefont {Wang}}, \bibinfo {author} {\bibfnamefont
  {A.}~\bibnamefont {Bansil}}, \bibinfo {author} {\bibfnamefont
  {F.}~\bibnamefont {Chou}}, \bibinfo {author} {\bibfnamefont {P.~P.}\
  \bibnamefont {Shibayev}}, \bibinfo {author} {\bibfnamefont {H.}~\bibnamefont
  {Lin}}, \bibinfo {author} {\bibfnamefont {S.}~\bibnamefont {Jia}},\ and\
  \bibinfo {author} {\bibfnamefont {M.~Z.}\ \bibnamefont {Hasan}},\ }\href
  {https://doi.org/10.1126/science.aaa9297} {\bibfield  {journal} {\bibinfo
  {journal} {Science}\ }\textbf {\bibinfo {volume} {349}},\ \bibinfo {pages}
  {613} (\bibinfo {year} {2015})}\BibitemShut {NoStop}%
\bibitem [{Fn1()}]{Fn1}%
  \BibitemOpen
  \href@noop {} {}\bibinfo {note} {In this argument, we used DFT calculations
  of $\mathrm{Nb}\mathrm{S}_2$ to estimate the unoccupied states. Since the
  unoccupied bands are located above $E_\mathrm{F}+2\ \mathrm{eV}$ as descibed
  in Fig.\ S12(e), these bands cannot appear even if the band shift at most 1
  eV, which we evaluated by XPS measurements, is considered.}\BibitemShut
  {Stop}%
\bibitem [{\citenamefont {Di~Sante}\ \emph {et~al.}(2017)\citenamefont
  {Di~Sante}, \citenamefont {Das}, \citenamefont {Bigi}, \citenamefont
  {Erg\"onenc}, \citenamefont {G\"urtler}, \citenamefont {Krieger},
  \citenamefont {Schmitt}, \citenamefont {Ali}, \citenamefont {Rossi},
  \citenamefont {Thomale}, \citenamefont {Franchini}, \citenamefont {Picozzi},
  \citenamefont {Fujii}, \citenamefont {Strocov}, \citenamefont {Sangiovanni},
  \citenamefont {Vobornik}, \citenamefont {Cava},\ and\ \citenamefont
  {Panaccione}}]{PhysRevLett.119.026403}%
  \BibitemOpen
  \bibfield  {author} {\bibinfo {author} {\bibfnamefont {D.}~\bibnamefont
  {Di~Sante}}, \bibinfo {author} {\bibfnamefont {P.~K.}\ \bibnamefont {Das}},
  \bibinfo {author} {\bibfnamefont {C.}~\bibnamefont {Bigi}}, \bibinfo {author}
  {\bibfnamefont {Z.}~\bibnamefont {Erg\"onenc}}, \bibinfo {author}
  {\bibfnamefont {N.}~\bibnamefont {G\"urtler}}, \bibinfo {author}
  {\bibfnamefont {J.~A.}\ \bibnamefont {Krieger}}, \bibinfo {author}
  {\bibfnamefont {T.}~\bibnamefont {Schmitt}}, \bibinfo {author} {\bibfnamefont
  {M.~N.}\ \bibnamefont {Ali}}, \bibinfo {author} {\bibfnamefont
  {G.}~\bibnamefont {Rossi}}, \bibinfo {author} {\bibfnamefont
  {R.}~\bibnamefont {Thomale}}, \bibinfo {author} {\bibfnamefont
  {C.}~\bibnamefont {Franchini}}, \bibinfo {author} {\bibfnamefont
  {S.}~\bibnamefont {Picozzi}}, \bibinfo {author} {\bibfnamefont
  {J.}~\bibnamefont {Fujii}}, \bibinfo {author} {\bibfnamefont {V.~N.}\
  \bibnamefont {Strocov}}, \bibinfo {author} {\bibfnamefont {G.}~\bibnamefont
  {Sangiovanni}}, \bibinfo {author} {\bibfnamefont {I.}~\bibnamefont
  {Vobornik}}, \bibinfo {author} {\bibfnamefont {R.~J.}\ \bibnamefont {Cava}},\
  and\ \bibinfo {author} {\bibfnamefont {G.}~\bibnamefont {Panaccione}},\
  }\href {https://doi.org/10.1103/PhysRevLett.119.026403} {\bibfield  {journal}
  {\bibinfo  {journal} {Phys. Rev. Lett.}\ }\textbf {\bibinfo {volume} {119}},\
  \bibinfo {pages} {026403} (\bibinfo {year} {2017})}\BibitemShut {NoStop}%
\bibitem [{\citenamefont {Matsui}\ \emph {et~al.}(2018)\citenamefont {Matsui},
  \citenamefont {Nishikawa}, \citenamefont {Daimon}, \citenamefont {Muntwiler},
  \citenamefont {Takizawa}, \citenamefont {Namba},\ and\ \citenamefont
  {Greber}}]{PhysRevB.97.045430}%
  \BibitemOpen
  \bibfield  {author} {\bibinfo {author} {\bibfnamefont {F.}~\bibnamefont
  {Matsui}}, \bibinfo {author} {\bibfnamefont {H.}~\bibnamefont {Nishikawa}},
  \bibinfo {author} {\bibfnamefont {H.}~\bibnamefont {Daimon}}, \bibinfo
  {author} {\bibfnamefont {M.}~\bibnamefont {Muntwiler}}, \bibinfo {author}
  {\bibfnamefont {M.}~\bibnamefont {Takizawa}}, \bibinfo {author}
  {\bibfnamefont {H.}~\bibnamefont {Namba}},\ and\ \bibinfo {author}
  {\bibfnamefont {T.}~\bibnamefont {Greber}},\ }\href
  {https://doi.org/10.1103/PhysRevB.97.045430} {\bibfield  {journal} {\bibinfo
  {journal} {Phys. Rev. B}\ }\textbf {\bibinfo {volume} {97}},\ \bibinfo
  {pages} {045430} (\bibinfo {year} {2018})}\BibitemShut {NoStop}%
\bibitem [{\citenamefont {Moser}(2017)}]{MOSER2017}%
  \BibitemOpen
  \bibfield  {author} {\bibinfo {author} {\bibfnamefont {S.}~\bibnamefont
  {Moser}},\ }\href
  {https://doi.org/https://doi.org/10.1016/j.elspec.2016.11.007} {\bibfield
  {journal} {\bibinfo  {journal} {Journal of Electron Spectroscopy and Related
  Phenomena}\ }\textbf {\bibinfo {volume} {214}},\ \bibinfo {pages} {29}
  (\bibinfo {year} {2017})}\BibitemShut {NoStop}%
\bibitem [{\citenamefont {Liang}\ \emph {et~al.}(2016)\citenamefont {Liang},
  \citenamefont {Zhou}, \citenamefont {Yu}, \citenamefont {Wang},\ and\
  \citenamefont {Weng}}]{PhysRevB.93.085427}%
  \BibitemOpen
  \bibfield  {author} {\bibinfo {author} {\bibfnamefont {Q.-F.}\ \bibnamefont
  {Liang}}, \bibinfo {author} {\bibfnamefont {J.}~\bibnamefont {Zhou}},
  \bibinfo {author} {\bibfnamefont {R.}~\bibnamefont {Yu}}, \bibinfo {author}
  {\bibfnamefont {Z.}~\bibnamefont {Wang}},\ and\ \bibinfo {author}
  {\bibfnamefont {H.}~\bibnamefont {Weng}},\ }\href
  {https://doi.org/10.1103/PhysRevB.93.085427} {\bibfield  {journal} {\bibinfo
  {journal} {Phys. Rev. B}\ }\textbf {\bibinfo {volume} {93}},\ \bibinfo
  {pages} {085427} (\bibinfo {year} {2016})}\BibitemShut {NoStop}%
\bibitem [{\citenamefont {Funada}\ \emph {et~al.}(2019)\citenamefont {Funada},
  \citenamefont {Yamakage}, \citenamefont {Yamashina},\ and\ \citenamefont
  {Kageyama}}]{Funada2019}%
  \BibitemOpen
  \bibfield  {author} {\bibinfo {author} {\bibfnamefont {K.}~\bibnamefont
  {Funada}}, \bibinfo {author} {\bibfnamefont {A.}~\bibnamefont {Yamakage}},
  \bibinfo {author} {\bibfnamefont {N.}~\bibnamefont {Yamashina}},\ and\
  \bibinfo {author} {\bibfnamefont {H.}~\bibnamefont {Kageyama}},\ }\href
  {https://doi.org/10.7566/JPSJ.88.044711} {\bibfield  {journal} {\bibinfo
  {journal} {Journal of the Physical Society of Japan}\ }\textbf {\bibinfo
  {volume} {88}},\ \bibinfo {pages} {044711} (\bibinfo {year}
  {2019})}\BibitemShut {NoStop}%
\bibitem [{\citenamefont {Yang}\ \emph {et~al.}(2017)\citenamefont {Yang},
  \citenamefont {Sun}, \citenamefont {Zhang}, \citenamefont {Shi},
  \citenamefont {Parkin},\ and\ \citenamefont {Yan}}]{Yang2017}%
  \BibitemOpen
  \bibfield  {author} {\bibinfo {author} {\bibfnamefont {H.}~\bibnamefont
  {Yang}}, \bibinfo {author} {\bibfnamefont {Y.}~\bibnamefont {Sun}}, \bibinfo
  {author} {\bibfnamefont {Y.}~\bibnamefont {Zhang}}, \bibinfo {author}
  {\bibfnamefont {W.-J.}\ \bibnamefont {Shi}}, \bibinfo {author} {\bibfnamefont
  {S.~S.~P.}\ \bibnamefont {Parkin}},\ and\ \bibinfo {author} {\bibfnamefont
  {B.}~\bibnamefont {Yan}},\ }\href {https://doi.org/10.1088/1367-2630/aa5487}
  {\bibfield  {journal} {\bibinfo  {journal} {New Journal of Physics}\ }\textbf
  {\bibinfo {volume} {19}},\ \bibinfo {pages} {015008} (\bibinfo {year}
  {2017})}\BibitemShut {NoStop}%
\bibitem [{\citenamefont {Kuroda}\ \emph {et~al.}(2017)\citenamefont {Kuroda},
  \citenamefont {Tomita}, \citenamefont {Suzuki}, \citenamefont {Bareille},
  \citenamefont {Nugroho}, \citenamefont {Goswami}, \citenamefont {Ochi},
  \citenamefont {Ikhlas}, \citenamefont {Nakayama}, \citenamefont {Akebi},
  \citenamefont {Noguchi}, \citenamefont {Ishii}, \citenamefont {Inami},
  \citenamefont {Ono}, \citenamefont {Kumigashira}, \citenamefont {Varykhalov},
  \citenamefont {Muro}, \citenamefont {Koretsune}, \citenamefont {Arita},
  \citenamefont {Shin}, \citenamefont {Kondo},\ and\ \citenamefont
  {Nakatsuji}}]{Kuroda2017}%
  \BibitemOpen
  \bibfield  {author} {\bibinfo {author} {\bibfnamefont {K.}~\bibnamefont
  {Kuroda}}, \bibinfo {author} {\bibfnamefont {T.}~\bibnamefont {Tomita}},
  \bibinfo {author} {\bibfnamefont {M.-T.}\ \bibnamefont {Suzuki}}, \bibinfo
  {author} {\bibfnamefont {C.}~\bibnamefont {Bareille}}, \bibinfo {author}
  {\bibfnamefont {A.~A.}\ \bibnamefont {Nugroho}}, \bibinfo {author}
  {\bibfnamefont {P.}~\bibnamefont {Goswami}}, \bibinfo {author} {\bibfnamefont
  {M.}~\bibnamefont {Ochi}}, \bibinfo {author} {\bibfnamefont {M.}~\bibnamefont
  {Ikhlas}}, \bibinfo {author} {\bibfnamefont {M.}~\bibnamefont {Nakayama}},
  \bibinfo {author} {\bibfnamefont {S.}~\bibnamefont {Akebi}}, \bibinfo
  {author} {\bibfnamefont {R.}~\bibnamefont {Noguchi}}, \bibinfo {author}
  {\bibfnamefont {R.}~\bibnamefont {Ishii}}, \bibinfo {author} {\bibfnamefont
  {N.}~\bibnamefont {Inami}}, \bibinfo {author} {\bibfnamefont
  {K.}~\bibnamefont {Ono}}, \bibinfo {author} {\bibfnamefont {H.}~\bibnamefont
  {Kumigashira}}, \bibinfo {author} {\bibfnamefont {A.}~\bibnamefont
  {Varykhalov}}, \bibinfo {author} {\bibfnamefont {T.}~\bibnamefont {Muro}},
  \bibinfo {author} {\bibfnamefont {T.}~\bibnamefont {Koretsune}}, \bibinfo
  {author} {\bibfnamefont {R.}~\bibnamefont {Arita}}, \bibinfo {author}
  {\bibfnamefont {S.}~\bibnamefont {Shin}}, \bibinfo {author} {\bibfnamefont
  {T.}~\bibnamefont {Kondo}},\ and\ \bibinfo {author} {\bibfnamefont
  {S.}~\bibnamefont {Nakatsuji}},\ }\href {https://doi.org/10.1038/nmat4987}
  {\bibfield  {journal} {\bibinfo  {journal} {Nature Materials}\ }\textbf
  {\bibinfo {volume} {16}},\ \bibinfo {pages} {1090} (\bibinfo {year}
  {2017})}\BibitemShut {NoStop}%
\bibitem [{\citenamefont {Liu}\ \emph {et~al.}(2019)\citenamefont {Liu},
  \citenamefont {Liang}, \citenamefont {Liu}, \citenamefont {Xu}, \citenamefont
  {Li}, \citenamefont {Chen}, \citenamefont {Pei}, \citenamefont {Shi},
  \citenamefont {Mo}, \citenamefont {Dudin}, \citenamefont {Kim}, \citenamefont
  {Cacho}, \citenamefont {Li}, \citenamefont {Sun}, \citenamefont {Yang},
  \citenamefont {Liu}, \citenamefont {Parkin}, \citenamefont {Felser},\ and\
  \citenamefont {Chen}}]{Liu2019}%
  \BibitemOpen
  \bibfield  {author} {\bibinfo {author} {\bibfnamefont {D.~F.}\ \bibnamefont
  {Liu}}, \bibinfo {author} {\bibfnamefont {A.~J.}\ \bibnamefont {Liang}},
  \bibinfo {author} {\bibfnamefont {E.~K.}\ \bibnamefont {Liu}}, \bibinfo
  {author} {\bibfnamefont {Q.~N.}\ \bibnamefont {Xu}}, \bibinfo {author}
  {\bibfnamefont {Y.~W.}\ \bibnamefont {Li}}, \bibinfo {author} {\bibfnamefont
  {C.}~\bibnamefont {Chen}}, \bibinfo {author} {\bibfnamefont {D.}~\bibnamefont
  {Pei}}, \bibinfo {author} {\bibfnamefont {W.~J.}\ \bibnamefont {Shi}},
  \bibinfo {author} {\bibfnamefont {S.~K.}\ \bibnamefont {Mo}}, \bibinfo
  {author} {\bibfnamefont {P.}~\bibnamefont {Dudin}}, \bibinfo {author}
  {\bibfnamefont {T.}~\bibnamefont {Kim}}, \bibinfo {author} {\bibfnamefont
  {C.}~\bibnamefont {Cacho}}, \bibinfo {author} {\bibfnamefont
  {G.}~\bibnamefont {Li}}, \bibinfo {author} {\bibfnamefont {Y.}~\bibnamefont
  {Sun}}, \bibinfo {author} {\bibfnamefont {L.~X.}\ \bibnamefont {Yang}},
  \bibinfo {author} {\bibfnamefont {Z.~K.}\ \bibnamefont {Liu}}, \bibinfo
  {author} {\bibfnamefont {S.~S.~P.}\ \bibnamefont {Parkin}}, \bibinfo {author}
  {\bibfnamefont {C.}~\bibnamefont {Felser}},\ and\ \bibinfo {author}
  {\bibfnamefont {Y.~L.}\ \bibnamefont {Chen}},\ }\href
  {https://doi.org/10.1126/science.aav2873} {\bibfield  {journal} {\bibinfo
  {journal} {Science}\ }\textbf {\bibinfo {volume} {365}},\ \bibinfo {pages}
  {1282} (\bibinfo {year} {2019})}\BibitemShut {NoStop}%
\bibitem [{\citenamefont {Saitoh}\ \emph {et~al.}(2005)\citenamefont {Saitoh},
  \citenamefont {Kobayashi}, \citenamefont {Fujimori}, \citenamefont
  {Yamamura}, \citenamefont {Koyano}, \citenamefont {Tsuji},\ and\
  \citenamefont {Katayama}}]{SAITOH2005}%
  \BibitemOpen
  \bibfield  {author} {\bibinfo {author} {\bibfnamefont {Y.}~\bibnamefont
  {Saitoh}}, \bibinfo {author} {\bibfnamefont {K.}~\bibnamefont {Kobayashi}},
  \bibinfo {author} {\bibfnamefont {A.}~\bibnamefont {Fujimori}}, \bibinfo
  {author} {\bibfnamefont {Y.}~\bibnamefont {Yamamura}}, \bibinfo {author}
  {\bibfnamefont {M.}~\bibnamefont {Koyano}}, \bibinfo {author} {\bibfnamefont
  {T.}~\bibnamefont {Tsuji}},\ and\ \bibinfo {author} {\bibfnamefont
  {S.}~\bibnamefont {Katayama}},\ }\href
  {https://doi.org/https://doi.org/10.1016/j.elspec.2005.01.186} {\bibfield
  {journal} {\bibinfo  {journal} {Journal of Electron Spectroscopy and Related
  Phenomena}\ }\textbf {\bibinfo {volume} {144-147}},\ \bibinfo {pages} {829}
  (\bibinfo {year} {2005})}\BibitemShut {NoStop}%
\bibitem [{\citenamefont {Medeiros}\ \emph {et~al.}(2014)\citenamefont
  {Medeiros}, \citenamefont {Stafstr\"om},\ and\ \citenamefont
  {Bj\"ork}}]{PhysRevB.89.041407}%
  \BibitemOpen
  \bibfield  {author} {\bibinfo {author} {\bibfnamefont {P.~V.~C.}\
  \bibnamefont {Medeiros}}, \bibinfo {author} {\bibfnamefont {S.}~\bibnamefont
  {Stafstr\"om}},\ and\ \bibinfo {author} {\bibfnamefont {J.}~\bibnamefont
  {Bj\"ork}},\ }\href {https://doi.org/10.1103/PhysRevB.89.041407} {\bibfield
  {journal} {\bibinfo  {journal} {Phys. Rev. B}\ }\textbf {\bibinfo {volume}
  {89}},\ \bibinfo {pages} {041407} (\bibinfo {year} {2014})}\BibitemShut
  {NoStop}%
\bibitem [{\citenamefont {Medeiros}\ \emph {et~al.}(2015)\citenamefont
  {Medeiros}, \citenamefont {Tsirkin}, \citenamefont {Stafstr\"om},\ and\
  \citenamefont {Bj\"ork}}]{PhysRevB.91.041116}%
  \BibitemOpen
  \bibfield  {author} {\bibinfo {author} {\bibfnamefont {P.~V.~C.}\
  \bibnamefont {Medeiros}}, \bibinfo {author} {\bibfnamefont {S.~S.}\
  \bibnamefont {Tsirkin}}, \bibinfo {author} {\bibfnamefont {S.}~\bibnamefont
  {Stafstr\"om}},\ and\ \bibinfo {author} {\bibfnamefont {J.}~\bibnamefont
  {Bj\"ork}},\ }\href {https://doi.org/10.1103/PhysRevB.91.041116} {\bibfield
  {journal} {\bibinfo  {journal} {Phys. Rev. B}\ }\textbf {\bibinfo {volume}
  {91}},\ \bibinfo {pages} {041116} (\bibinfo {year} {2015})}\BibitemShut
  {NoStop}%
\bibitem [{\citenamefont {Lee}\ \emph {et~al.}(2013)\citenamefont {Lee},
  \citenamefont {Yamada-Takamura},\ and\ \citenamefont {Ozaki}}]{Lee2013}%
  \BibitemOpen
  \bibfield  {author} {\bibinfo {author} {\bibfnamefont {C.-C.}\ \bibnamefont
  {Lee}}, \bibinfo {author} {\bibfnamefont {Y.}~\bibnamefont
  {Yamada-Takamura}},\ and\ \bibinfo {author} {\bibfnamefont {T.}~\bibnamefont
  {Ozaki}},\ }\href {https://doi.org/10.1088/0953-8984/25/34/345501} {\bibfield
   {journal} {\bibinfo  {journal} {Journal of Physics: Condensed Matter}\
  }\textbf {\bibinfo {volume} {25}},\ \bibinfo {pages} {345501} (\bibinfo
  {year} {2013})}\BibitemShut {NoStop}%
\bibitem [{\citenamefont {Kondo}\ \emph {et~al.}(2007)\citenamefont {Kondo},
  \citenamefont {Khasanov}, \citenamefont {Karpinski}, \citenamefont {Kazakov},
  \citenamefont {Zhigadlo}, \citenamefont {Ohta}, \citenamefont {Fretwell},
  \citenamefont {Palczewski}, \citenamefont {Koll}, \citenamefont {Mesot},
  \citenamefont {Rotenberg}, \citenamefont {Keller},\ and\ \citenamefont
  {Kaminski}}]{PhysRevLett.98.157002}%
  \BibitemOpen
  \bibfield  {author} {\bibinfo {author} {\bibfnamefont {T.}~\bibnamefont
  {Kondo}}, \bibinfo {author} {\bibfnamefont {R.}~\bibnamefont {Khasanov}},
  \bibinfo {author} {\bibfnamefont {J.}~\bibnamefont {Karpinski}}, \bibinfo
  {author} {\bibfnamefont {S.~M.}\ \bibnamefont {Kazakov}}, \bibinfo {author}
  {\bibfnamefont {N.~D.}\ \bibnamefont {Zhigadlo}}, \bibinfo {author}
  {\bibfnamefont {T.}~\bibnamefont {Ohta}}, \bibinfo {author} {\bibfnamefont
  {H.~M.}\ \bibnamefont {Fretwell}}, \bibinfo {author} {\bibfnamefont {A.~D.}\
  \bibnamefont {Palczewski}}, \bibinfo {author} {\bibfnamefont {J.~D.}\
  \bibnamefont {Koll}}, \bibinfo {author} {\bibfnamefont {J.}~\bibnamefont
  {Mesot}}, \bibinfo {author} {\bibfnamefont {E.}~\bibnamefont {Rotenberg}},
  \bibinfo {author} {\bibfnamefont {H.}~\bibnamefont {Keller}},\ and\ \bibinfo
  {author} {\bibfnamefont {A.}~\bibnamefont {Kaminski}},\ }\href
  {https://doi.org/10.1103/PhysRevLett.98.157002} {\bibfield  {journal}
  {\bibinfo  {journal} {Phys. Rev. Lett.}\ }\textbf {\bibinfo {volume} {98}},\
  \bibinfo {pages} {157002} (\bibinfo {year} {2007})}\BibitemShut {NoStop}%
\bibitem [{\citenamefont {Kang}\ \emph {et~al.}(2020)\citenamefont {Kang},
  \citenamefont {Ye}, \citenamefont {Fang}, \citenamefont {You}, \citenamefont
  {Levitan}, \citenamefont {Han}, \citenamefont {Facio}, \citenamefont
  {Jozwiak}, \citenamefont {Bostwick}, \citenamefont {Rotenberg}, \citenamefont
  {Chan}, \citenamefont {McDonald}, \citenamefont {Graf}, \citenamefont
  {Kaznatcheev}, \citenamefont {Vescovo}, \citenamefont {Bell}, \citenamefont
  {Kaxiras}, \citenamefont {van~den Brink}, \citenamefont {Richter},
  \citenamefont {Prasad~Ghimire}, \citenamefont {Checkelsky},\ and\
  \citenamefont {Comin}}]{Kang2020}%
  \BibitemOpen
  \bibfield  {author} {\bibinfo {author} {\bibfnamefont {M.}~\bibnamefont
  {Kang}}, \bibinfo {author} {\bibfnamefont {L.}~\bibnamefont {Ye}}, \bibinfo
  {author} {\bibfnamefont {S.}~\bibnamefont {Fang}}, \bibinfo {author}
  {\bibfnamefont {J.-S.}\ \bibnamefont {You}}, \bibinfo {author} {\bibfnamefont
  {A.}~\bibnamefont {Levitan}}, \bibinfo {author} {\bibfnamefont
  {M.}~\bibnamefont {Han}}, \bibinfo {author} {\bibfnamefont {J.~I.}\
  \bibnamefont {Facio}}, \bibinfo {author} {\bibfnamefont {C.}~\bibnamefont
  {Jozwiak}}, \bibinfo {author} {\bibfnamefont {A.}~\bibnamefont {Bostwick}},
  \bibinfo {author} {\bibfnamefont {E.}~\bibnamefont {Rotenberg}}, \bibinfo
  {author} {\bibfnamefont {M.~K.}\ \bibnamefont {Chan}}, \bibinfo {author}
  {\bibfnamefont {R.~D.}\ \bibnamefont {McDonald}}, \bibinfo {author}
  {\bibfnamefont {D.}~\bibnamefont {Graf}}, \bibinfo {author} {\bibfnamefont
  {K.}~\bibnamefont {Kaznatcheev}}, \bibinfo {author} {\bibfnamefont
  {E.}~\bibnamefont {Vescovo}}, \bibinfo {author} {\bibfnamefont {D.~C.}\
  \bibnamefont {Bell}}, \bibinfo {author} {\bibfnamefont {E.}~\bibnamefont
  {Kaxiras}}, \bibinfo {author} {\bibfnamefont {J.}~\bibnamefont {van~den
  Brink}}, \bibinfo {author} {\bibfnamefont {M.}~\bibnamefont {Richter}},
  \bibinfo {author} {\bibfnamefont {M.}~\bibnamefont {Prasad~Ghimire}},
  \bibinfo {author} {\bibfnamefont {J.~G.}\ \bibnamefont {Checkelsky}},\ and\
  \bibinfo {author} {\bibfnamefont {R.}~\bibnamefont {Comin}},\ }\href
  {https://doi.org/10.1038/s41563-019-0531-0} {\bibfield  {journal} {\bibinfo
  {journal} {Nature Materials}\ }\textbf {\bibinfo {volume} {19}},\ \bibinfo
  {pages} {163} (\bibinfo {year} {2020})}\BibitemShut {NoStop}%
\bibitem [{\citenamefont {Iwasawa}\ \emph {et~al.}(2018)\citenamefont
  {Iwasawa}, \citenamefont {Schr\"oter}, \citenamefont {Masui}, \citenamefont
  {Tajima}, \citenamefont {Kim},\ and\ \citenamefont
  {Hoesch}}]{PhysRevB.98.081112}%
  \BibitemOpen
  \bibfield  {author} {\bibinfo {author} {\bibfnamefont {H.}~\bibnamefont
  {Iwasawa}}, \bibinfo {author} {\bibfnamefont {N.~B.~M.}\ \bibnamefont
  {Schr\"oter}}, \bibinfo {author} {\bibfnamefont {T.}~\bibnamefont {Masui}},
  \bibinfo {author} {\bibfnamefont {S.}~\bibnamefont {Tajima}}, \bibinfo
  {author} {\bibfnamefont {T.~K.}\ \bibnamefont {Kim}},\ and\ \bibinfo {author}
  {\bibfnamefont {M.}~\bibnamefont {Hoesch}},\ }\href
  {https://doi.org/10.1103/PhysRevB.98.081112} {\bibfield  {journal} {\bibinfo
  {journal} {Phys. Rev. B}\ }\textbf {\bibinfo {volume} {98}},\ \bibinfo
  {pages} {081112} (\bibinfo {year} {2018})}\BibitemShut {NoStop}%
\bibitem [{\citenamefont {Iwasawa}\ \emph {et~al.}(2019)\citenamefont
  {Iwasawa}, \citenamefont {Dudin}, \citenamefont {Inui}, \citenamefont
  {Masui}, \citenamefont {Kim}, \citenamefont {Cacho},\ and\ \citenamefont
  {Hoesch}}]{PhysRevB.99.140510}%
  \BibitemOpen
  \bibfield  {author} {\bibinfo {author} {\bibfnamefont {H.}~\bibnamefont
  {Iwasawa}}, \bibinfo {author} {\bibfnamefont {P.}~\bibnamefont {Dudin}},
  \bibinfo {author} {\bibfnamefont {K.}~\bibnamefont {Inui}}, \bibinfo {author}
  {\bibfnamefont {T.}~\bibnamefont {Masui}}, \bibinfo {author} {\bibfnamefont
  {T.~K.}\ \bibnamefont {Kim}}, \bibinfo {author} {\bibfnamefont
  {C.}~\bibnamefont {Cacho}},\ and\ \bibinfo {author} {\bibfnamefont
  {M.}~\bibnamefont {Hoesch}},\ }\href
  {https://doi.org/10.1103/PhysRevB.99.140510} {\bibfield  {journal} {\bibinfo
  {journal} {Phys. Rev. B}\ }\textbf {\bibinfo {volume} {99}},\ \bibinfo
  {pages} {140510} (\bibinfo {year} {2019})}\BibitemShut {NoStop}%
\bibitem [{\citenamefont {Nakayama}\ \emph {et~al.}(2019)\citenamefont
  {Nakayama}, \citenamefont {Souma}, \citenamefont {Trang}, \citenamefont
  {Takane}, \citenamefont {Chen}, \citenamefont {Avila}, \citenamefont
  {Takahashi}, \citenamefont {Sasaki}, \citenamefont {Segawa}, \citenamefont
  {Asensio}, \citenamefont {Ando},\ and\ \citenamefont {Sato}}]{Nakayama2019}%
  \BibitemOpen
  \bibfield  {author} {\bibinfo {author} {\bibfnamefont {K.}~\bibnamefont
  {Nakayama}}, \bibinfo {author} {\bibfnamefont {S.}~\bibnamefont {Souma}},
  \bibinfo {author} {\bibfnamefont {C.~X.}\ \bibnamefont {Trang}}, \bibinfo
  {author} {\bibfnamefont {D.}~\bibnamefont {Takane}}, \bibinfo {author}
  {\bibfnamefont {C.}~\bibnamefont {Chen}}, \bibinfo {author} {\bibfnamefont
  {J.}~\bibnamefont {Avila}}, \bibinfo {author} {\bibfnamefont
  {T.}~\bibnamefont {Takahashi}}, \bibinfo {author} {\bibfnamefont
  {S.}~\bibnamefont {Sasaki}}, \bibinfo {author} {\bibfnamefont
  {K.}~\bibnamefont {Segawa}}, \bibinfo {author} {\bibfnamefont {M.~C.}\
  \bibnamefont {Asensio}}, \bibinfo {author} {\bibfnamefont {Y.}~\bibnamefont
  {Ando}},\ and\ \bibinfo {author} {\bibfnamefont {T.}~\bibnamefont {Sato}},\
  }\href {https://doi.org/10.1021/acs.nanolett.9b00875} {\bibfield  {journal}
  {\bibinfo  {journal} {Nano Letters}\ }\textbf {\bibinfo {volume} {19}},\
  \bibinfo {pages} {3737} (\bibinfo {year} {2019})}\BibitemShut {NoStop}%
\bibitem [{\citenamefont {Xu}\ \emph {et~al.}(2020)\citenamefont {Xu},
  \citenamefont {Mao}, \citenamefont {Wang}, \citenamefont {Li}, \citenamefont
  {Chen}, \citenamefont {Xia}, \citenamefont {Li}, \citenamefont {Pei},
  \citenamefont {Zhang}, \citenamefont {Zheng}, \citenamefont {Huang},
  \citenamefont {Zhang}, \citenamefont {Cui}, \citenamefont {Liang},
  \citenamefont {Xia}, \citenamefont {Su}, \citenamefont {Jung}, \citenamefont
  {Cacho}, \citenamefont {Wang}, \citenamefont {Li}, \citenamefont {Xu},
  \citenamefont {Guo}, \citenamefont {Yang}, \citenamefont {Liu}, \citenamefont
  {Chen},\ and\ \citenamefont {Jiang}}]{XU2020}%
  \BibitemOpen
  \bibfield  {author} {\bibinfo {author} {\bibfnamefont {L.}~\bibnamefont
  {Xu}}, \bibinfo {author} {\bibfnamefont {Y.}~\bibnamefont {Mao}}, \bibinfo
  {author} {\bibfnamefont {H.}~\bibnamefont {Wang}}, \bibinfo {author}
  {\bibfnamefont {J.}~\bibnamefont {Li}}, \bibinfo {author} {\bibfnamefont
  {Y.}~\bibnamefont {Chen}}, \bibinfo {author} {\bibfnamefont {Y.}~\bibnamefont
  {Xia}}, \bibinfo {author} {\bibfnamefont {Y.}~\bibnamefont {Li}}, \bibinfo
  {author} {\bibfnamefont {D.}~\bibnamefont {Pei}}, \bibinfo {author}
  {\bibfnamefont {J.}~\bibnamefont {Zhang}}, \bibinfo {author} {\bibfnamefont
  {H.}~\bibnamefont {Zheng}}, \bibinfo {author} {\bibfnamefont
  {K.}~\bibnamefont {Huang}}, \bibinfo {author} {\bibfnamefont
  {C.}~\bibnamefont {Zhang}}, \bibinfo {author} {\bibfnamefont
  {S.}~\bibnamefont {Cui}}, \bibinfo {author} {\bibfnamefont {A.}~\bibnamefont
  {Liang}}, \bibinfo {author} {\bibfnamefont {W.}~\bibnamefont {Xia}}, \bibinfo
  {author} {\bibfnamefont {H.}~\bibnamefont {Su}}, \bibinfo {author}
  {\bibfnamefont {S.}~\bibnamefont {Jung}}, \bibinfo {author} {\bibfnamefont
  {C.}~\bibnamefont {Cacho}}, \bibinfo {author} {\bibfnamefont
  {M.}~\bibnamefont {Wang}}, \bibinfo {author} {\bibfnamefont {G.}~\bibnamefont
  {Li}}, \bibinfo {author} {\bibfnamefont {Y.}~\bibnamefont {Xu}}, \bibinfo
  {author} {\bibfnamefont {Y.}~\bibnamefont {Guo}}, \bibinfo {author}
  {\bibfnamefont {L.}~\bibnamefont {Yang}}, \bibinfo {author} {\bibfnamefont
  {Z.}~\bibnamefont {Liu}}, \bibinfo {author} {\bibfnamefont {Y.}~\bibnamefont
  {Chen}},\ and\ \bibinfo {author} {\bibfnamefont {M.}~\bibnamefont {Jiang}},\
  }\href {https://doi.org/https://doi.org/10.1016/j.scib.2020.07.032}
  {\bibfield  {journal} {\bibinfo  {journal} {Science Bulletin}\ }\textbf
  {\bibinfo {volume} {65}},\ \bibinfo {pages} {2086} (\bibinfo {year}
  {2020})}\BibitemShut {NoStop}%
\bibitem [{\citenamefont {Sirica}\ \emph {et~al.}(2020)\citenamefont {Sirica},
  \citenamefont {Vilmercati}, \citenamefont {Bondino}, \citenamefont {Pis},
  \citenamefont {Nappini}, \citenamefont {Mo}, \citenamefont {Fedorov},
  \citenamefont {Das}, \citenamefont {Vobornik}, \citenamefont {Fujii},
  \citenamefont {Li}, \citenamefont {Sapkota}, \citenamefont {Parker},
  \citenamefont {Mandrus},\ and\ \citenamefont {Mannella}}]{Sirica2020}%
  \BibitemOpen
  \bibfield  {author} {\bibinfo {author} {\bibfnamefont {N.}~\bibnamefont
  {Sirica}}, \bibinfo {author} {\bibfnamefont {P.}~\bibnamefont {Vilmercati}},
  \bibinfo {author} {\bibfnamefont {F.}~\bibnamefont {Bondino}}, \bibinfo
  {author} {\bibfnamefont {I.}~\bibnamefont {Pis}}, \bibinfo {author}
  {\bibfnamefont {S.}~\bibnamefont {Nappini}}, \bibinfo {author} {\bibfnamefont
  {S.-K.}\ \bibnamefont {Mo}}, \bibinfo {author} {\bibfnamefont {A.~V.}\
  \bibnamefont {Fedorov}}, \bibinfo {author} {\bibfnamefont {P.~K.}\
  \bibnamefont {Das}}, \bibinfo {author} {\bibfnamefont {I.}~\bibnamefont
  {Vobornik}}, \bibinfo {author} {\bibfnamefont {J.}~\bibnamefont {Fujii}},
  \bibinfo {author} {\bibfnamefont {L.}~\bibnamefont {Li}}, \bibinfo {author}
  {\bibfnamefont {D.}~\bibnamefont {Sapkota}}, \bibinfo {author} {\bibfnamefont
  {D.~S.}\ \bibnamefont {Parker}}, \bibinfo {author} {\bibfnamefont {D.~G.}\
  \bibnamefont {Mandrus}},\ and\ \bibinfo {author} {\bibfnamefont
  {N.}~\bibnamefont {Mannella}},\ }\href
  {https://doi.org/10.1038/s42005-020-0333-3} {\bibfield  {journal} {\bibinfo
  {journal} {Communications Physics}\ }\textbf {\bibinfo {volume} {3}},\
  \bibinfo {pages} {65} (\bibinfo {year} {2020})}\BibitemShut {NoStop}%
\bibitem [{\citenamefont {Ishida}\ and\ \citenamefont
  {Shin}(2018)}]{Ishida2018}%
  \BibitemOpen
  \bibfield  {author} {\bibinfo {author} {\bibfnamefont {Y.}~\bibnamefont
  {Ishida}}\ and\ \bibinfo {author} {\bibfnamefont {S.}~\bibnamefont {Shin}},\
  }\href {https://doi.org/10.1063/1.5007226} {\bibfield  {journal} {\bibinfo
  {journal} {Review of Scientific Instruments}\ }\textbf {\bibinfo {volume}
  {89}},\ \bibinfo {pages} {043903} (\bibinfo {year} {2018})}\BibitemShut
  {NoStop}%
\bibitem [{\citenamefont {Burns}\ and\ \citenamefont
  {Glazer}(2013)}]{BURNS2013}%
  \BibitemOpen
  \bibfield  {author} {\bibinfo {author} {\bibfnamefont {G.}~\bibnamefont
  {Burns}}\ and\ \bibinfo {author} {\bibfnamefont {A.}~\bibnamefont {Glazer}},\
  }\href {https://doi.org/https://doi.org/10.1016/B978-0-12-394400-9.00001-0}
  {\emph {\bibinfo {title} {Space Groups for Solid State Scientists (Third
  Edition)}}}\ (\bibinfo  {publisher} {Academic Press},\ \bibinfo {address}
  {Oxford},\ \bibinfo {year} {2013})\BibitemShut {NoStop}%
\end{thebibliography}%

\end{document}


\title{Supplemental Material:\\ Large anomalous Hall effect induced by weak ferromagnetism \\in the noncentrosymmetric antiferromagnet $\mathbf{Co}\mathbf{Nb}_3\mathbf{S}_6$}

\author{Hiroaki Tanaka}
\affiliation{Institute for Solid State Physics, The University of Tokyo, Kashiwa, Chiba 277-8581, Japan}

\author{Shota Okazaki}
\affiliation{Materials and Structures Laboratory, Tokyo Institute of Technology, Yokohama, Kanagawa 226-8503, Japan}

\author{Kenta Kuroda}
\email{kuroken224@hiroshima-u.ac.jp}
\affiliation{Institute for Solid State Physics, The University of Tokyo, Kashiwa, Chiba 277-8581, Japan}

\author{Ryo Noguchi}
\affiliation{Institute for Solid State Physics, The University of Tokyo, Kashiwa, Chiba 277-8581, Japan}
\affiliation{Center for Correlated Electron Systems, Institute for Basic Science, Seoul 08826, Republic of Korea}
\affiliation{Department of Physics and Astronomy, Seoul National University, Seoul 08826, Republic of Korea}

\author{Yosuke Arai}
\affiliation{Institute for Solid State Physics, The University of Tokyo, Kashiwa, Chiba 277-8581, Japan}

\author{Susumu Minami}
\affiliation{Department of Physics, The University of Tokyo, Tokyo 113-0033, Japan}

\author{Shinichiro Ideta}
\affiliation{UVSOR Facility, Institute for Molecular Science, Okazaki, Aichi 444-8585, Japan}

\author{Kiyohisa Tanaka}
\affiliation{UVSOR Facility, Institute for Molecular Science, Okazaki, Aichi 444-8585, Japan}

\author{Donghui Lu}
\affiliation{Stanford Synchrotron Radiation Lightsource, SLAC National Accelerator Laboratory, Menlo Park, CA 94025, United States}

\author{Makoto Hashimoto}
\affiliation{Stanford Synchrotron Radiation Lightsource, SLAC National Accelerator Laboratory, Menlo Park, CA 94025, United States}

\author{Viktor Kandyba}
\affiliation{Elettra Sincrotrone Trieste, 34149 Basovizza, Trieste, Italy}

\author{Mattia Cattelan}
\affiliation{Elettra Sincrotrone Trieste, 34149 Basovizza, Trieste, Italy}

\author{Alexei Barinov}
\affiliation{Elettra Sincrotrone Trieste, 34149 Basovizza, Trieste, Italy}

\author{Takayuki Muro}
\affiliation{Japan Synchrotron Radiation Research Institute (JASRI), Sayo, Hyogo 679-5198, Japan}

\author{Takao Sasagawa}
\email{sasagawa@msl.titech.ac.jp}
\affiliation{Materials and Structures Laboratory, Tokyo Institute of Technology, Yokohama, Kanagawa 226-8503, Japan}

\author{Takeshi Kondo}
\email{kondo1215@issp.u-tokyo.ac.jp}
\affiliation{Institute for Solid State Physics, The University of Tokyo, Kashiwa, Chiba 277-8581, Japan}
\affiliation{Trans-scale Quantum Science Institute, The University of Tokyo, Tokyo 113-0033, Japan}

\date{\today}

\renewcommand{\thefigure}{S\arabic{figure}}
\renewcommand{\thetable}{S\Roman{table}}
\titleformat*{\section}{\bfseries}

\maketitle

\section*{Note 1: Ordinal and anomalous Hall resistivities of materials}
In this report, we use the two-dimensional resistivity tensor $\rho_{ij}$ defined by
\begin{equation}
\left(\begin{array}{c}E_x\\E_y\end{array}\right)=\left(\begin{array}{cc}\rho_{xx}&-\rho_{xy}\\ \rho_{xy} & \rho_{yy}\end{array}\right)\left(\begin{array}{c}j_x\\j_y\end{array}\right),
\end{equation}
where $E_i$ and $j_i$ are the electric field and current density.

Table \ref{Table: R0_Rs} summarizes Hall coefficients $R_0$ and $R_s$ for some ferromagnetic materials.
Sometimes we estimated values from their graphs and calculated the coefficients.
We noted the measurement temperature if the coefficients strongly depended on the temperature.

\begin{table}[hb]
	\centering
	\caption{\label{Table: R0_Rs} Hall coefficients $R_0$ and $R_s$ for ferromagnetic materials.}
	\begin{tabular}{c|ccc|c}\hline\hline
	Material & $R_0\ (\mathrm{m}^3/\mathrm{C})$ & $R_s\ (\mathrm{m}^3/\mathrm{C})$ & $|R_s/R_0|$ & Ref.\\\hline
	ferromagnetic Fe & & & larger than 200 & [16] \\
	$\mathrm{Fe}_3\mathrm{Sn}_2$ (100 K) & $-2.5\times10^{-10}$ & $8.7\times10^{-9}$ & 35 & [17] \\
	$\mathrm{Fe}_3\mathrm{Sn}_2$ (200 K) & $-1.4\times10^{-9}$ & $2.8\times10^{-8}$ & 20 & [18] \\
	$\mathrm{Co}_3\mathrm{Sn}_2\mathrm{S}_2$ (150 K) & $3.8\times10^{-9}$ & $4.6\times10^{-6}$ & $8.3\times10^2$ & [19] \\
	$\mathrm{Fe}\mathrm{Ta}_3\mathrm{S}_6$ & $2.7\times10^{-9}$ & $-2.0\times10^{-7}$ & 74 & [20] \\
	$\mathrm{Fe}\mathrm{Ta}_4\mathrm{S}_8$ (80 K) & $4.5\times10^{-10}$ & $2.1\times10^{-9}$ & $4.7$ & [21]\\
	$\mathrm{Co}\mathrm{Nb}_3\mathrm{S}_6$ & $2.4\times10^{-9}$ & $-1.41\times10^{-4}$ & $5.9\times10^4$ & [11]\\
	$\mathrm{Co}\mathrm{Nb}_3\mathrm{S}_6$ & $2.27\times10^{-8}$ & $-1.47\times10^{-4}$ & $6.5\times10^3$ & This work \\\hline\hline
	\end{tabular}
\end{table}
\clearpage

\section*{Note 2: Magnetic structure of $\mathbf{Co}\mathbf{Nb}_3\mathbf{S}_6$}

\begin{figure}[hb]
\includegraphics{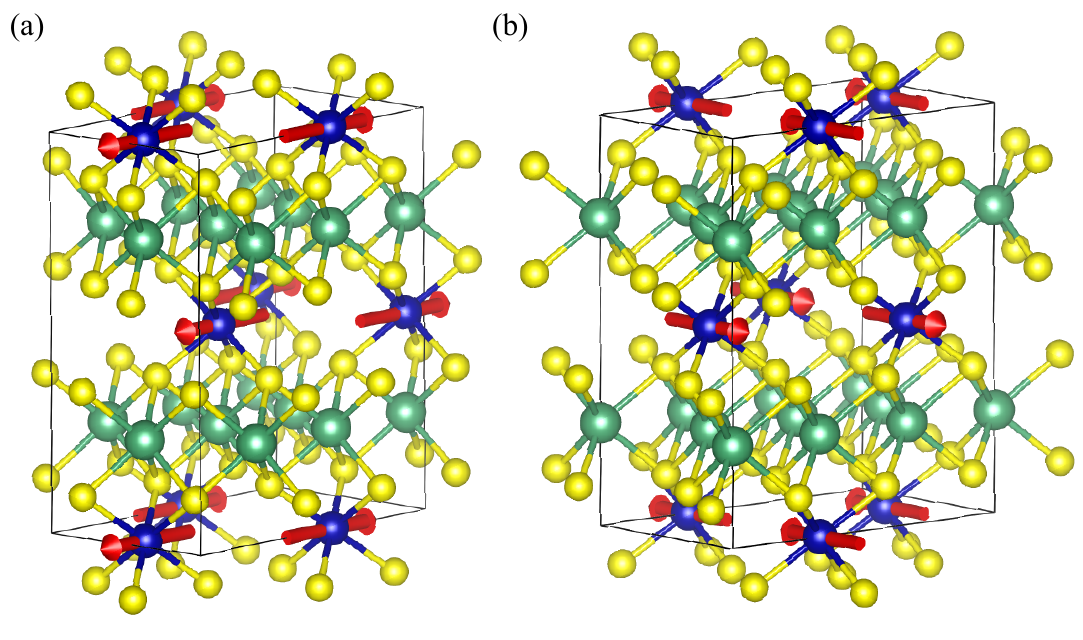}
\caption{\label{Fig: Magnetic_structure}Magnetic structure of $\mathrm{Co}\mathrm{Nb}_3\mathrm{S}_6$. (a) Magnetic structure revealed by neutron scattering measurements [8]. Since the measurements could not determine the in-plane directions of magnetic moments, we use the direction registered in the MAGNDATA database [9]. (b) Another magnetic structure that can cause ``crystal time-reversal symmetry breaking" [23].}
\end{figure}

Figure \ref{Fig: Magnetic_structure} represents magnetic structures of $\mathrm{Co}\mathrm{Nb}_3\mathrm{S}_6$.
The neutron scattering measurements [8] revealed that the magnetic structure is collinearly antiferromagnetic.
Parallel or antiparallel relationships between adjacent magnetic moments are represented in Fig.\ \ref{Fig: Magnetic_structure}(a).

On the other hand, another magnetic structure like Fig.\ \ref{Fig: Magnetic_structure}(b) has been reported to cause emergent time-reversal symmetry breaking.
When we only see the magnetic moments of Fig.\ \ref{Fig: Magnetic_structure}(b), this structure is invariant under the time-reversal followed by a translation along the out-of-plane direction.
However, the electronic structure based on this magnetic structure is changed by these operations because they move nonmagnetic S atoms.
The time-reversal symmetry breaking of this type has been proposed as ``crystal time-reversal symmetry breaking" in Ref.\ [23].

We note that magnetic structures of Figs.\ \ref{Fig: Magnetic_structure}(a) and (b) are different.
In Fig.\ \ref{Fig: Magnetic_structure}(a), one Co layer includes magnetic moments of two different directions, while in Fig.\ \ref{Fig: Magnetic_structure}(b) one Co layer includes magnetic moments of only one direction.
The magnetic structure of Fig.\ \ref{Fig: Magnetic_structure}(a) is invariant under the time-reversal followed by a translation along the diagonal direction on the stacking plane, even if nonmagnetic Nb and S atoms are considered.

\section*{Note 3: Preparation and characterization of single crystals}

Polycrystals of $\mathrm{Fe}\mathrm{Nb}_3\mathrm{S}_6$, $\mathrm{Co}\mathrm{Nb}_3\mathrm{S}_6$, $\mathrm{Ni}\mathrm{Nb}_3\mathrm{S}_6$, and sulfur deficient $\mathrm{Co}\mathrm{Nb}_3\mathrm{S}_{6-x}$ were synthesized by the solid-state reaction using the elemental powders (3N for Fe, Co, and Ni, 5Nup for S and I) with the compositional ratios (e.g.\ the sulfur deficiency $x$ was set to 0.18) in evacuated quartz tubes at 900 ${}^\circ \mathrm{C}$ for 5 days.
Subsequent chemical vapor transport crystal growth was carried out using the polycrystals with 10 wt\% of iodine in evacuated quartz tubes for a week at 950 ${}^\circ \mathrm{C}$ for the source zone and 850 ${}^\circ \mathrm{C}$ for the growth zone. The obtained crystals were hexagonal platelets with $\sim2\times2\times0.1\ \mathrm{mm}^3$. 

The crystal structure was analyzed by both powder and single-crystal X-ray diffraction (XRD) using the Bruker D2 PHASER with the LynxExe 1D detector and the Rigaku XtaLAB mini, respectively. The chemical analysis based on X-ray fluorescence (XRF) was performed using an X-ray analytical and imaging microscope (HORIBA XGT-5000).

\begin{figure}[hb]
\includegraphics{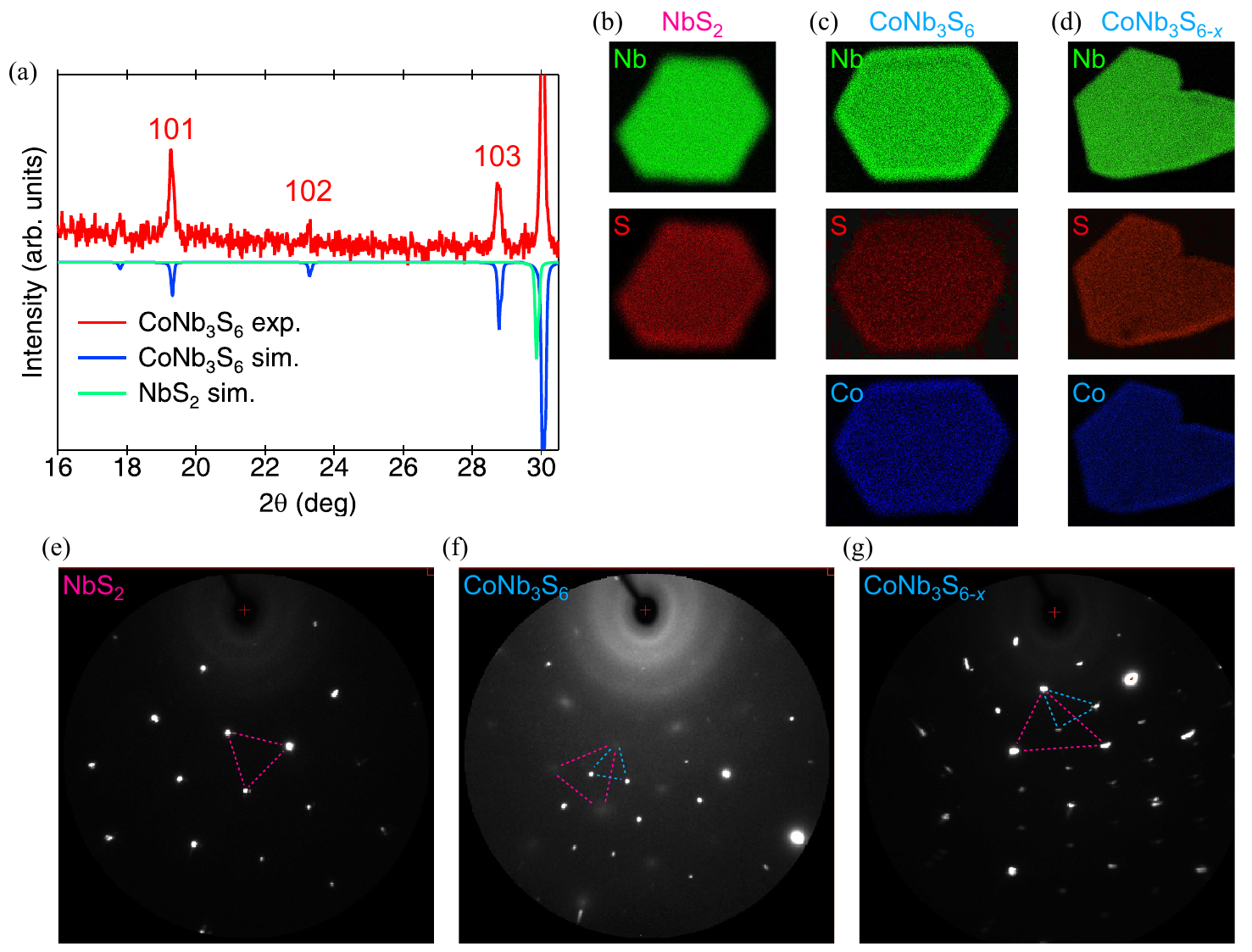}
\caption{\label{Fig: Characterization}Characterization of $\mathrm{Nb}\mathrm{S}_2$, $\mathrm{Co}\mathrm{Nb}_3\mathrm{S}_6$, and $\mathrm{Co}\mathrm{Nb}_3\mathrm{S}_{6-x}$ crystals. (a) Powder x-ray diffraction spectrum of $\mathrm{Co}\mathrm{Nb}_3\mathrm{S}_6$ polycrystal (red line) and simulated spectra of $\mathrm{Co}\mathrm{Nb}_3\mathrm{S}_6$ and $\mathrm{Nb}\mathrm{S}_2$ (blue and green respectively). (b)-(d) X-ray fluorescence spectroscopy measurements of $\mathrm{Nb}\mathrm{S}_2$, $\mathrm{Co}\mathrm{Nb}_3\mathrm{S}_6$ and $\mathrm{Co}\mathrm{Nb}_3\mathrm{S}_{6-x}$ single crystals. (e)-(g) X-ray diffraction patterns of  $\mathrm{Nb}\mathrm{S}_2$, $\mathrm{Co}\mathrm{Nb}_3\mathrm{S}_6$ and $\mathrm{Co}\mathrm{Nb}_3\mathrm{S}_{6-x}$ single crystals. Pink and light blue guides correspond to $\mathrm{Nb}\mathrm{S}_2$ and $\mathrm{Co}\mathrm{Nb}_3\mathrm{S}_6$ crystals in the reciprocal space, respectively.}
\end{figure}

Results of characterization measurements are summarized in Fig.\ \ref{Fig: Characterization}.
Powder XRD measurements of $\mathrm{Co}\mathrm{Nb}_3\mathrm{S}_6$, showed additional peaks which don't appear in the simulated $\mathrm{Nb}\mathrm{S}_2$ spectrum [Fig.\ \ref{Fig: Characterization}(a)].
XRF measurements of single crystals revealed that Nb, S, and Co atoms were distributed uniformly [Figs.\ \ref{Fig: Characterization}(b)-(d)].
The analyzed sulfur-deficiency in $\mathrm{Co}\mathrm{Nb}_3\mathrm{S}_{6-x}$ was $x = 0.11-0.12$ within the accuracy of the XRF measurements.
XRD patterns [Figs.\ \ref{Fig: Characterization}(e)-(g)] showed sharp peaks, and for $\mathrm{Co}\mathrm{Nb}_3\mathrm{S}_{6(-x)}$ we observed triangular peak patterns with shorter periodicity due to the larger unit cell than $\mathrm{Nb}\mathrm{S}_2$.
These XRD measurements revealed that the lattice constants are $a_1=b_1=3.338\ \textrm{\AA},\ c=11.474\ \textrm{\AA}$ for $\mathrm{Nb}\mathrm{S}_2$ and $a_2=b_2=5.849\ \textrm{\AA},\ c=11.954\ \textrm{\AA}$ for $\mathrm{Co}\mathrm{Nb}_3\mathrm{S}_6$ [see the main text for the notation].

The magnetization curves under magnetic fields parallel to the crystal $c$-axis were measured using a SQUID magnetometer (Quantum Design MPMS-XL5).
 
The temperature and field dependence of the in-plane electrical transport properties (the current and the field parallel to the $a$-axis and the $c$-axis of $\mathrm{Co}\mathrm{Nb}_3\mathrm{S}_6$, respectively) was measured by the six-probe method using the Quantum Design PPMS. 

\clearpage

\section*{Note 4: ARPES and XPS measurements}
VUV-ARPES measurements were performed at BL5U of UVSOR and BL5-2 of Stanford Synchrotron Radiation Lightsource (SSRL).
The 90 and 120 eV VUV lights were used, and the energy resolutions were 45 meV for UVSOR BL5U and 35 meV for SSRL BL5-2.
Nano-VUV-ARPES measurements were performed at Spectromicroscopy-3.2L beamline of Elettra Sincrotrone Trieste [25].
The 74 eV VUV light was used, and the resolution was 190 meV.
SX-ARPES and XPS measurements were performed at BL25SU of SPring-8 [26, 27].
The photon energies ranged from 370 to 590 eV for SX-ARPES measurements, and the 600 eV SX light was used for XPS measurements.
The energy resolution was 75 meV.

In all ARPES experiments, the samples were cleaved \textit{in situ} along the (001) plane at an ultrahigh vacuum better than $\sim3\times10^{-8}\ \mathrm{Pa}$.

In nano-VUV-ARPES measurements, the temperature was $30\ \mathrm{K}(>T_\mathrm{N}\simeq28\ \mathrm{K})$ due to the machine limitation, but this is non-critical because the ARPES spectra did not change drastically at the border of $T_\mathrm{N}$ [Note 8].

\section*{Note 5: X-ray photoemission spectroscopy measurements of $\mathbf{Nb}\mathbf{S}_2$ and $\mathbf{Co}\mathbf{Nb}_3\mathbf{S}_6$}

\begin{figure}
\includegraphics{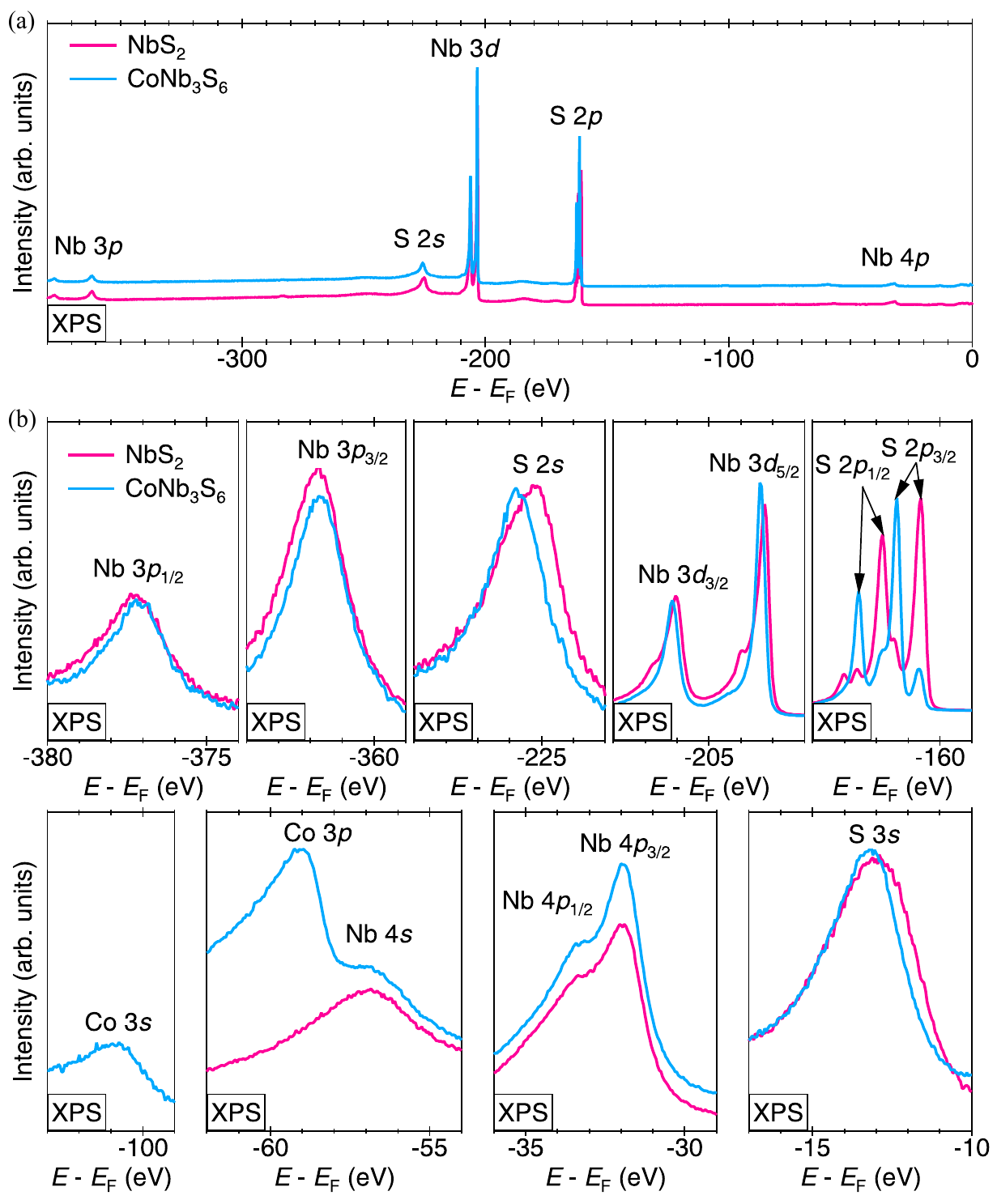}
\caption{\label{Fig: XPS_All} XPS measurements of $\mathrm{Nb}\mathrm{S}_2$ and $\mathrm{Co}\mathrm{Nb}_3\mathrm{S}_6$ using the 600 eV SX light. (a) Overall XPS spectra of $\mathrm{Nb}\mathrm{S}_2$ (pink) and $\mathrm{Co}\mathrm{Nb}_3\mathrm{S}_6$ (light blue). Some labels representing the element and orbital are attached to the corresponding peaks. (b) Core level spectra from Nb, S, and Co atoms within the energy range of (a). In each panel, spectra of $\mathrm{Nb}\mathrm{S}_2$ and $\mathrm{Co}\mathrm{Nb}_3\mathrm{S}_6$ are magnified independently for easy understanding.}
\end{figure}

X-ray photoemission spectroscopy (XPS) measurements of $\mathrm{Nb}\mathrm{S}_2$ and $\mathrm{Co}\mathrm{Nb}_3\mathrm{S}_6$ taken by the 600 eV SX light are summarized in Fig.\ \ref{Fig: XPS_All}.
Comparing the XPS spectra of $\mathrm{Nb}\mathrm{S}_2$ and $\mathrm{Co}\mathrm{Nb}_3\mathrm{S}_6$, we observed additional peaks of Co $3s$ and $3p$ orbitals only in $\mathrm{Co}\mathrm{Nb}_3\mathrm{S}_6$.
For S orbitals, peak positions for $\mathrm{Co}\mathrm{Nb}_3\mathrm{S}_6$ are lower by 0.5-1 eV than those for $\mathrm{Nb}\mathrm{S}_2$, due to electron doping by Co cations.
On the other hand, Nb peaks are shifted only slightly, representing that this electron doping changes both the Fermi level and local electronic structure around Nb and S atoms.
In addition, the change of the local electronic structure in the $\mathrm{Nb}\mathrm{S}_2$ layers removes small peaks in Nb $3d$ core level spectra.
These changes are consistent with previous researches of $\mathrm{Fe}_x\mathrm{Nb}\mathrm{S}_2\ (x=1/3,\ 1/4)$ [46] and $\mathrm{Cr}\mathrm{Nb}_3\mathrm{S}_6$ [34].
We note that small peaks in S $2p$ core level are due to the surface electronic structure as discussed in Ref.\ [46], so the changes of these peaks cannot be discussed only based on the change of the bulk electronic structure by the Co intercalation.

\clearpage

\section*{Note 6: Other VUV-ARPES measurements of $\mathbf{Nb}\mathbf{S}_2$ and $\mathbf{Co}\mathbf{Nb}_3\mathbf{S}_6$}

\begin{figure}[hb]
\includegraphics{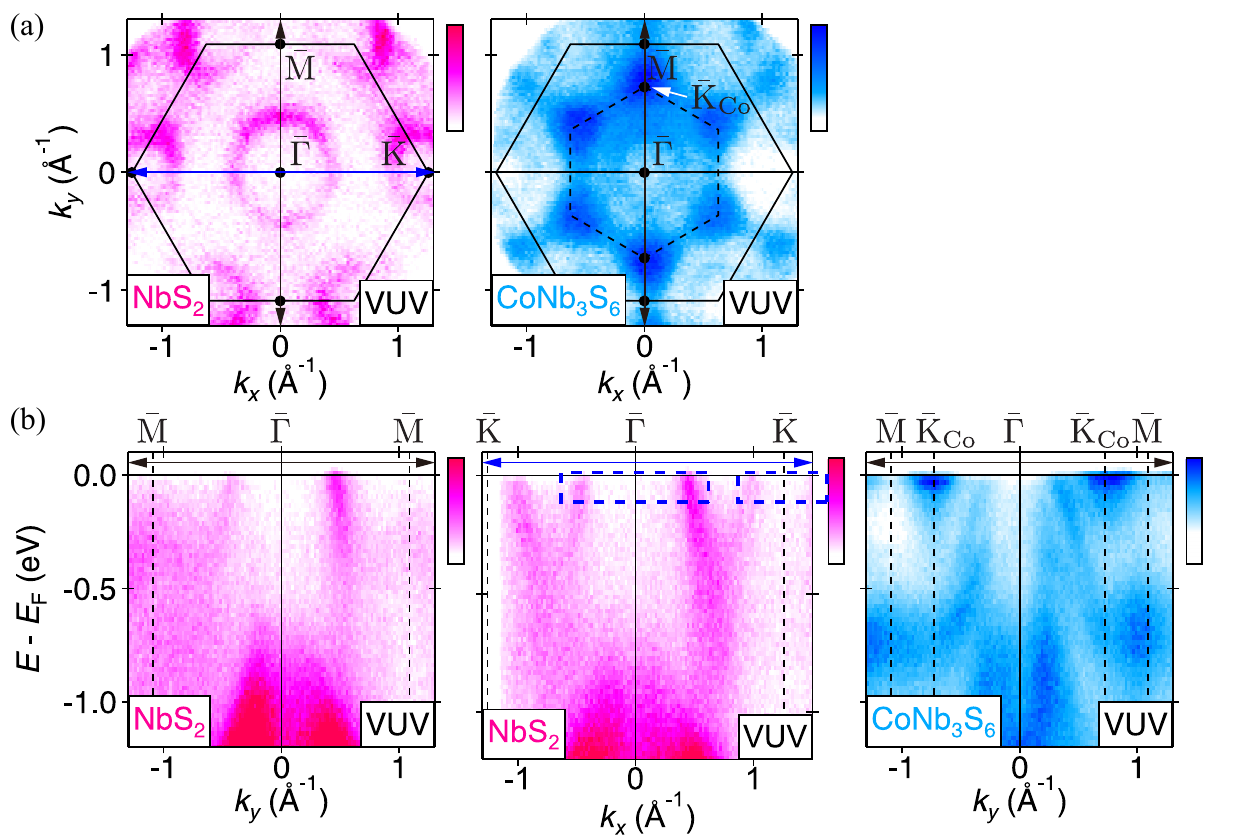}
\caption{\label{Fig: VUV_UVSOR} VUV-ARPES measurements of $\mathrm{Nb}\mathrm{S}_2$ and $\mathrm{Co}\mathrm{Nb}_3\mathrm{S}_6$ taken by the 120 eV VUV light at UVSOR BL5U. (a) Fermi surfaces of $\mathrm{Nb}\mathrm{S}_2$ (left panel) and $\mathrm{Co}\mathrm{Nb}_3\mathrm{S}_6$ (right panel). The solid and dashed hexagons represent the Brillouin zones of $\mathrm{Nb}\mathrm{S}_2$ and $\mathrm{Co}\mathrm{Nb}_3\mathrm{S}_6$, respectively. (b) Band dispersions along the $k_y$ and $k_x$ directions [black and blue arrows in (a), respectively].}
\end{figure}

Figure \ref{Fig: VUV_UVSOR} summarizes the result of VUV-ARPES measurements of $\mathrm{Nb}\mathrm{S}_2$ and $\mathrm{Co}\mathrm{Nb}_3\mathrm{S}_6$ at UVSOR BL5U.
Although the signal-to-noise ratio is poorer than the data taken at SSRL BL5-2 [Figs.\ 1(b1)-(c2)], these measurements also observed the additional band dispersion near the Fermi level in the electronic structure of $\mathrm{Co}\mathrm{Nb}_3\mathrm{S}_6$.

The blue dashed rectangles in Fig.\ \ref{Fig: VUV_UVSOR} center panel show that the pockets around the $\bar{\Gamma}$ and $\bar{K}$ points are hole-type.

\section*{Note 7: SX-ARPES measurements of $\mathbf{Co}\mathbf{Nb}_3\mathbf{S}_6$}

\begin{figure}
\includegraphics[width=160mm]{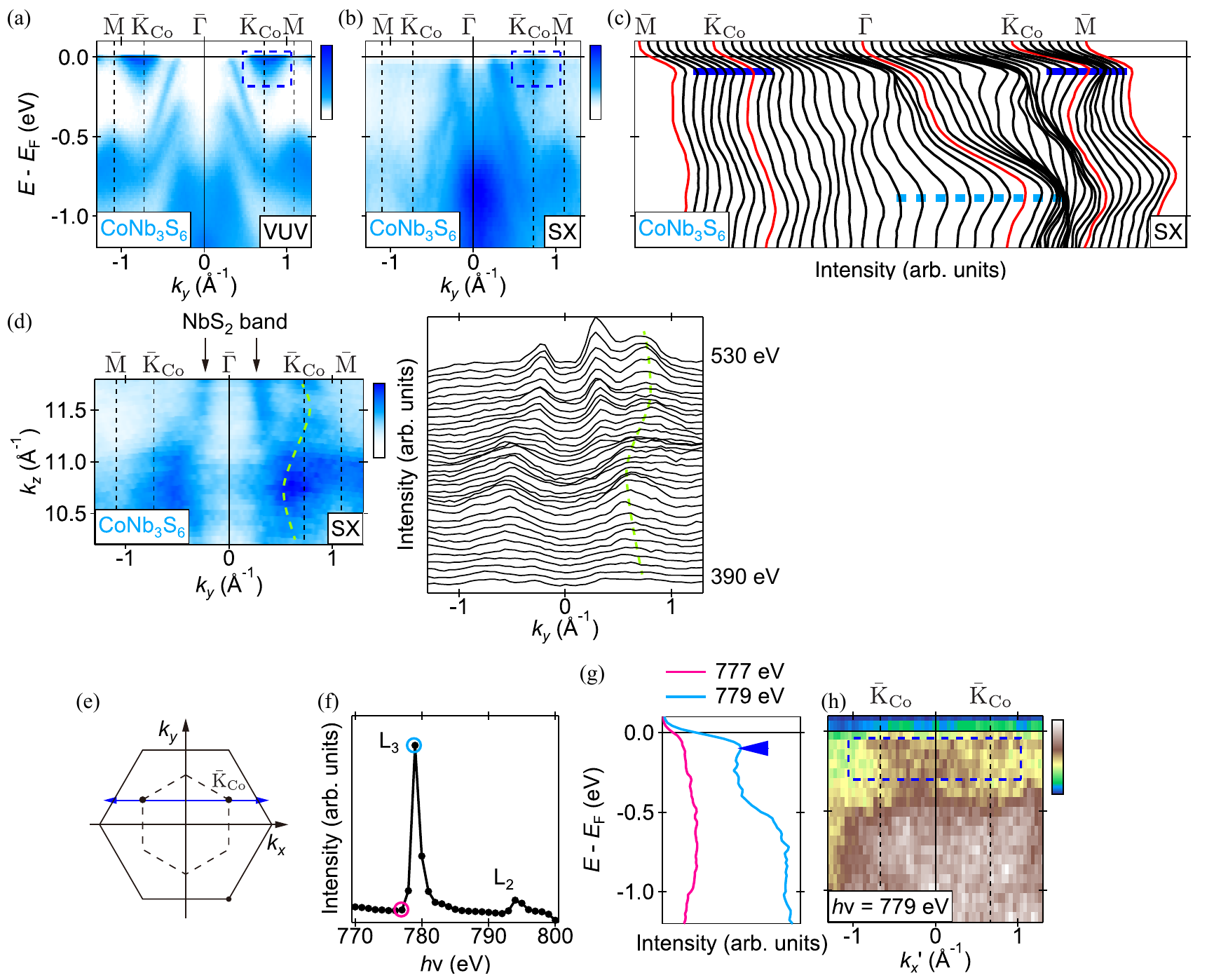}
\caption{\label{Fig: SX} SX-ARPES measurements of $\mathrm{Co}\mathrm{Nb}_3\mathrm{S}_6$ and comparison with VUV-ARPES measurements. (a), (b) Band dispersion of $\mathrm{Co}\mathrm{Nb}_3\mathrm{S}_6$ along the $k_y$ direction taken by the 120 eV VUV light and 525 eV SX light respectively. The corresponding band dispersions are marked as blue and pink dashed rectangles. The panel (a) is the same as Fig.\ 1(c2). (c) Energy distribution curves extracted from (b). The blue lines highlight spectrum peaks corresponding to the additional band dispersion. (d) Fermi surface along the $k_y$ and $k_z$ directions of $\mathrm{Co}\mathrm{Nb}_3\mathrm{S}_6$ obtained by $h\nu$ dependence measurements ($h\nu=390 \sim 530\ \mathrm{eV}$). The green dashed curves are guides to the eyes for the Co band. (e) Brillouin zones of $\mathrm{Nb}\mathrm{S}_2$ (solid lines) and $\mathrm{Co}\mathrm{Nb}_3\mathrm{S}_6$ (dashed lines). Blue arrow represents the cut along which the following ARPES measurements were performed. (f) $h\nu$ dependence of the photoemission intensity near $\mathrm{L}_2$ and $\mathrm{L}_3$ edges of Co. The light blue and pink circles show the energies on- and off-resonance, respectively. The intergration along the energy axis is performed between $E_\mathrm{F}$ and $E_\mathrm{F}-2.6\ \mathrm{eV}$. (f) Integrated energy distribution curves at energies on (779 eV) and off (777 eV) Co resonance. (h) Band dispersion at energy on Co resonance (779 eV). The blue dashed rectangle represents the intensity enhancement corresponding to the blue arrow in (g).}
\end{figure}

Figures \ref{Fig: SX}(b) and (c) represent bulk-sensitive SX-ARPES measurements of $\mathrm{Co}\mathrm{Nb}_3\mathrm{S}_6$.
The additional band dispersion appears in the same position as VUV-ARPES measurements [Fig.\ \ref{Fig: SX}(a)], so we claim that this dispersion comes from the bulk electronic structure of $\mathrm{Co}\mathrm{Nb}_3\mathrm{S}_6$.
$h\nu$ dependence measurements shows weakly oscillating feature of the additional band along the $k_z$ direction [Fig.\ \ref{Fig: SX}(d)], which also indicates that this band has bulk origin.

Fig.\ \ref{Fig: SX}(f) shows two peak structures of the integrated photoemission intensity near $\mathrm{L}_3$ (779 eV) and $\mathrm{L}_2$ (794 eV) edges.
In Fig.\ \ref{Fig: SX}(f), the integration is performed between $E_\mathrm{F}$ and $E_\mathrm{F}-2.6\ \mathrm{eV}$ and in the momentum range represented by the blue arrow in Fig.\ \ref{Fig: SX}(e).
The energy distribution curve integrated along the blue arrow in Fig.\ \ref{Fig: SX}(e) at the energy on resonance have a peak structure near the Fermi level, highlighted by the blue arrow in Fig.\ \ref{Fig: SX}(g), which was not observed in off-resonance measurements.
Also, the intensity difference between on- and off-resonance becomes larger below $E_\mathrm{F}-0.5\ \mathrm{eV}$.
These results have the correspondence with newly observed Co bands, indicated by the blue or light blue curves in Figs.\ 1(c3) and \ref{Fig: SX}(c).

\clearpage

\section*{Note 8: Temperature dependence measurements of $\mathbf{Co}\mathbf{Nb}_3\mathbf{S}_{6-x}$}

\begin{figure}[hb]
\includegraphics[width=160 mm]{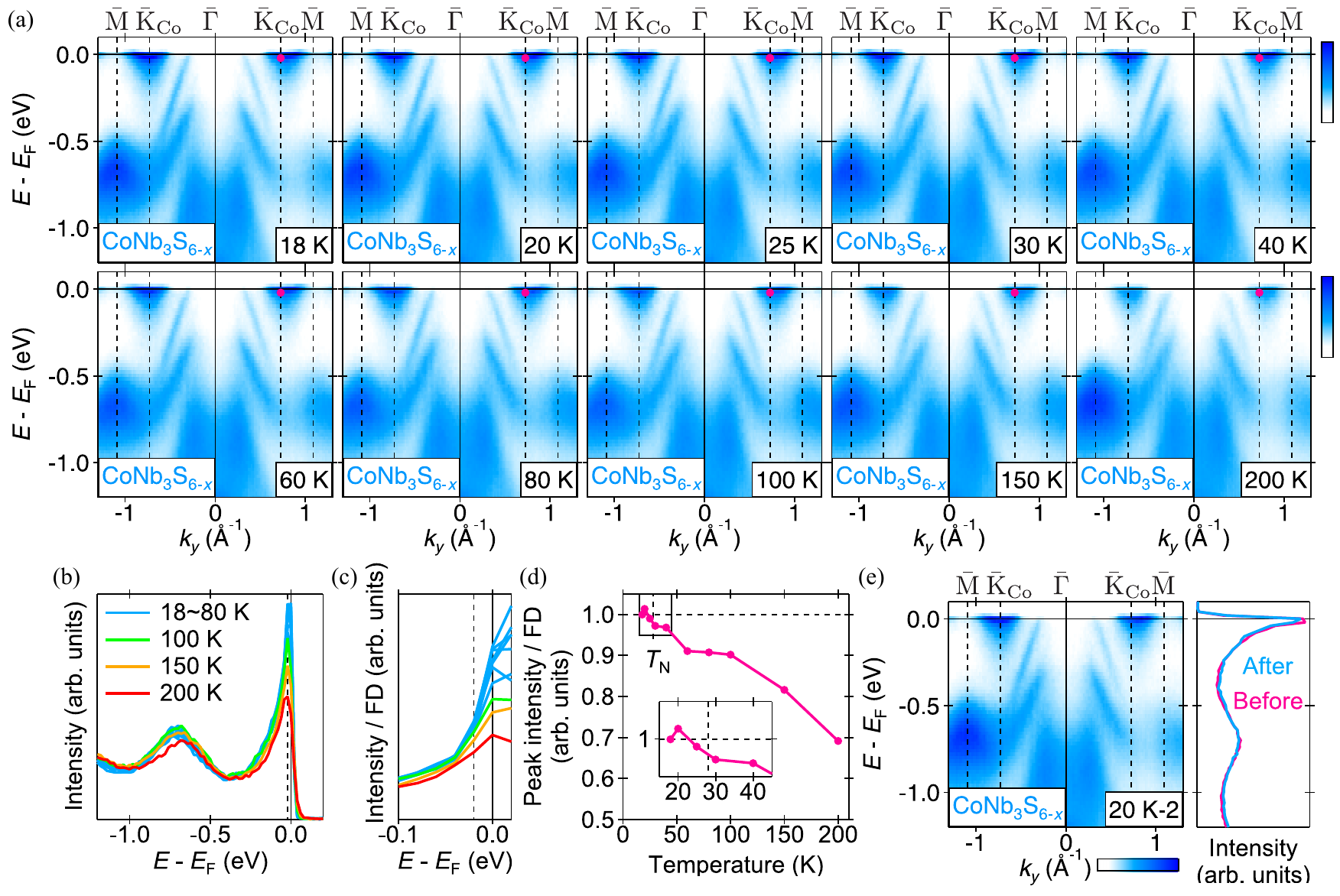}
\caption{\label{Fig: Temp_dependence} Temperature dependence of the electronic structure of $\mathrm{Co}\mathrm{Nb}_3\mathrm{S}_{6-x}$. (a) Band dispersions measured by increasing the temperature from 18 K to 200 K. The photon energy was set to 120 eV. (b) Energy distribution curves (EDCs) at the $\bar{K}_\mathrm{Co}$ point in (a). The black dashed line represents the peak position ($E_\mathrm{F}-0.02\ \mathrm{eV}$). (c) Temperature dependence of EDCs near the Fermi level, divided by the convolution of the Fermi-Dirac (FD) distribution and the Gauss function corresponding to the thermal fluctuation and finite energy resolution, respectively. (d) Temperature dependence of the divided intensity at $E_\mathrm{F}-0.02\ \mathrm{eV}$. The inset shows the behavior across $T_\mathrm{N}$. (e) Band dispersion measured after cooling the sample from 200 K to 20 K. The right panel shows the EDCs at the $\bar{K}_\mathrm{Co}$ point at 20 K, before and after heating.}
\end{figure}

We performed temperature dependence measurements using $\mathrm{Co}\mathrm{Nb}_3\mathrm{S}_{6-x}$.
Increasing the temperature higher than $T_\mathrm{N}\simeq28\ \mathrm{K}$, we observed that the intensity of the additional band near the Fermi level became weaker [Figs.\ \ref{Fig: Temp_dependence}(a), (b)].
The Fermi-Dirac distribution function and finite energy resolution should be responsible for this weakening, but the weakening still remained after the correction of these effects [Figs.\ \ref{Fig: Temp_dependence}(c) and (d)].
Since the intensity near the Fermi level recovered when the temperature became low [Fig.\ \ref{Fig: Temp_dependence}(e)], we could exclude the possibility of surface aging due to the heating.
Therefore, this weakening may be due to the disordered magnetic structure above $T_\mathrm{N}$.
Although we only have the temperature dependence of $\mathrm{Co}\mathrm{Nb}_3\mathrm{S}_{6-x}$, we think that $\mathrm{Co}\mathrm{Nb}_3\mathrm{S}_6$ will show the similar behavior, because the electronic structures of them are almost the same [Figs.\ 1(c2) and 3(c)].

We note that the amplitude change across $T_\mathrm{N}$ [Fig.\ \ref{Fig: Temp_dependence}(d) inset] is very small, but it is not unusual in ARPES measurements of magnetic materials; for example, ferromagnetic $\mathrm{Cr}\mathrm{Nb}_3\mathrm{S}_6$, with the same crystal structure as $\mathrm{Co}\mathrm{Nb}_3\mathrm{S}_6$, does not show the disappearance of the Zeeman splitting, but only shows the suppression of some small peaks across $T_\mathrm{C}$ [56].

\clearpage

\section*{Note 9: Comparison between $\mathbf{Nb}\mathbf{S}_2$ and $M\mathbf{Nb}_3\mathbf{S}_6\ (M=\mathbf{Fe},\ \mathbf{Co},\ \mathbf{Ni})$}

\begin{figure}[hb]
\includegraphics{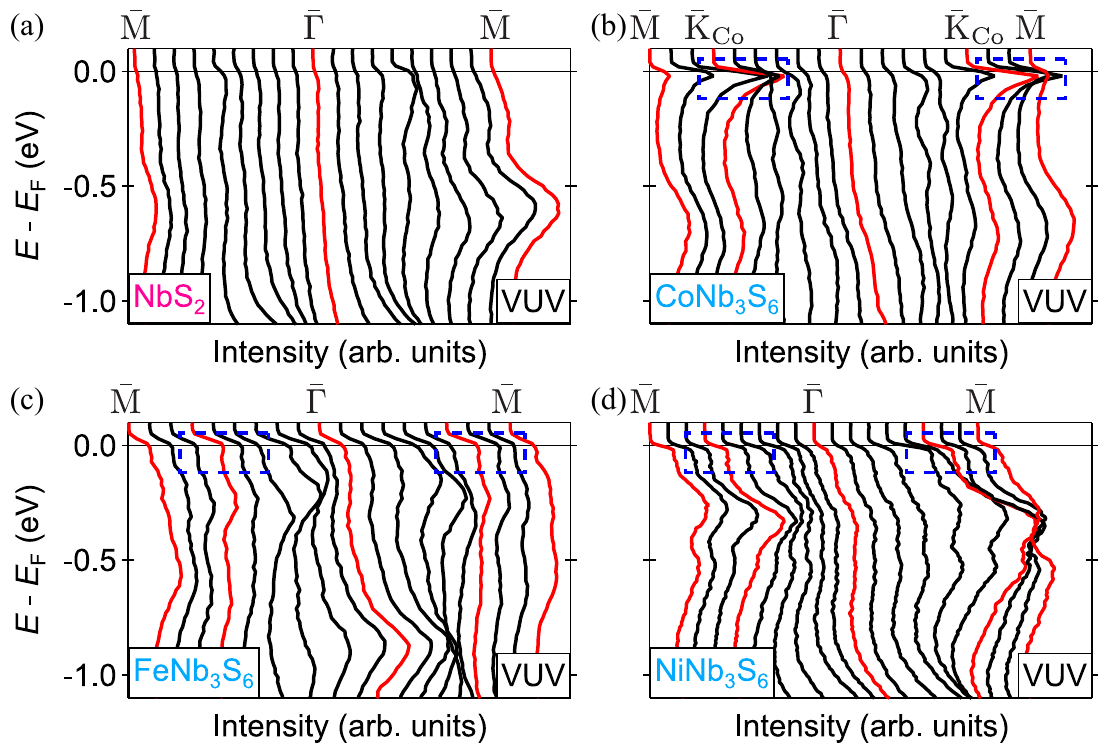}
\caption{\label{Fig: MNb3S6_EDCs} Energy distribution curves (EDCs) of $\mathrm{Nb}\mathrm{S}_2$ and $M\mathrm{Nb}_3\mathrm{S}_6\ (M=\mathrm{Fe},\ \mathrm{Co},\ \mathrm{Ni})$ extracted from Figs. 1(b2)-(e2). Spectrum peaks corresponding to the additional band dispersion in $\mathrm{Co}\mathrm{Nb}_3\mathrm{S}_6$ are marked by the blue dashed rectangles, and these rectangles are also placed in (c) and (d).}
\end{figure}

Figure \ref{Fig: MNb3S6_EDCs} shows the energy distribution curves (EDCs) of $\mathrm{Nb}\mathrm{S}_2$ and $M\mathrm{Nb}_3\mathrm{S}_6\ (M=\mathrm{Fe},\ \mathrm{Co},\ \mathrm{Ni})$ taken by 120 eV VUV light.
In the EDCs of $\mathrm{Co}\mathrm{Nb}_3\mathrm{S}_6$ [Fig.\ \ref{Fig: MNb3S6_EDCs}(b)], sharp peaks appear near the Fermi level, corresponding to the additional band dispersion discussed above.
On the other hand, the EDCs of $\mathrm{Fe}\mathrm{Nb}_3\mathrm{S}_6$ and $\mathrm{Ni}\mathrm{Nb}_3\mathrm{S}_6$ do not have such peak structures.
Although there seem to be small peaks near the Fermi level in Fig.\ \ref{Fig: MNb3S6_EDCs}(c), we cannot derive any strong conclusion due to the poor quality of our data.
Anyway, these small features of $\mathrm{Fe}\mathrm{Nb}_3\mathrm{S}_6$ and $\mathrm{Ni}\mathrm{Nb}_3\mathrm{S}_6$ are completely different from sharp peaks in $\mathrm{Co}\mathrm{Nb}_3\mathrm{S}_6$.
From this viewpoint, we claim that this additional band dispersion is unique to $\mathrm{Co}\mathrm{Nb}_3\mathrm{S}_6$.
Also, we note that additional occupied states near $E_\mathrm{F}-0.3\ \mathrm{eV}$ in $\mathrm{Ni}\mathrm{Nb}_3\mathrm{S}_6$ [blue dashed curves in Fig.\ 1(e3)] may have correspondence with the unique band in $\mathrm{Co}\mathrm{Nb}_3\mathrm{S}_6$, considering that a Ni atom has one more electron than a Co atom does.

\begin{figure}
\includegraphics{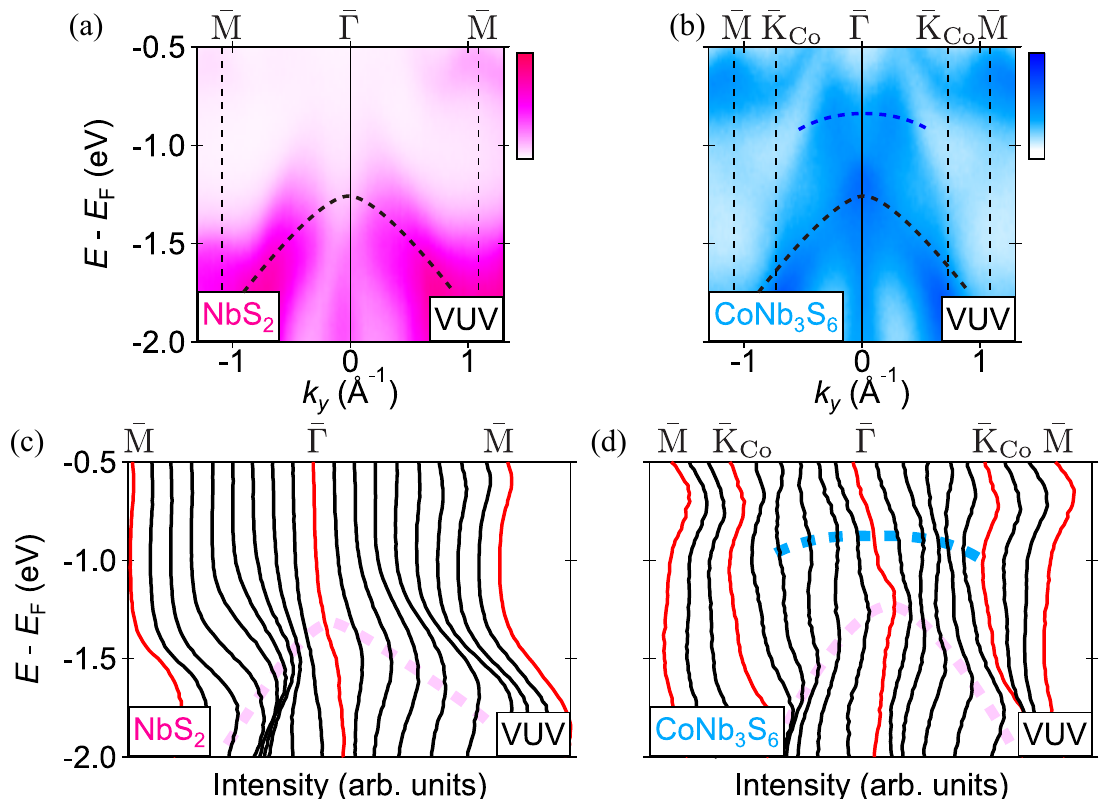}
\caption{\label{Fig: NS_CNS_high} (a), (b) Band dispersions of $\mathrm{Nb}\mathrm{S}_2$ and $\mathrm{Co}\mathrm{Nb}_3\mathrm{S}_6$ at high binding energies. The black and blue dashed curves correspond to pink and light blue ones in (c) and (d), originating from $\mathrm{Nb}\mathrm{S}_2$ and the Co intercalation, respectively. (c), (d) EDCs of (a) and (b), respectively.}
\end{figure}

Figures \ref{Fig: NS_CNS_high}(c) and (d) compare EDCs of $\mathrm{Nb}\mathrm{S}_2$ and $\mathrm{Co}\mathrm{Nb}_3\mathrm{S}_6$ at high binding energies.
An inverted V-shaped band, highlighted by the pink dashed curves, exists in both panels and tail of this band appears in Figs.\ 1(b2) and (c2).
In addition, a weakly dispersing band exists in $\mathrm{Co}\mathrm{Nb}_3\mathrm{S}_6$, as highlighted by the light blue dashed curve.

\begin{figure}
\includegraphics{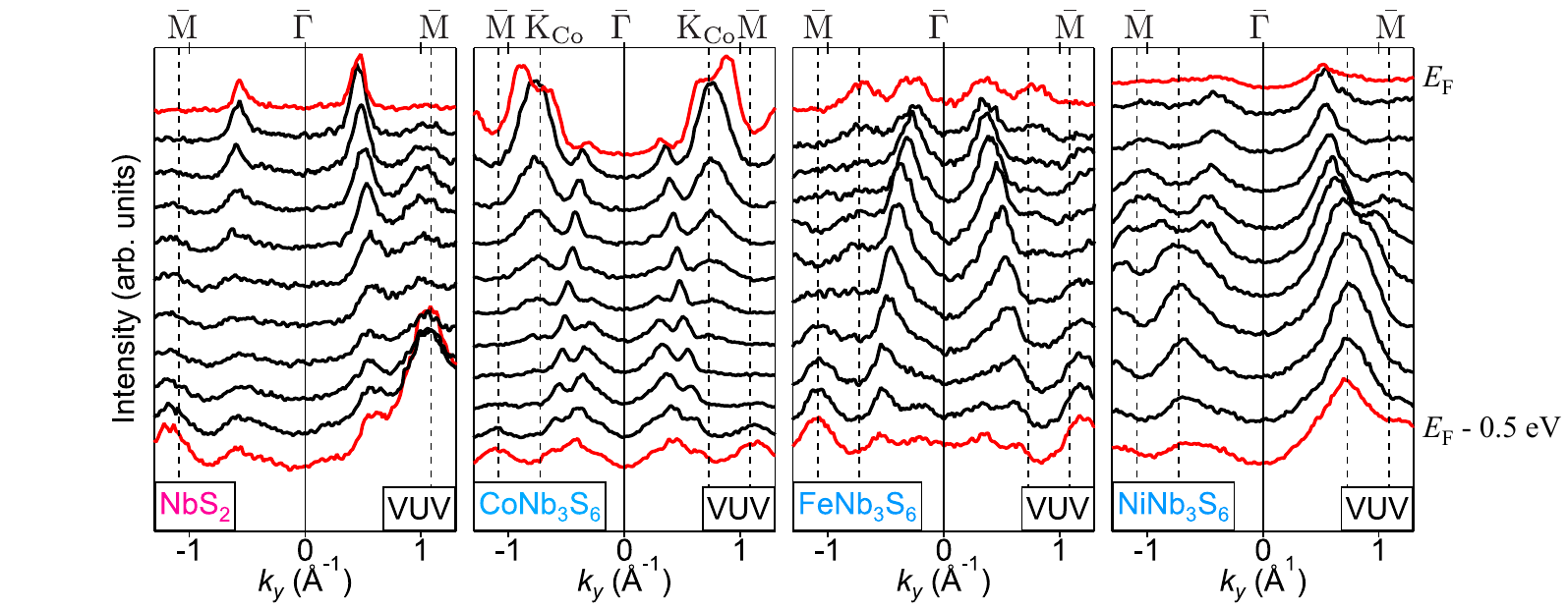}
\caption{\label{Fig: MNb3S6_MDCs} Momentum distribution curves (MDCs) of $\mathrm{Nb}\mathrm{S}_2$ and $M\mathrm{Nb}_3\mathrm{S}_6$.}
\end{figure}

\section*{Note 10: Nano-VUV-ARPES measurements of $\mathbf{Co}\mathbf{Nb}_3\mathbf{S}_6$}

\begin{figure}[hb]
\includegraphics{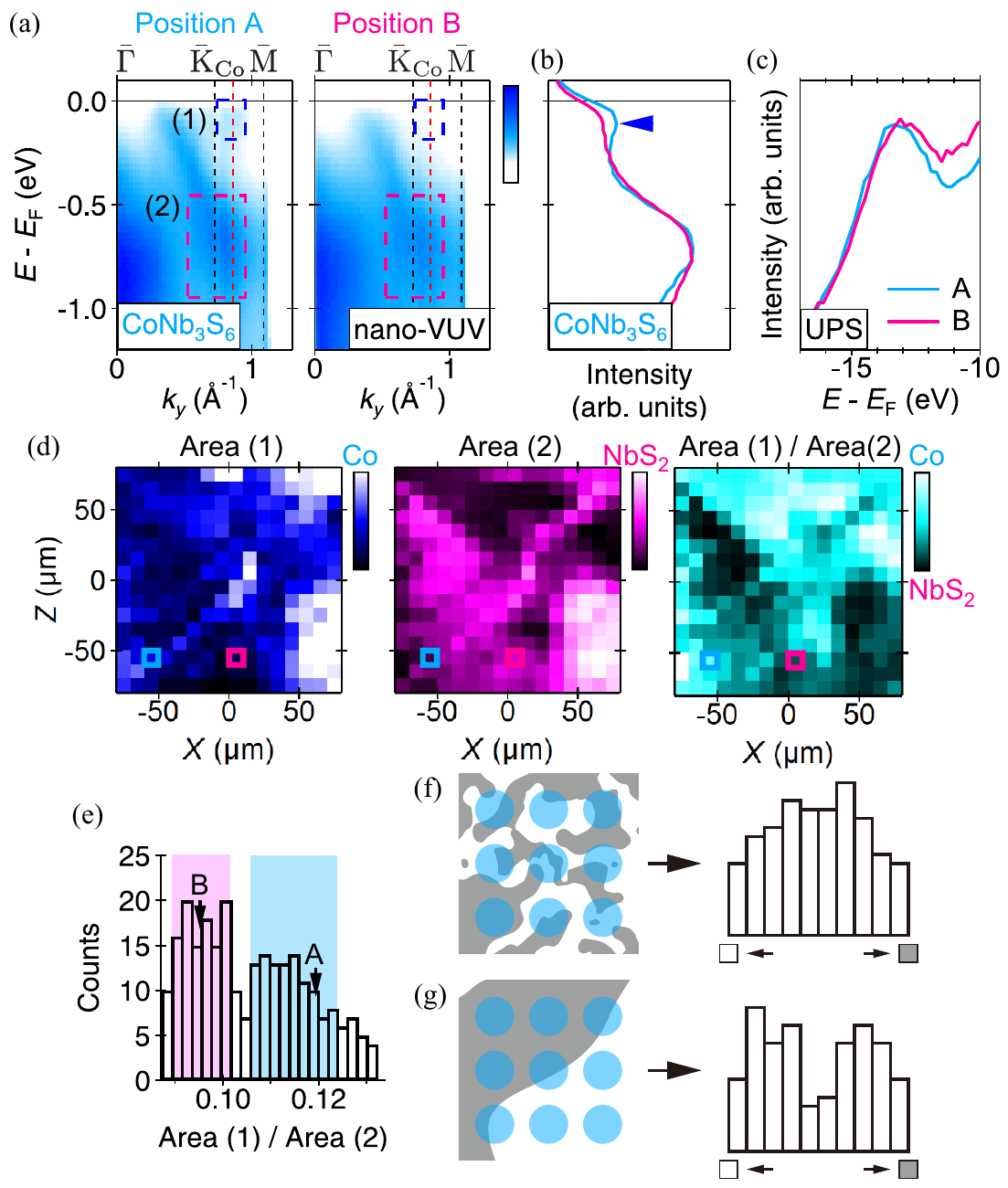}
\caption{\label{Fig: nano_ARPES_All} Nano-VUV-ARPES measurements of $\mathrm{Co}\mathrm{Nb}_3\mathrm{S}_6$ ($h\nu=74\ \mathrm{eV}$). (a) Band dispersions along the $k_y$ direction at two different positions (Positions A and B). Two areas corresponding to the band dispersions of Co atoms and $\mathrm{Nb}\mathrm{S}_2$ layers are marked by the blue and pink dashed rectangles and hereafter called Areas (1) and (2), respectively. (b) Energy distribution curves (EDCs) of (a) at $k_y=0.86\ \textrm{\AA}$ [red dashed lines in (a)]. The EDC at Position A (light blue) is multiplied by 1.1 to highlight the spectrum peak near the Fermi level. (c) Core level spectra at Positions A and B. The spectrum at Position A (light blue) is multiplied by 1.2 to make the comparison easier. (d) Real-space maps of the intensities of Area (1) (left panel), Area (2) (center panel), and Area (1) divided by Area (2) (right panel). Areas (1) and (2) are defined in (a), and the light blue and pink squares in each panel represent Positions A and B used in (a), respectively. (e) Histogram of the normalized intensities in (d) right panel. The two arrows represent the values at Positions A and B. (f), (g) Schematics of real-space map measurements and expected histograms when the spot size is as large as the domain size and when the spot is small enough comparing to the domain size, respectively.}
\end{figure}

Since $\mathrm{Co}\mathrm{Nb}_3\mathrm{S}_6$ exposes two termination surfaces (Co and $\mathrm{Nb}\mathrm{S}_2$ layers), surface-sensitive VUV-ARPES enables us to distinguish them, as reported for some materials [50, 51].
Using this advantage and nano-focused light, we selectively detected Co layers and extracted the effect of the Co intercalation on the electronic structure.
Figure \ref{Fig: nano_ARPES_All}(a) shows band dispersions taken at two different positions (Positions A and B).
Intensities in the blue dashed rectangles [Area (1)] are different between the two positions.
This difference is more clearly displayed by the energy distribution curves [Fig.\ \ref{Fig: nano_ARPES_All}(b)]: band dispersion taken at Position A has a peak structure near the Fermi level, as highlighted by the blue arrow, while at Position B the peak is wholly suppressed.
Since Area (1) corresponds to the band dispersion of Co atoms, this difference in Positions A and B reflects two termination surfaces; Co and $\mathrm{Nb}\mathrm{S}_2$ layers, respectively.
Deeper S $3s$ core level peak at Position A than that at Position B [Fig.\ \ref{Fig: nano_ARPES_All}(c)] also supports this discrimination, because the comparison of this core level peak in $\mathrm{Nb}\mathrm{S}_2$ and $\mathrm{Co}\mathrm{Nb}_3\mathrm{S}_6$ shows the consistent behavior [Fig.\ \ref{Fig: XPS_All} right bottom panel].

We obtained the real-space map using this difference and clearly distinguished two cleavage surfaces [Fig.\ \ref{Fig: nano_ARPES_All}(d)], demonstrating the advantage of nano-ARPES technique [52-54].
In the real-space maps of the intensities of Areas (1) and (2), we observed that positions with the weak intensity of Area (1) [black area in Fig.\ \ref{Fig: nano_ARPES_All}(d) left panel] have the strong intensity of Area (2) [pink or white area in Fig.\ \ref{Fig: nano_ARPES_All}(d) center panel] and vice versa.
This result means that the intensities of Areas (1) and (2) have a relationship with Co and $\mathrm{Nb}\mathrm{S}_2$ layers on the top surface of the $\mathrm{Co}\mathrm{Nb}_3\mathrm{S}_6$ crystal, respectively.
These results also support our argument that the unique band near the Fermi level is associated with intercalated Co atoms.

The histogram of the real-space map [Fig.\ \ref{Fig: nano_ARPES_All}(e)] shows two-peak structure reflecting two cleavage surfaces.
This two-peak structure certifies that our nano-VUV-ARPES successfully distinguished two termination surface domains.
If the spot size is as large as the domain size, ARPES measurements will capture the electronic structure of both domains more than likely, and then the histogram is expected to show a one-peak structure [Fig.\ \ref{Fig: nano_ARPES_All}(f)].
On the other hand, if the spot size is small enough, ARPES measurements will hardly capture both domains, and then the histogram will have two peaks [Fig.\ \ref{Fig: nano_ARPES_All}(g)].

\clearpage

\section*{Note 11: Measurement position dependence in VUV-ARPES}

\begin{figure}[hb]
\includegraphics{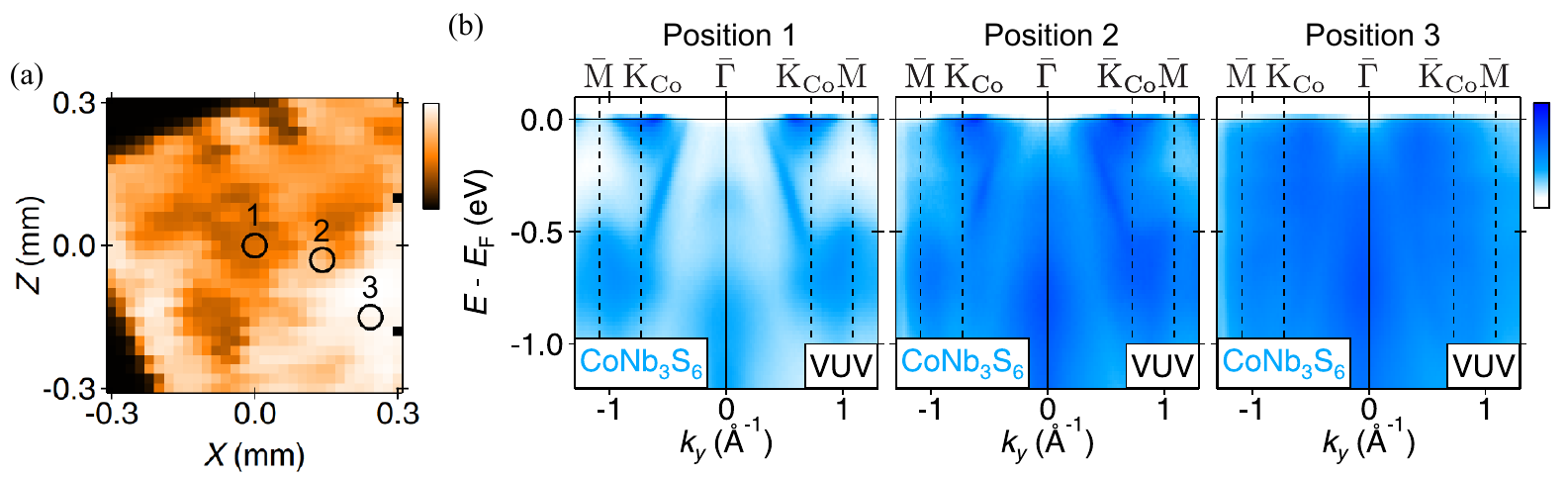}
\caption{\label{Fig: VUV_RealSpace} Measurement position dependence in VUV-ARPES at SSRL BL5-2 ($h\nu=90\ \mathrm{eV}$). (a) Real-space map of the integrated intensity in VUV-ARPES measurements of $\mathrm{Co}\mathrm{Nb}_3\mathrm{S}_6$. Labels 1-3 represent three positions used for measurements. (b) Band dispersions along the $k_y$ direction measured at the three deferent positions in (a).}
\end{figure}

Figure \ref{Fig: VUV_RealSpace} represents measurement position dependence which we observed in VUV-ARPES at SSRL BL5-2.
We performed the same band dispersion measurements at three different positions represented in Fig.\ \ref{Fig: VUV_RealSpace}(a), and then got band dispersions like Fig.\ \ref{Fig: VUV_RealSpace}(b).
Although the integrated intensities are similar among these positions, the clearness of the band dispersion is completely different from each other.
We used Position 1 to get clear electronic structure of $\mathrm{Co}\mathrm{Nb}_3\mathrm{S}_6$ like Figs.\ 1(c1) and (c2).

\clearpage

\section*{Note 12: Broken inversion symmetry in $\mathbf{Co}\mathbf{Nb}_3\mathbf{S}_6$}
\begin{figure}[hb]
\includegraphics{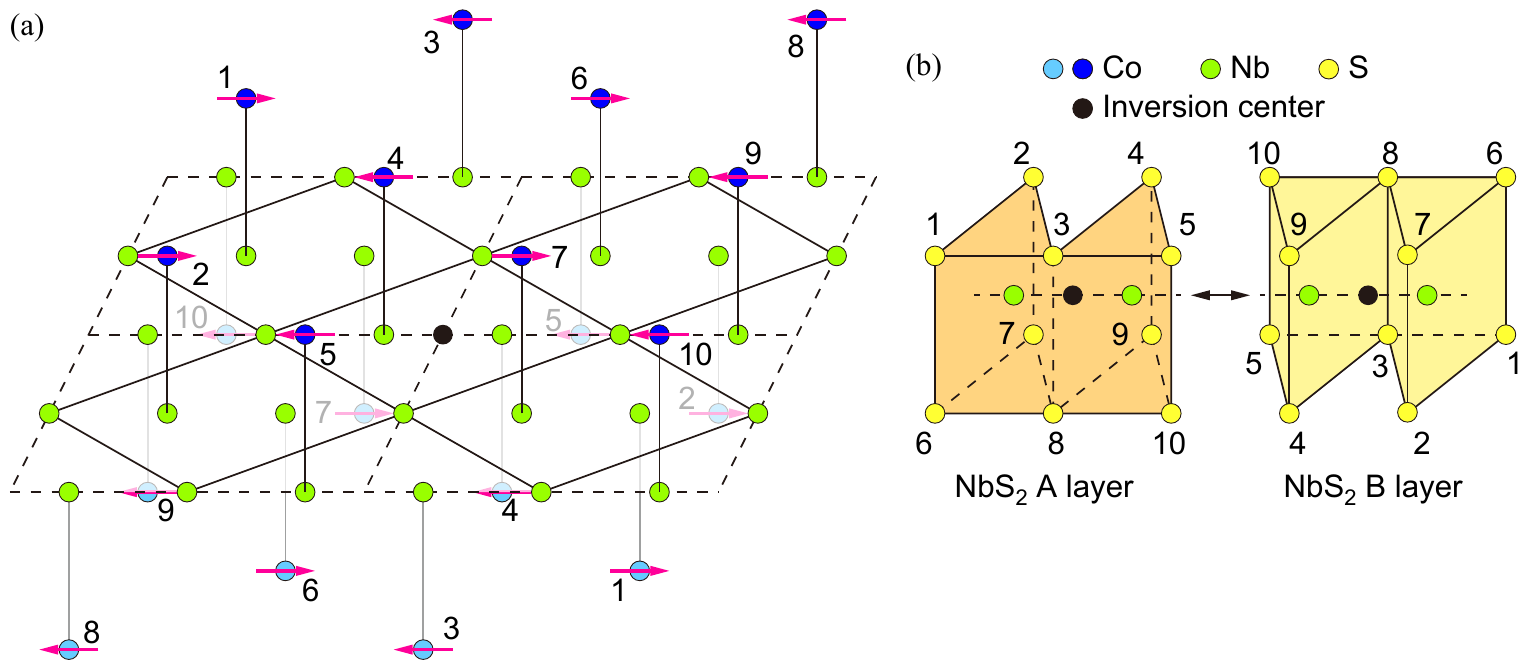}
\caption{\label{Fig: Broken_inversion} Property of the broken inversion symmetry in $\mathrm{Co}\mathrm{Nb}_3\mathrm{S}_6$. (a) Atomic configuration and magnetic moments of Nb and Co atoms. This structure is invariant under the inversion operation for the black circle in the center because Co atoms with the same numbers are exchanged with each other. (b) Schematic of two different $\mathrm{Nb}\mathrm{S}_2$ layers. They are exchanged with each other by the same inversion operation as (a).}
\end{figure}

Here we discuss how the inversion symmetry is broken in $\mathrm{Co}\mathrm{Nb}_3\mathrm{S}_6$.
If we only see Nb and Co atoms and their magnetic moments [Fig.\ \ref{Fig: Broken_inversion}(a)], this structure has inversion center points.
However, since this inversion operation changes the position of S atoms, the whole structure breaks the inversion symmetry.
$2H$-$\mathrm{Nb}\mathrm{S}_2$ has two $\mathrm{Nb}\mathrm{S}_2$ layers of different orientations, and the inversion operation represented in Fig.\ \ref{Fig: Broken_inversion}(a) changes a $\mathrm{Nb}\mathrm{S}_2$ layer to the other orientation.
In other words, the inversion symmetry is broken in $\mathrm{Co}\mathrm{Nb}_3\mathrm{S}_6$ because $2H$-$\mathrm{Nb}\mathrm{S}_2$ has a $AB$-stacked structure.
Therefore, we can consider the difference of wavefunctions on $A$ and $B$ layers as a criterion of the degree of the broken inversion symmetry.

\section*{Note 13: ARPES spectra analysis of $AB$-stacked materials based on a tight-binding model}

\begin{figure}[hb]
\includegraphics[width=160mm]{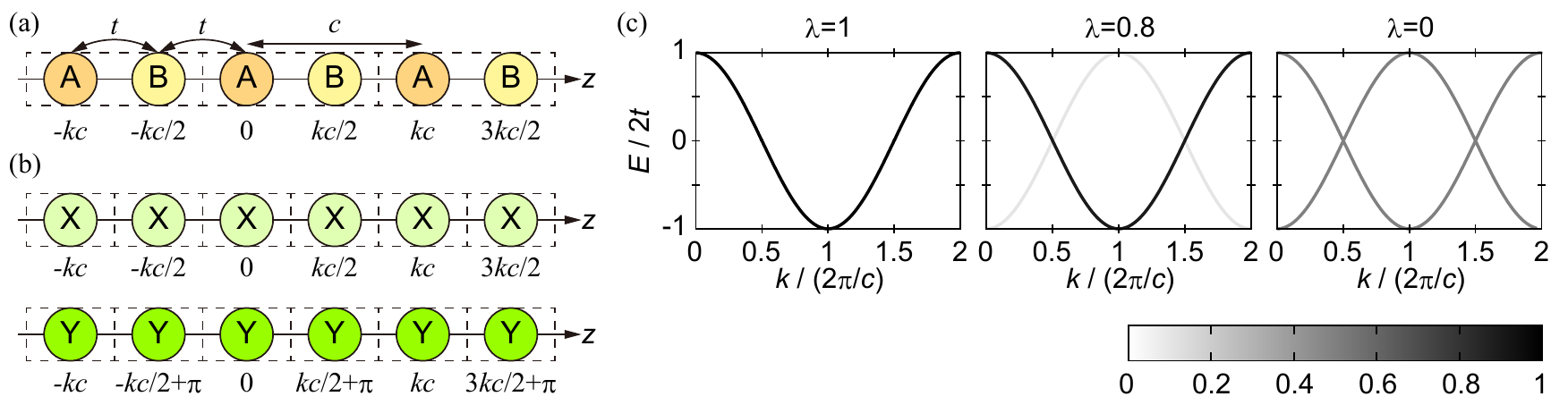}
\caption{\label{Fig: AB_tight_binding} ARPES spectra analysis of $AB$-stacked materials. (a) One-dimensional tight-binding model including two atoms ($A$ and $B$). The bottom row represents the phase for each atom corresponding to the Bloch wavevector $k$. (b) Two possible bases with periodicity $c/2$ corresponding to the Bloch wavevector $k$ when the considered periodicity is $c$. (c) Band dispersion of the tight-binding model unfolded to the periodicity $4\pi/c$. These three panels correspond to schematics in Figs.\ 2(a)-(c).}
\end{figure}

In this Note we discuss the ARPES spectra of $AB$-stacked materials along the $k_z$ (out-of-plane) direction based on the one-dimensional tight-binding model represented in Fig.\ \ref{Fig: AB_tight_binding}(a).
We define two bases for the model corresponding to the Bloch wavevector $k$ as
\begin{gather}
\Psi_A(z,k)=\sum_{n\in\mathbb{Z}}\exp(inkc)\psi_A(z-nc)\\
\Psi_B(z,k)=\sum_{n\in\mathbb{Z}}\exp\Bigl(i(n+1/2)kc)\Bigr)\psi_B\Bigl(z-(n+1/2)c\Bigr),
\end{gather}
where $\psi_A(z)$ and $\psi_B(z)$ are wavefunctions localized around $z=0$.
We assume that the diagonal components of the Hamiltonian matrix are the same, and then the Hamiltonian matrix for the Bloch wavevector $k$ becomes
\begin{equation}
\mathcal{H}(k)=\begin{pmatrix}E_0 & 2t\cos(kc/2)\\2t\cos(kc/2) & E_0\end{pmatrix},
\end{equation}
where $t$ is the hopping integral between $A$ and $B$.
Hereafter we set the zero point of the eigenenergy so that $E_0$ can be neglected.
The Hamiltonian has two eigenstates with energies $E(k)=\pm2t\cos(kc/2)$.

Next, the band dispersion is unfolded to the periodicity $4\pi/c$ to estimate the ARPES spectra [47-49].
The Bloch wavevectors $k$ and $k+2\pi/c$ are inequivalent in the reciprocal space of periodicity $4\pi/c$, so the eigenstate $\Psi(z,k)=p\Psi_A(z)+q\Psi_B(z)$ can be decomposed into two wavefunctions $\Psi_X(z)$ and $\Psi_Y(z)$ in Fig.\ \ref{Fig: AB_tight_binding}(b).
In the form of equations, this argument becomes
\begin{gather}
\Psi(z,k)=p\Psi_A(z)+q\Psi_B(z)=x\Psi_X(z)+y\Psi_Y(z)\\
\Psi_X(z,k)=\frac{1}{\sqrt{2}}\sum_{m\in\mathbb{Z}}\exp(imkc/2)\psi_X(z-mc/2)\\
\Psi_Y(z,k)=\frac{1}{\sqrt{2}}\sum_{m\in\mathbb{Z}}\exp(im(k+2\pi/c)/2)\psi_Y(z-mc/2),
\end{gather}
and the goal is to obtain $x,\ y,\ \psi_X(z),\ \textrm{and}\ \psi_Y(z)$.
After some calculations, we get
\begin{gather}
x\psi_X(z)=\frac{1}{\sqrt{2}}\Bigl(p\psi_A(z)+q\psi_B(z)\Bigr)\\
y\psi_Y(z)=\frac{1}{\sqrt{2}}\Bigl(p\psi_A(z)-q\psi_B(z)\Bigr).
\end{gather}
Therefore, the unfolded spectrum weights are
\begin{gather}
k:\ |x|^2=\frac{1}{2}\Bigl(1+2pq\lambda\Bigr)\\
k+\frac{2\pi}{c}:\ |y|^2=\frac{1}{2}\Bigl(1-2pq\lambda\Bigr)\\
\lambda=\int\psi_A^*(z)\psi_B(z)\mathrm{d}z.
\end{gather}
The parameter $\lambda$ can be between 0 and 1 and represents how wavefunctions $\psi_A(z)$ and $\psi_B(z)$ are similar.

Based on this argument, the unfolded band dispersion are drawn in cases of $\lambda=1,\ 0.8,\ 0$ [Fig.\ \ref{Fig: AB_tight_binding}(c)].
We observed that the spectrum weight depends on $\lambda$, as discussed in the main text.

\section*{Note 14: SX-ARPES measurements and DFT calculations of $\mathbf{Nb}\mathbf{S}_2$}
In our DFT calculations, the unit cell was fixed to the experimentally obtained one, and calculations were performed after optimizing atomic positions with an $11\times11\times11$ $k$ mesh.

\begin{figure}
	\includegraphics{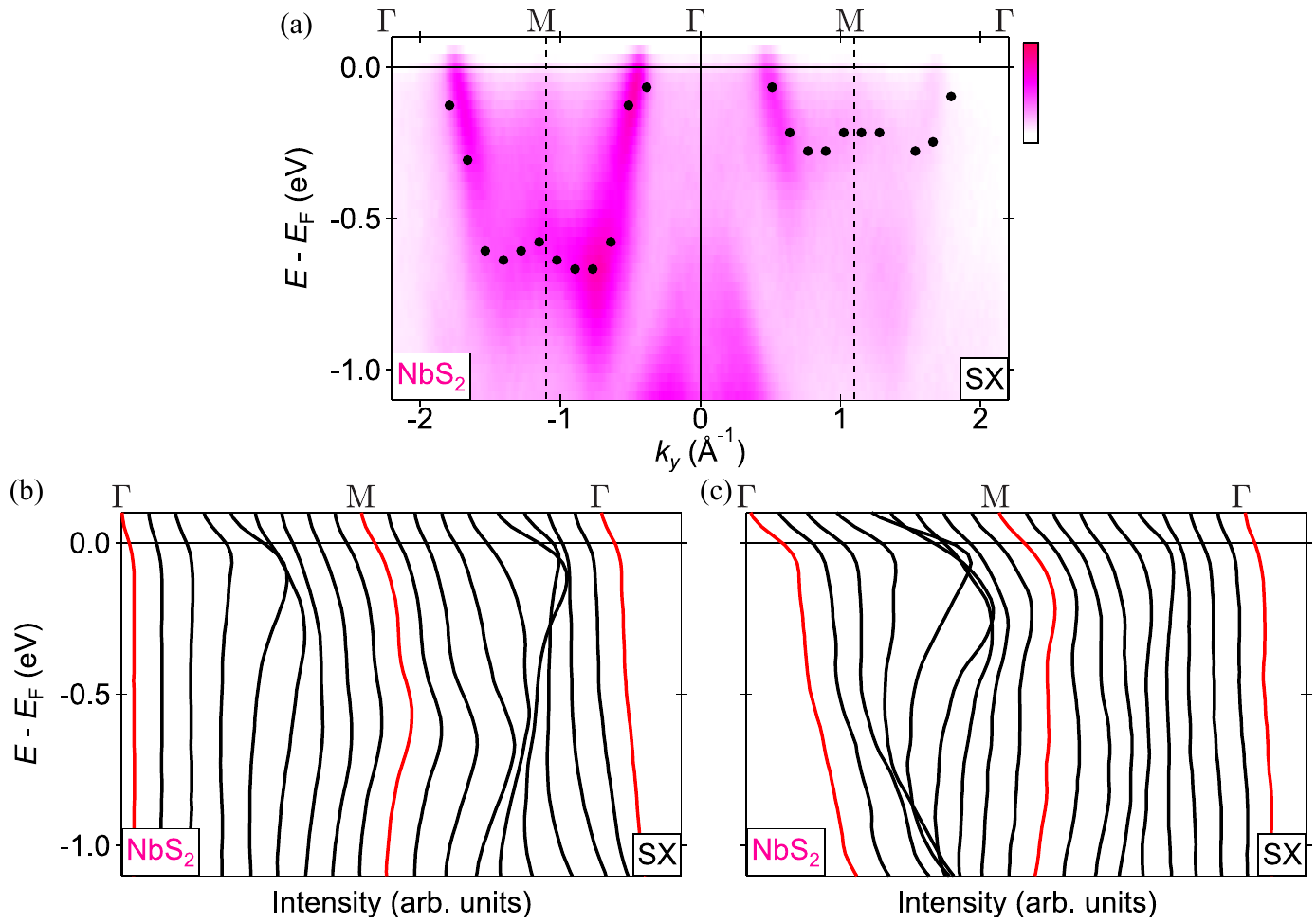}
	\caption{\label{Fig: NbS2_525} SX-ARPES measurements ($h\nu=525\ \mathrm{eV}$) of $\mathrm{Nb}\mathrm{S}_2$. (a) Band dispersion of $\mathrm{Nb}\mathrm{S}_2$ along the $k_y$ direction. (b), (c) energy distribution curves of (a) in $k_y<0$ and $k_y>0$ regions respectively. We note that the intensity scales used in (b) and (c) are different.}
\end{figure}

Figure \ref{Fig: NbS2_525} represents the band dispersion obtained by the 525 eV SX light ($k_z=22\times2\pi/c$).
Although two inverted M-shaped bands appear near the Fermi level, we only observed the lower one in $k_y<0$ region and the upper one in $k_y>0$ region, probably due to weak transition probability.
Therefore, we symmetrized peak positions extracted from energy distribution curves both of $k_y<0$ and $k_y>0$ regions and obtained the band dispersion consistent with our DFT calculations [Fig.\ 2(f)].

\begin{figure}
\includegraphics[width=160mm]{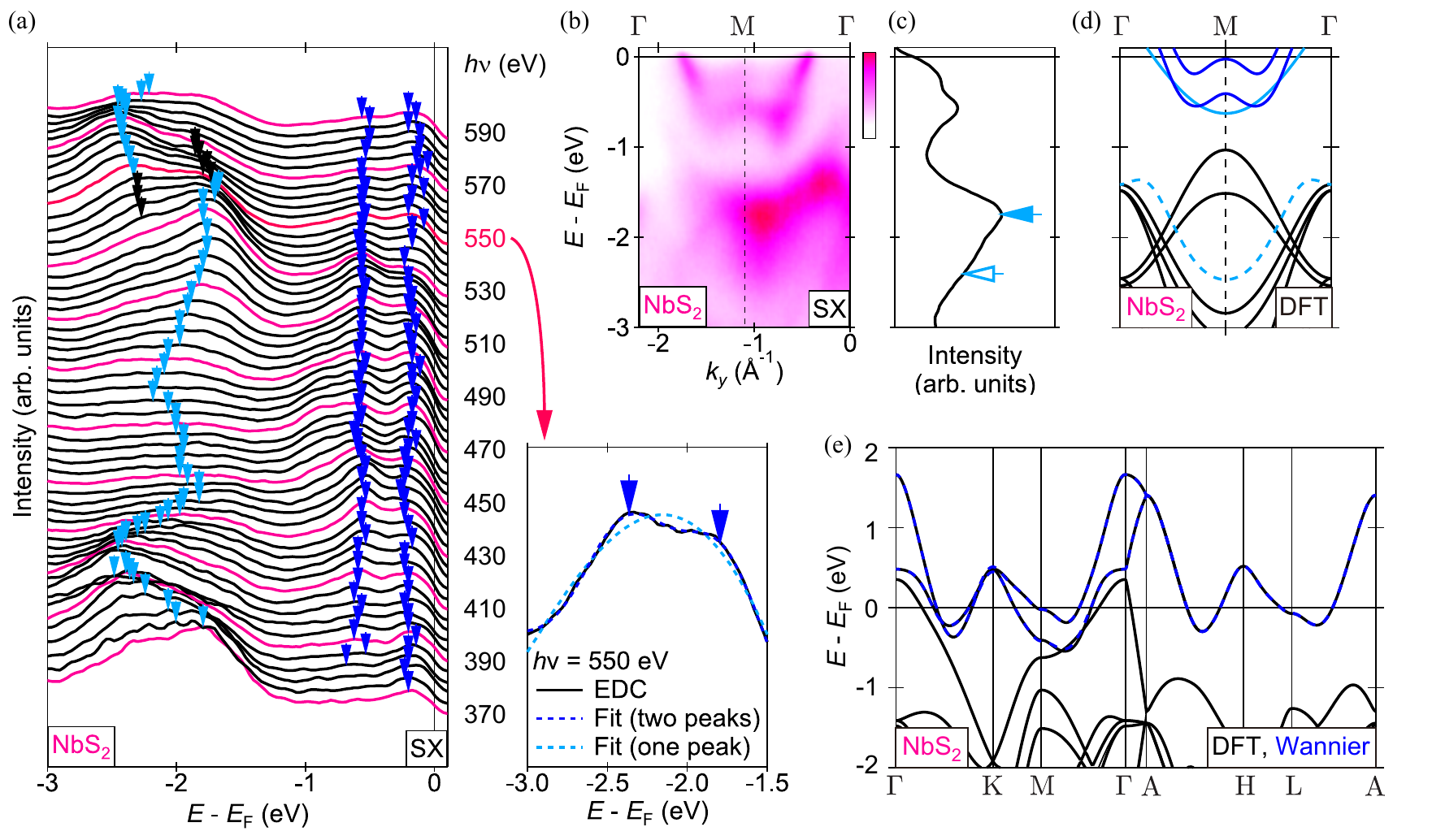}
\caption{\label{Fig: NbS2_hnDC_Wannier} SX-ARPES measurements and DFT calculations of $\mathrm{Nb}\mathrm{S}_2$. (a) Energy distribution curves (EDCs) at $\bar{M}$ point taken by the 370-590 eV SX lights. Peak positions are highlighted by blue, light blue, and black arrows and put in Fig.\ 2(d). The right bottom panel represents the EDC at 550 eV and the fitting curves, and the blue arrows represent peak positions in two-peak fitting. (b) Band dispersions of $\mathrm{Nb}\mathrm{S}_2$ along the $k_y$ direction obtained by the 525 eV SX light ($k_z=22\times2\pi/c$) (c) Energy distribution curve of (b) at the $M$ point. (d) DFT band dispersions of $\mathrm{Nb}\mathrm{S}_2$.  Dispersions corresponding to those observed by SX-ARPES are highlighted by the blue and light blue solid curves. The light blue dashed curve is the dispersion forming a pair with the solid one, which we did not observe in (b) and (c). (e) DFT band dispersion of $\mathrm{Nb}\mathrm{S}_2$ (black curves). Band dispersions of maximally localized Wannier functions obtained by Wannier90 code are overlaid by blue dashed curves.}
\end{figure}

Figure \ref{Fig: NbS2_hnDC_Wannier}(a) represents SX-ARPES measurements of $\mathrm{Nb}\mathrm{S}_2$ before the conversion from the photon energy $h\nu$ to the out-of-plane wavevector $k_z$.
We observed several peaks in each energy distribution curve highlighted by arrows.
Two peaks appear in each EDC near the Fermi level, as marked by the blue arrows in Fig.\ \ref{Fig: NbS2_hnDC_Wannier}(a). Since two peaks are close, we extracted peak positions by fitting a two-peak function (sum of two gauss functions and linear background) to each EDC. They correspond to the inverted M-shaped bands in the band dispersion along the $k_y$ direction [Figs.\ 2(f), \ref{Fig: NbS2_525}(a), and \ref{Fig: NbS2_hnDC_Wannier}(b)].
These bands appear also in the DFT band dispersion, as highlighted by blue curves in Figs.\ 2(f) and \ref{Fig: NbS2_hnDC_Wannier}(d), which form a pair originating from the bilayer structure of $\mathrm{Nb}\mathrm{S}_2$.

Around $E-E_\mathrm{F}=-2\ \mathrm{eV}$, we generally observed only one peak, marked by light blue arrows in Figs.\ \ref{Fig: NbS2_hnDC_Wannier}(a) and (c).
In Fig.\ \ref{Fig: NbS2_hnDC_Wannier}(c), the filled arrow marks the peak with the lowest binding energy in the $k_z$ dispersion, while the other one of the paired dispersion, which should have the highest binding energy $\sim 2.5\ \mathrm{eV}$, does not appear, as marked by the open arrow.
However, some spectra in Fig.\ \ref{Fig: NbS2_hnDC_Wannier}(a) are very broad, maybe indicating the existence of multiple peaks; indeed, for example, the broad EDC spectrum at 550 eV is well fitted by a two-peak function rather than a one-peak function [right bottom panel of Fig.\ \ref{Fig: NbS2_hnDC_Wannier}(a)].
We performed two-peak fitting from 534 eV to 566 eV spectra, which are particularly broad, and highlighted peak positions by light blue for the major peaks and black for the minor peaks.
In the band dispersions of DFT calculations [Fig.\ \ref{Fig: NbS2_hnDC_Wannier}(d)], these two bands are highlighted by light blue; the dispersion which we observed in Figs.\ \ref{Fig: NbS2_hnDC_Wannier}(b) and (c) is represented by the solid curve, while the other one, which we did not observe, is represented by the dashed curve.
We note that the correspondence between the light blue bands may be incorrect due to the poor resolution of SX-ARPES, but this difference has minor effect on our argument of the electronic structure near the Fermi level and the anomalous Hall effect of $\mathrm{Co}\mathrm{Nb}_3\mathrm{S}_6$.

Two band dispersions near the Fermi level were used for the calculation of the maximally localized Wannier functions (MLWFs).
The band dispersions and shapes of the MLWFs are drawn as blue curves in Fig.\ \ref{Fig: NbS2_hnDC_Wannier}(e) and in Fig.\ 2(e), respectively.

\clearpage

\section*{Note 15: Other magnetic and magnetotransport properties of $M\mathbf{Nb}_3\mathbf{S}_6\ (M=\mathbf{Fe},\ \mathbf{Co},\ \mathbf{Ni})$}

\begin{figure}[ht]
\includegraphics[width=160 mm]{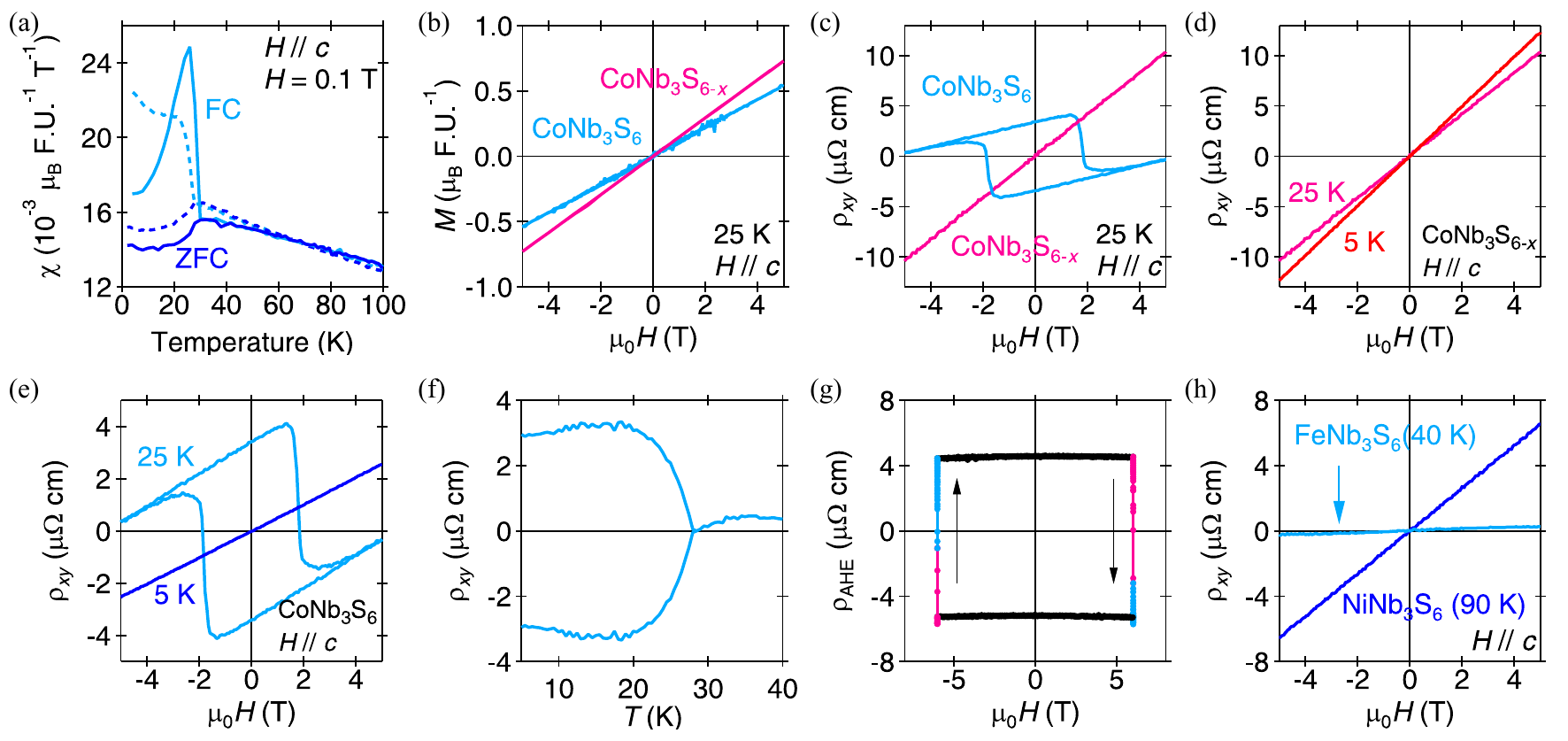}
\caption{\label{Fig: MNb3S6_magnetotransport} Supplemental data of the magnetic and magnetotransport properties of $M\mathrm{Nb}_3\mathrm{S}_6\ (M=\mathrm{Fe},\ \mathrm{Co},\ \mathrm{Ni})$. (a) Magnetic susceptibility of $\mathrm{Co}\mathrm{Nb}_3\mathrm{S}_6$ measured by two different pieces. Solid lines represent the same data as Fig.\ 3(a1) and dashed lines represent the data of another crystal piece. (b) $M$-$H$ curves of $\mathrm{Co}\mathrm{Nb}_3\mathrm{S}_6$ and $\mathrm{Co}\mathrm{Nb}_3\mathrm{S}_{6-x}$, containing the linear term of in-plane antiferromagnetic moments and the out-of-plane weak ferromagnetic moment. (c) $\rho_{xy}$-$H$ curves of $\mathrm{Co}\mathrm{Nb}_3\mathrm{S}_6$ and $\mathrm{Co}\mathrm{Nb}_3\mathrm{S}_{6-x}$. (d), (e) $\rho_{xy}$-$H$ curves of $\mathrm{Co}\mathrm{Nb}_3\mathrm{S}_{6-x}$ and $\mathrm{Co}\mathrm{Nb}_3\mathrm{S}_{6}$ respectively, measured at 5 and 25 K. (f) Temperature dependence of the Hall resistivity $\rho_{xy}$ of $\mathrm{Co}\mathrm{Nb}_3\mathrm{S}_6$ measured after a field-cooled process. (g) Hysteresis curve of $\rho_{xy}$ of $\mathrm{Co}\mathrm{Nb}_3\mathrm{S}_6$ at 5 K. The pink and light blue curves represent increasing (5 to 27 K) and decreasing (27 to 5 K) processes, respectively. (h) $\rho_{xy}$-$H$ curves of $\mathrm{Fe}\mathrm{Nb}_3\mathrm{S}_6$ and $\mathrm{Ni}\mathrm{Nb}_3\mathrm{S}_6$, measured at a slightly lower temperature than each $T_\mathrm{N}$.}
\end{figure}

Figure \ref{Fig: MNb3S6_magnetotransport} represents supplemental data of the magnetic and magnetotransport properties.
The disappearance of the AHE in $\mathrm{Co}\mathrm{Nb}_3\mathrm{S}_{6-x}$ is intrinsic because the disappearance was independent of the temperature [Fig.\ \ref{Fig: MNb3S6_magnetotransport}(d)].
When the temperature is much lower than $T_\mathrm{N}$, the AHE looks disappeared in $\mathrm{Co}\mathrm{Nb}_3\mathrm{S}_6$ [Fig.\ \ref{Fig: MNb3S6_magnetotransport}(e)].
However, it is just because the coercive field for the weak ferromagnetic moment is too large, as pointed out by Ref.\ [22].
To avoid this problem, we need to measure the temperature dependence of the anomalous Hall resistivity after a field-cooled process.
This measurement successfully reproduced the result of the previous research [Fig.\ \ref{Fig: MNb3S6_magnetotransport}(f)].
Also, we observed the hysteresis of the AHE at 5 K as represented in Fig.\ \ref{Fig: MNb3S6_magnetotransport}(g), temporarily raising the temperature to 27 K to flip the ferromagnetic moment.
Therefore, we conclude that the AHE of $\mathrm{Co}\mathrm{Nb}_3\mathrm{S}_6$ intrinsically appears at any temperature lower than $T_\mathrm{N}$.
We performed $\rho_{xy}$-$H$ measurements for $\mathrm{Fe}\mathrm{Nb}_3\mathrm{S}_6$ and $\mathrm{Ni}\mathrm{Nb}_3\mathrm{S}_6$, and did not find the signature of the AHE [Fig.\ \ref{Fig: MNb3S6_magnetotransport}(h)].

\clearpage

\section*{Note 16: Comparison between $\mathbf{Co}\mathbf{Nb}_3\mathbf{S}_6$ and $\mathbf{Co}\mathbf{Nb}_3\mathbf{S}_{6-x}$}

\begin{figure}[hb]
\includegraphics{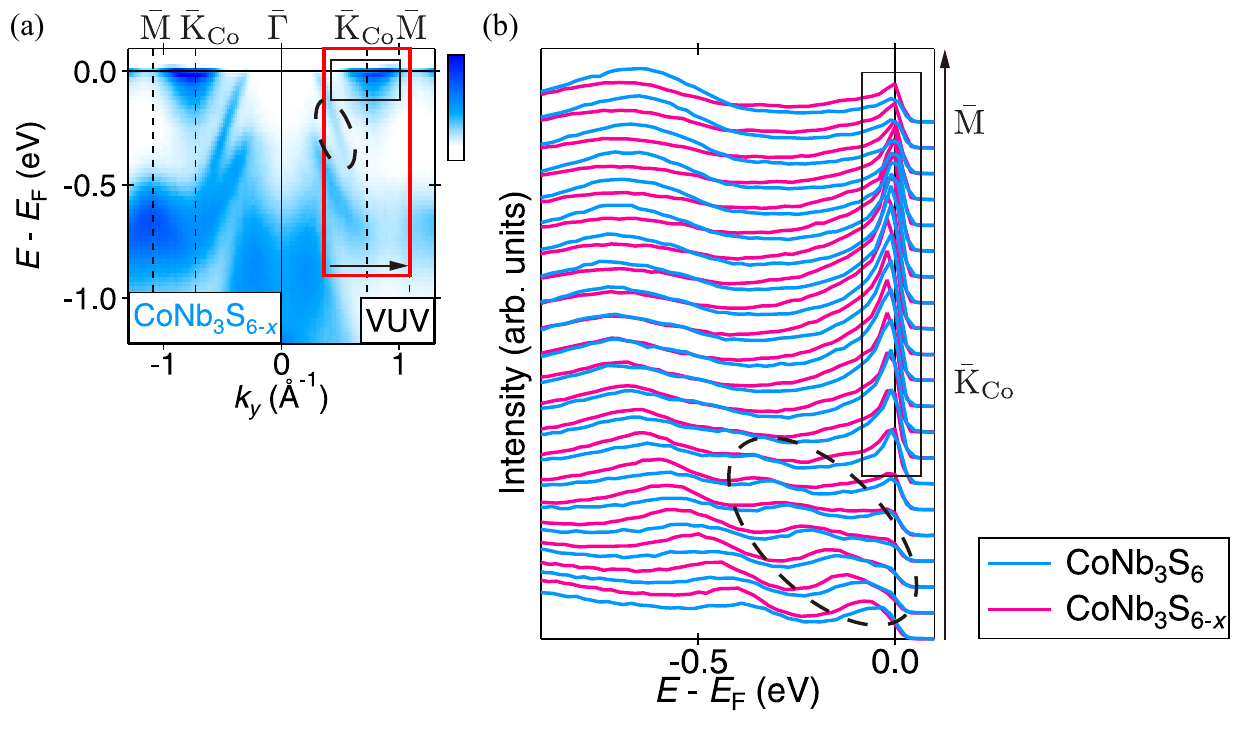}
\caption{\label{Fig: CoNb3S6-x_supplemental} Comparison between $\mathrm{Co}\mathrm{Nb}_3\mathrm{S}_6$ and $\mathrm{Co}\mathrm{Nb}_3\mathrm{S}_{6-x}$. (a) Band dispersion of $\mathrm{Co}\mathrm{Nb}_3\mathrm{S}_{6-x}$ [same as Fig.\ 3(c)]. (b) Energy distribution curves of $\mathrm{Co}\mathrm{Nb}_3\mathrm{S}_6$ and $\mathrm{Co}\mathrm{Nb}_3\mathrm{S}_{6-x}$ in the red rectangular area in (a). The observed band shift is highlighted by the black dashed oval. The dashed ovals and solid rectangles in (a) and (b) correspond each other.}
\end{figure}

Figure \ref{Fig: CoNb3S6-x_supplemental}(b) compares band dispersions of $\mathrm{Co}\mathrm{Nb}_3\mathrm{S}_6$ and $\mathrm{Co}\mathrm{Nb}_3\mathrm{S}_{6-x}$ in $k_y>0$ area represented by the red rectangle in Fig.\ \ref{Fig: CoNb3S6-x_supplemental}(a).
Similar band shifts are observed in both Figs.\ 3(d) and \ref{Fig: CoNb3S6-x_supplemental}(b), so we conclude that this change is essential, not due to the systematic error in the position of $k_y=0$.

\section*{Note 17: Effect of broken time-reversal symmetry on Kramers-Weyl points}

\begin{figure}
\includegraphics[width=160mm]{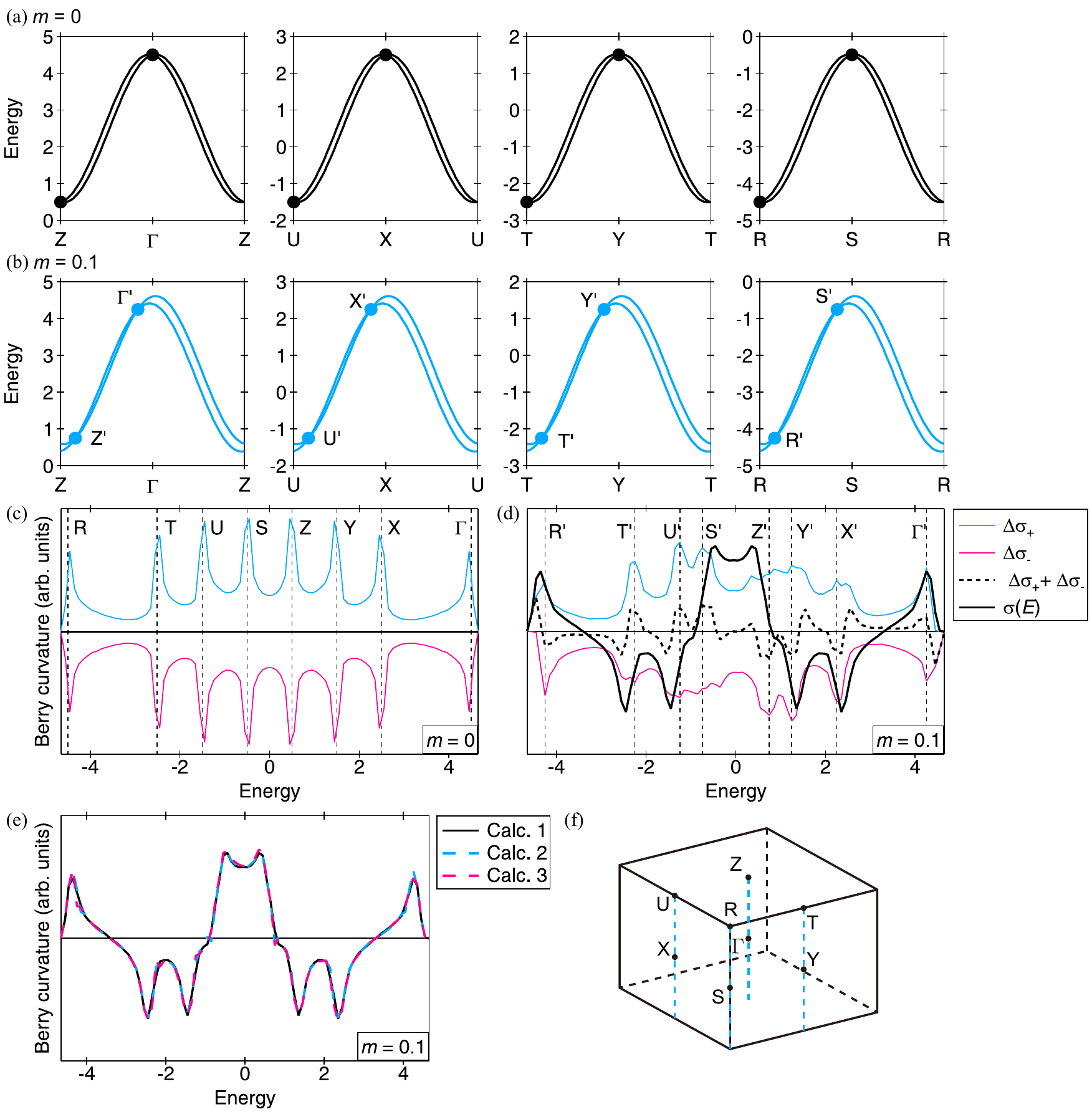}
\caption{\label{Fig: KW_Tight_binding} Band dispersion and berry curvature of the tight-binding model for Kramers-Weyl points. (a), (b) Band dispersion along high-symmetry lines when $m$ is equal to 0 or 0.1, respectively. Degenerate points near TRIMs are labeled by attaching ${}^\prime$. (c), (d) Integrated $k_z$ components of Berry curvatures. The light blue, pink, and black (dashed) lines represent the integration of positive, negative, and all components of Berry curvatures at $\mathbf{k}$ points having specific energy. The black solid curve represents the summation of Berry curvature from the lowest energy. Dashed lines and labels at the top represent positions of (Kramers-)Weyl points. (e) Integrated $k_z$ components of Berry curvatures calculated by three different numerical calculations. (f) Orthorhombic Brillouin zone and labels of high-symmetry points.}
\end{figure}

We use the tight-binding model of space group 222 in Ref.\ [24] and add a magnetic-field term to discuss the effect of broken time-reversal symmetry on Kramers-Weyl points;
\begin{equation}
\mathcal{H}(\mathbf{k})=\sum_{i=x,y,z}\Bigl[t_i\cos(k_i)+s_i\sin(k_i)\sigma_i\Bigr]+m\sigma_z,
\end{equation}
where $\sigma_i$ are the Pauli matrices and $t_i\not=t_j$ and $s_i\not=s_j$ hold when $i\not=j$.
In numerical calculations, $(t_x,\ t_y,\ t_z)=(1,\ 1.5,\ 2)$ and $(s_x,\ s_y,\ s_z)=(0.1,\ 0.15,\ 0.2)$ were used as the hopping terms.
For the summation of Berry curvature, $k_i=2\pi(n_i+1/2)/N\ (0\le n_i < N)$ mesh was used.
The mesh size $N$ was set to 100 in most calculations and 101 in \textsf{Calc. 3} of Fig.\ \ref{Fig: KW_Tight_binding}(e).
The offset $1/2$ is to avoid the time-reversal invariant momenta (TRIMs), and it was removed in \textsf{Calc. 2} of Fig.\ \ref{Fig: KW_Tight_binding}(e).

When $m$ is equal to zero, which means no magnetic field is applied and the system is invariant under the time-reversal, two bands degenerate at TRIMs [Fig.\ \ref{Fig: KW_Tight_binding}(a)].
When a weak magnetic field is applied, degenerated points move slightly from TRIMs, forming Weyl points [Fig.\ \ref{Fig: KW_Tight_binding}(b)].

Graphs of the integrated Berry curvature show some peak structures near (Kramers-)Weyl points when only positive or negative components are summarized [light blue and pink curves in Figs.\ \ref{Fig: KW_Tight_binding}(c), (d)].
However, when $m$ is equal to zero, the positive and negative components are completely canceled because of the time-reversal symmetry [Fig.\ \ref{Fig: KW_Tight_binding}(c)].
The integration of all components becomes nonzero when $m$ is nonzero, as shown in Fig.\ \ref{Fig: KW_Tight_binding}(d).
We performed these integral calculations by using three different $\mathbf{k}$ meshes and confirmed that the numerical error is negligible [Fig.\ \ref{Fig: KW_Tight_binding}(e)].

